\documentclass[reqno,10pt,letterpaper]{amsart}

\usepackage{lipsum}
\usepackage{amsmath}
\usepackage{amssymb}
\usepackage{amsthm}
\usepackage{mathrsfs}
\usepackage{accents}
\usepackage{calc}
\usepackage{arydshln}
\usepackage{upgreek}
\usepackage{slashed}
\usepackage{xifthen}
\usepackage{graphicx}
\usepackage[inline]{enumitem}
\usepackage{stmaryrd}

\usepackage{xcolor}
\definecolor{winered}{rgb}{0.6,0,0}
\definecolor{lessblue}{rgb}{0,0,0.7}

\usepackage[pdftex,colorlinks=true,linkcolor=winered,citecolor=lessblue,urlcolor=lessblue,breaklinks=true,bookmarksopen=true]{hyperref}

\makeatletter
\newcommand{\myitem}[2]{\item[\rm(#2)]\def\@currentlabel{#2}\label{#1}}
\makeatother

\usepackage{titletoc}

\makeatletter

\def\@tocline#1#2#3#4#5#6#7{
\begingroup
  \par
    \parindent\z@ \leftskip#3 \relax \advance\leftskip\@tempdima\relax
                  \rightskip\@pnumwidth plus 4em \parfillskip-\@pnumwidth
    \ifcase #1 % sections
       \vskip 0.6em \hskip 0em % add a little vspace before
       \or
       \or \hskip 0em % subsections
       \or \hskip 1em % subsubsections
    \fi%
    #6
    \nobreak\relax{\leavevmode\leaders\hbox{\,.}\hfill}
    \hbox to\@pnumwidth {\@tocpagenum{#7}}
  \par
\endgroup
}

 \def\l@section{\@tocline{0}{0pt}{0pc}{}{}}

\renewcommand{\tocsection}[3]{%
  \indentlabel{\@ifnotempty{#2}{ % for numbered sections
    \ignorespaces\bfseries{#2. #3}}}
  \indentlabel{\@ifempty{#2}{\ignorespaces\bfseries{#3}}{}} % for unnumbered sections
    \vspace{1.5pt}}

\renewcommand{\tocsubsection}[3]{%
  \indentlabel{\@ifnotempty{#2}{
    \ignorespaces#2. #3}}
  \indentlabel{\@ifempty{#2}{\ignorespaces #3}{}}
    \vspace{1.5pt}}

\renewcommand{\tocsubsubsection}[3]{%
  \indentlabel{\@ifnotempty{#2}{
    \ignorespaces#2. #3}}
  \indentlabel{\@ifempty{#2}{\ignorespaces #3}{}}
    \vspace{1.5pt}}

\makeatother

\makeatletter
\def\@nomenstarted{0}
\newlength{\@nomenoldtabcolsep}

\newcommand{\nomenstart}
  {%
    \def\@nomenstarted{1}%
    \setlength{\@nomenoldtabcolsep}{\tabcolsep}%
    \setlength{\tabcolsep}{3.5pt}%
    \begin{longtable}{p{0.11\textwidth} p{0.86\textwidth}}%found by hand
  }

\newcommand{\nomenitem}[2]{%
    \ifcase\@nomenstarted%
      \or % if nomenstarted=1, do nothing
      \or \\ % if nomenstarted=2, add newline to previous one
    \fi%
    #1\,{\leavevmode\leaders\hbox{\,.}\hfill} & #2%
    \def\@nomenstarted{2}%
  }%
\newcommand{\nomenend}
  {\\%
      \end{longtable}%
      \setlength{\tabcolsep}{\@nomenoldtabcolsep}%
      \def\@nomenstarted{0}%
  }
\makeatother

\makeatletter
\newcommand{\BIG}{\bBigg@{3.5}}
\newcommand{\vast}{\bBigg@{4}}
\newcommand{\Vast}{\bBigg@{5}}
\newcommand{\VAST}[1]{\bBigg@{#1}}
\makeatother

\allowdisplaybreaks

\numberwithin{equation}{section}
\numberwithin{figure}{section}
\newtheorem{thm}{Theorem}[section]

\newtheorem{prop}[thm]{Proposition}
\newtheorem{lemma}[thm]{Lemma}

\newtheorem{conj}[thm]{Conjecture}
\newtheorem{prob}[thm]{Problem}

\newtheorem*{thm*}{Theorem}
\newtheorem*{prop*}{Proposition}
\newtheorem*{cor*}{Corollary}
\newtheorem*{conj*}{Conjecture}

\theoremstyle{definition}
\newtheorem{definition}[thm]{Definition}

\theoremstyle{remark}
\newtheorem{rmk}[thm]{Remark}

\makeatletter
\newcommand{\fakephantomsection}{%
  \Hy@MakeCurrentHref{\@currenvir.\the\Hy@linkcounter}
  \Hy@raisedlink{\hyper@anchorstart{\@currentHref}\hyper@anchorend}%
  \Hy@GlobalStepCount\Hy@linkcounter%
}
\makeatother

\newcommand{\mc}{\mathcal}
\newcommand{\cA}{\mc A}

\newcommand{\cC}{\mc C}
\newcommand{\cD}{\mc D}

\newcommand{\cF}{\mc F}

\newcommand{\cL}{\mc L}
\newcommand{\cM}{\mc M}

\newcommand{\cO}{\mc O}

\newcommand{\cV}{\mc V}

\newcommand{\cX}{\mc X}

\newcommand{\C}{\mathbb{C}}
\newcommand{\N}{\mathbb{N}}
\newcommand{\R}{\mathbb{R}}

\newcommand{\Sph}{\mathbb{S}}

\newcommand{\fm}{\mathfrak{m}}
\newcommand{\fp}{\mathfrak{p}}

\newcommand{\ft}{\mathfrak{t}}

\newcommand{\slg}{\slashed{g}{}}

\newcommand{\slDelta}{\slashed{\Delta}{}}
\newcommand{\slnabla}{\slashed{\nabla}{}}

\newcommand{\ran}{\operatorname{ran}}

\renewcommand{\Re}{\operatorname{Re}}
\renewcommand{\Im}{\operatorname{Im}}

\newcommand{\mathspan}{\operatorname{span}}
\newcommand{\supp}{\operatorname{supp}}

\newcommand{\ext}{{\rm ext}}
\newcommand{\aug}{{\rm aug}}

\newcommand{\QNM}{{\mathrm{QNM}}}

\newcommand{\AdS}{{\mathrm{AdS}}}

\newcommand{\NH}{{\mathrm{NH}}}

\newcommand{\eps}{\epsilon}

\newcommand{\hra}{\hookrightarrow}
\newcommand{\la}{\langle}

\newcommand{\ol}{\overline}
\newcommand{\pa}{\partial}
\newcommand{\dd}{{\mathrm d}}
\newcommand{\ra}{\rangle}

\newcommand{\wh}{\widehat}
\newcommand{\wt}{\widetilde}
\newcommand{\xra}{\xrightarrow}

\newcommand{\pfstep}[1]{$\bullet$\ \underline{\textit{#1}}}

\newcommand{\bop}{{\mathrm{b}}}

\newcommand{\qop}{{\mathrm{q}}}

\newcommand{\res}{{\mathrm{res}}}

\newcommand{\cp}{{\mathrm{c}}}

\newcommand{\Diff}{\mathrm{Diff}}

\newcommand{\Vb}{\cV_\bop}

\newcommand{\Vq}{\cV_\qop}

\newcommand{\Diffb}{\Diff_\bop}

\newcommand{\Diffq}{\Diff_\qop}

\newcommand{\loc}{{\mathrm{loc}}}
\newcommand{\CI}{\cC^\infty}

\newcommand{\CIc}{\cC^\infty_\cp}

\newcommand{\Hb}{H_{\bop}}

\newcommand{\Hbext}{\bar H_{\bop}}

\newcommand{\Hbsupp}{\dot H_{\bop}}

\newcommand{\Ric}{\mathrm{Ric}}

\newcommand{\bhm}{\fm}

\newcommand{\openbigpmatrix}[1]
  {%
    \def\@bigpmatrixsize{#1}%
    \addtolength{\arraycolsep}{-#1}%
    \begin{pmatrix}%
  }
\newcommand{\closebigpmatrix}
  {%
    \end{pmatrix}%
    \addtolength{\arraycolsep}{\@bigpmatrixsize}%
  }

\newlength{\enummargin}

\newcommand{\usref}[1]{{\upshape\ref{#1}}}

\DeclareGraphicsExtensions{.mps}
\newcommand{\inclfig}[1]{\includegraphics{#1-mps.pdf}}

\makeatletter
\newcommand*{\fwbw}[1]{\expandafter\@fwbw\csname c@#1\endcsname}
\newcommand*{\@fwbw}[1]{\ifcase #1 \or {\rm fw}\or {\rm bw}\fi}
\AddEnumerateCounter{\fwbw}{\@fwbw}
\makeatother

\begin{document}

%%%%%%%%%%%%%%%%%%%%%%%%%%%%%%%%%%%%%%%%%%%%%%%%%%%%%%%%%%%%%%%%%%%%%%
% title page
\title[QNMs of near-extremal RNdS]{Quasinormal modes of near-extremal Reissner--Nordstr\"om--de~Sitter spacetimes}

\date{\today}

\begin{abstract}
  We study quasinormal modes (QNMs) for the Klein--Gordon equation on Reissner--Nordstr\"om--de~Sitter black holes with near-extremal charge. We locate all QNMs of size $\cO(\kappa_{\rm C})$ where $\kappa_{\rm C}$ is the surface gravity of the Cauchy horizon (which vanishes at extremality): they are well-approximated by $\kappa_{\rm C}$ times QNMs of the near-horizon geometry $\AdS^2\times\Sph^2$ of the extremal limit.
\end{abstract}

% 83C57: black holes
% 35P20: asymptotic distribution of eigenvalues and eigenfunctions
% 35L05: wave equation
% 35B40: asymptotic behavior of solutions

\author{Peter Hintz}
\address{Department of Mathematics, ETH Z\"urich, R\"amistrasse 101, 8092 Z\"urich, Switzerland}
\email{peter.hintz@math.ethz.ch}

\maketitle

%%%%%%%%%%%%%%%%%%%%%%%%%%%%%%%%%%%%%%%%%%%%%%%%%%%%%%%%%%%%%%%%%%%%%%
\section{Introduction}
\label{SI}

\subsection{Setup and main result.} 
\begin{figure}[!ht]
\centering
\inclfig{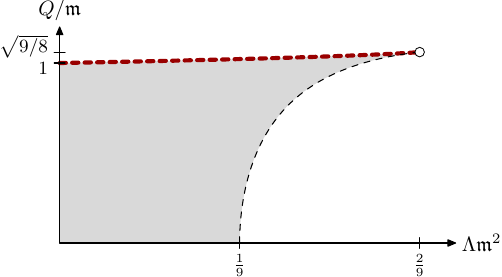}
\caption{Parameter space of subextremal RNdS black holes (computed using \cite[Proposition~3.2]{HintzKNdSStability}). At the thick dashed line at the top, the charge is extremal but the mass is not; thus $r_{\rm C}=r_{\rm e}<r_{\rm c}$. (We exclude the circle on the top right, where $r_{\rm C}=r_{\rm e}=r_{\rm c}$.) We study RNdS black holes with parameters in a small neighborhood of this dashed line.}
\label{FigIParam}
\end{figure}

\begin{figure}[!ht]
\centering
\inclfig{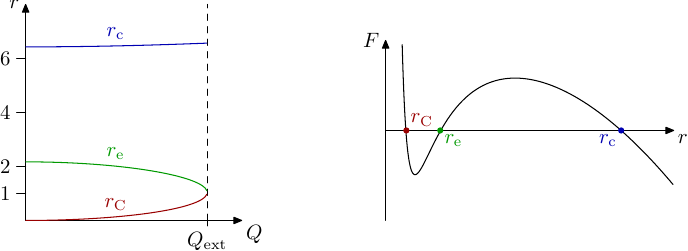}
\caption{\textit{On the left:} the radii $r_{\rm C},r_{\rm e},r_{\rm c}$ as functions of the charge $Q\in[0,Q_\ext]$ for $\Lambda=0.05$, $\bhm=1$ where $Q_\ext\approx 1.00893$ is the extremal charge. \textit{On the right:} the graph of $F$ for the near-extremal parameters $\Lambda=0.05$, $\bhm=1$, $Q=0.9$.}
\label{FigIParam2}
\end{figure}

The Reissner--Nordstr\"om--de~Sitter (RNdS) solution of the Einstein--Maxwell equations, with cosmological constant $\Lambda>0$, describes a spherically symmetric black hole with mass $\bhm>0$ and charge $Q$. The underlying geometry is described by the Lorentzian manifold $(\cM,g)$ where
\begin{equation}
\label{EqIMetric}
  \cM=\R_t \times \cX,\quad \cX=(r_{\rm e},r_{\rm c})_r \times \Sph^2,\quad g=-F(r)\,\dd t^2 + F(r)^{-1}\,\dd r^2 + r^2\slg;
\end{equation}
here $\slg$ is the standard metric on the unit 2-sphere, and $0<r_{\rm e}<r_{\rm c}$ are the largest two roots of the function
\begin{equation}
\label{EqIF}
  F(r) = 1-\frac{2\bhm}{r}+\frac{Q^2}{r^2} - \frac{\Lambda r^2}{3}.
\end{equation}
We assume here that the parameters $\Lambda,\bhm,Q$ are \emph{subextremal}. In the case $Q\neq 0$ this means that $F$ has three distinct positive roots
\begin{equation}
\label{EqIRadii}
  0 < r_{\rm C} < r_{\rm e} < r_{\rm c}
\end{equation}
which are, in this order, the area radius of the Cauchy, event, and cosmological horizon; see the right panel of Figure~\ref{FigIParam2}. (For $Q=0$, there is no Cauchy horizon.) See Figure~\ref{FigIParam} for the parameter space of subextremal RNdS black holes, parameterized using the dimensionless quantities $\Lambda\bhm^2$ and $Q/\bhm$. In this paper we are interested in black holes which have near-extremal charges (but not near-extremal masses). This corresponds to the relationship $r_{\rm C}\approx r_{\rm e}<r_{\rm c}$; see the left panel of Figure~\ref{FigIParam2}. From now on, we parameterize subextremal RNdS black holes using the radii~\eqref{EqIRadii}.

The coordinate singularities of $g$ at $r=r_{\rm e},r_{\rm c}$ can be removed by passing to a new time coordinate
\begin{subequations}
\begin{equation}
\label{EqItstar}
  t_*=t-T(r),\quad T'(r)=\frac{\tilde T(r)}{F(r)}, \quad \tilde T(r):=2\frac{r-r_{\rm e}}{r_{\rm c}-r_{\rm e}}.
\end{equation}
The level sets of $t_*$ are transversal to the future event and cosmological horizon. (The key feature is that $\tilde T(r_{\rm e})=-1$, $\tilde T(r_{\rm c})=+1$.) In the coordinates $t_*,r$ then, the metric
\begin{equation}
\label{EqItstarMet}
  g = -F(r)\,\dd t_*^2 - 2\tilde T(r)\,\dd t_*\,\dd r + \frac{1-\tilde T(r)^2}{F(r)}\,\dd r^2 + r^2\slg
\end{equation}
\end{subequations}
extends real analytically to
\[
  M = \R_{t_*} \times X,\quad X := [r_-,r_+]\times\Sph^2,
\]
where we set $r_-=\frac{r_{\rm C}+r_{\rm e}}{2}$ and fix any $r_+>r_{\rm c}$. (The analytic continuation exists for $r\in(r_{\rm C},\infty)$.)

In this paper, we study the set
\[
  \QNM(r_{\rm C},r_{\rm e},r_{\rm c}) \subset \C
\]
of \emph{quasinormal modes} (QNMs) (or \emph{resonances}) for the wave equation
\begin{equation}
\label{EqIBox}
  \Box_g\psi = 0
\end{equation}
on nearly extremally charged RNdS backgrounds. The set $\QNM(r_{\rm C},r_{\rm e},r_{\rm c})$ consists of all complex numbers $\sigma\in\C$ for which there exists $0\neq u\in\CI(X)$ (a \emph{resonant state}) such that
\begin{equation}
\label{EqIBoxSol}
  \Box_g(e^{-i\sigma t_*}u) = 0.
\end{equation}
(Equivalently, $e^{-i\sigma t}\tilde u$ solves~\eqref{EqIBox} where $\tilde u=e^{i\sigma T(r)}u$. The smoothness of $u$ amounts to $\tilde u$ being \emph{outgoing} at the event and cosmological horizons.) Thus, $-\Im\sigma$ is the exponential rate of decay of the \emph{mode solution} $e^{-i\sigma t_*}u$. We always have
\[
  0 \in \QNM(r_{\rm C},r_{\rm e},r_{\rm c})
\]
since $\Box_g 1=0$. The set $\QNM(r_{\rm C},r_{\rm e},r_{\rm c})$ is discrete, as was shown by Besset \cite{BessetRNdSDecay} (this also follows from results in \cite{HintzKNdSStability} combined with Vasy's method \cite{VasyMicroKerrdS}). Solutions $\psi=\psi(t_*,x)$ of the wave equation~\eqref{EqIBox} with smooth initial data admit \emph{resonance} (or \emph{QNM}) \emph{expansions} of the form
\[
  \psi(t_*,x) = \sum_j e^{-i\sigma_j t_*}u_j(x) + \tilde\psi(t_*,x)
\]
(ignoring the possibility of higher multiplicities) where the $\sigma_j$ and $u_j$ are QNMs and resonant states, and $\tilde\psi$ has faster exponential decay in $t_*$ than the last term one chooses to include in the sum (sorted by the exponential rates of decay $-\Im\sigma_j$).

Fixing $r_{\rm e}<r_{\rm c}$, we show that in the extremal charge limit $r_{\rm C}\nearrow r_{\rm e}$, the set $\QNM(r_{\rm C},r_{\rm e},r_{\rm c})$ contains complex numbers $\sigma$, depending continuously on $r_{\rm C}$, whose imaginary part tends to $0$. Such families of modes are called \emph{zero-damped} \cite{YangZhangZimmermanNicholsBertiChenQNMNearXKerr,YangZimmermanZenginogluZhangBertiChenQNMNearXKerr}. More generally, we consider the set $\QNM(r_{\rm C},r_{\rm e},r_{\rm c},\mu)$ of quasinormal modes for the Klein--Gordon equation
\begin{equation}
\label{EqIKG}
  (\Box_g+\mu)\psi = 0,\quad \mu\geq 0.
\end{equation}
In our main result, we in fact determine \emph{all} QNMs of size $\cO(\kappa_{\rm C})$ where $\kappa_{\rm C}:=\frac12|F'(r_{\rm C})|$ is the \emph{surface gravity} of the Cauchy horizon. The latter is equal to\footnote{See~\eqref{EqGSurfGrav} for the calculation.} $\kappa_{\rm C}=(r_{\rm e}-r_{\rm C})\frac{\varkappa}{2 r_{\rm e}^2}+\cO((r_{\rm e}-r_{\rm C})^2)$ in the extremal charge limit where $\varkappa$ is given in~\eqref{EqIvarkappa}.

\begin{thm}[Main result, abridged version]
\label{ThmI}
  Fix $0<r_{\rm e}<r_{\rm c}$ and define the quantity
  \begin{equation}
  \label{EqIvarkappa}
    \varkappa := \frac{r_{\rm c}^2+2 r_{\rm e} r_{\rm c}-3 r_{\rm e}^2}{r_{\rm c}^2+2 r_{\rm e} r_{\rm c}+3 r_{\rm e}^2} \in (0,1).
  \end{equation}
  For $\mu\geq 0$ and $\ell\in\N_0$, define
  \[
    \lambda_\ell^+(\mu) := \frac12\biggl(1+\sqrt{1+4\frac{\ell(\ell+1)+r_{\rm e}^2\mu}{\varkappa}}\,\biggr),
  \]
  and define the set of \emph{QNMs for the massive scalar wave equation on the near-horizon geometry} by\footnote{We do not make the dependence of this set on $r_{\rm e},r_{\rm c}$ explicit in the notation.}
  \[
    \QNM_\NH(\mu) := \{ -i(\lambda_\ell^+(\mu)+n) \colon \ell,n\in\N_0 \}.
  \]
  Let $C_0>0$ with $C_0\neq\lambda_\ell^+(\mu)+n$ for all $\ell,n\in\N_0$. Then, in the Hausdorff distance sense,
  \begin{equation}
  \label{EqIConv}
  \begin{split}
    &\Bigl\{ \frac{\varsigma}{\kappa_{\rm C}} \colon \varsigma\in\QNM(r_{\rm C},r_{\rm e},r_{\rm c},\mu),\ |\varsigma|\leq C_0\kappa_{\rm C} \Bigr\} \\
    &\qquad \xra{r_{\rm C}\nearrow\,r_{\rm e}}
      \begin{cases}
        \phantom{\{0\}\cup{}}\QNM_\NH(\mu) \cap \{|\sigma|<C_0\}, & \mu>0, \\
        \{0\}\cup\QNM_\NH(\mu) \cap \{|\sigma|<C_0\}, & \mu=0.
      \end{cases}
  \end{split}
  \end{equation}
  For small $r_{\rm e}-r_{\rm C}$, the set on the left is contained in $i\R$.
\end{thm}

The proof of Theorem~\ref{ThmI} is given in \S\ref{SPf} (see in particular Proposition~\ref{PropPfPY}, Theorem~\ref{ThmPfPN}, Proposition~\ref{PropPf0N}, and Theorem~\ref{ThmPf0Y}). We establish the following more precise results.
\begin{enumerate}
\item The convergence of QNMs in~\eqref{EqIConv} holds \emph{with multiplicity}.
\item Let us restrict attention to functions (and resonant states) of the form $u(r,\omega)=u_0(r)Y_\ell(\omega)$ where $Y_\ell$ is a degree $\ell$ spherical harmonic.\footnote{Due to the spherical symmetry of the RNdS metric, one can project resonant states onto degree $\ell$ modes for any $\ell\in\N_0$.} Then the element $\sigma:=-i(\lambda_\ell^+(\mu)+n)\in\QNM_\NH(\mu)$ has multiplicity $2\ell+1$ (Theorem~\ref{ThmNHQNM}). Moreover, for $\mu>0$, and also for $\mu=0$ and $\ell\geq 1$ (see Remark~\ref{RmkPfP0ell}), the resonant state corresponding to the QNM $\approx\kappa_{\rm C}\sigma$ is well-approximated by the function~\eqref{EqNHQNMResState} (with $z=2\frac{r-r_{\rm C}}{r_{\rm e}-r_{\rm C}}-1$ and using the notation~\eqref{EqNHQNMPoly}) which is localized $\cO(r_{\rm e}-r_{\rm C})$-close to $r=r_{\rm e}$ (Theorem~\ref{ThmPfPN}). In the case $\mu=0,\ell=0$, a similar statement holds upon subtracting appropriate constants from the resonant state (Theorem~\ref{ThmPf0Y}).
\end{enumerate}

\begin{figure}[!ht]
\centering
\inclfig{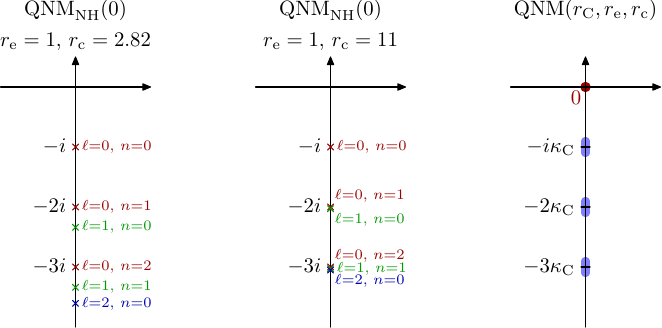}
\caption{\textit{On the left:} the set $\QNM_\NH(0)$ of near-horizon QNMs for $r_{\rm e}=1$, $r_{\rm c}=2.82$; the corresponding extremal RNdS parameters satisfy $\Lambda\bhm^2\approx 0.14$. The values of $\ell,n$ identify the QNM $-i(\lambda^+_\ell(0)+n)$. \textit{In the middle:} the set $\QNM_\NH(0)$ for $r_{\rm e}=1$, $r_{\rm c}=11$, and thus $\Lambda\bhm^2\approx 0.02$. \textit{On the right:} illustration of~\eqref{EqIConv} for $\mu=0$. The QNMs of near-extremal RNdS are equal to $\kappa_{\rm C}$ times small perturbations of the near-horizon QNMs (indicated by the blue intervals), while the red QNM $0$ is independent of the RNdS parameters. (For scalar field masses $\mu>0$, there is no QNM at $0$.)}
\label{FigI}
\end{figure}

\begin{figure}[!ht]
\centering
\inclfig{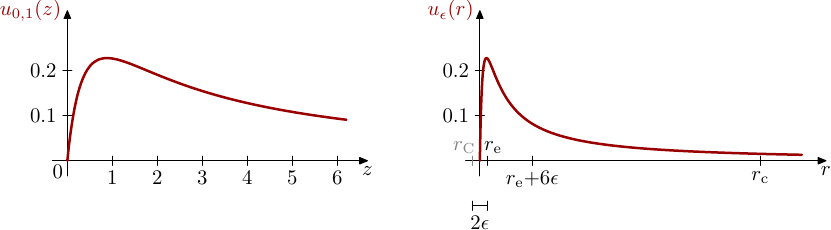}
\caption{We fix $r_{\rm e}=1$, $r_{\rm c}=2.82$, $\mu=0.1$, and consider $\ell=0$, $n=1$. \textit{On the left:} the resonant state $u_{0,1}(z)$ (see~\eqref{EqNHQNMResState}) for the massive wave equation on the near-horizon geometry corresponding to the near-horizon QNM $-i(\lambda_0^+(\mu)+1)\approx -2.138$. \textit{On the right:} illustration of the resonant state $u_\eps$ for the Klein--Gordon equation on RNdS with parameters $r_{\rm C}=r_{\rm e}-2\eps$, $\eps=0.05\ll r_{\rm e}$. We are showing here the approximation $u_{0,1}(2\frac{r-r_{\rm C}}{r_{\rm e}-r_{\rm C}}-1)$ of $u_\eps$.}
\label{FigI2}
\end{figure}

See Figures~\ref{FigI} and \ref{FigI2}. In the case of massless scalar fields ($\mu=0$), Theorem~\ref{ThmI} confirms the numerical observations regarding \emph{near-extremal (NE) QNMs} in \cite{CardosoCostaDestounisHintzJansenSCC} for the spherical harmonic degree $\ell=0$. For $\ell\geq 1$ however, our result implies that the prediction in \cite[Equation~(13)]{CardosoCostaDestounisHintzJansenSCC} that the QNMs are given by $-i(\ell+n+1)\kappa_{\rm C}$ in the extremal limit is inaccurate even to leading order in the near-extremality parameter $r_{\rm e}-r_{\rm C}$. Our results are consistent with the more precise heuristics based on matched asymptotic expansions for near-extremal Kerr--Newman--de~Sitter (KNdS) black holes in \cite[\S{3.3.2}]{DaveyDiasGilSCCKNdS}; our arguments can be regarded as providing a rigorous justification (for the RNdS sub-family of KNdS) for various approximations made there.

The existence of zero-damped modes for the Klein--Gordon equation with \emph{conformal mass} $\mu=\frac{{\rm scal}_g}{6}=\frac{2\Lambda}{3}$ was proved by Joykutty \cite{JoykuttyXZeroDamped}.\footnote{The particular choice of $\mu$ plays a key role in several places of \cite{JoykuttyXZeroDamped}: a radial inversion exchanges the almost-extremal event and the subextremal cosmological horizon \cite[\S{4.1}]{JoykuttyXZeroDamped}; and in the de~Sitter limit of a rescaling of the resulting spacetime, the dual resonant states (also called co-resonant states or co-modes) are supported on the de~Sitter horizon \cite[Proposition~2.3]{JoykuttyXZeroDamped} (see also \cite[\S{III.B}]{HintzXiedS}).} Joykutty obtained similar results on nearly extremally rotating Kerr--de~Sitter spacetimes in his thesis \cite{JoykuttyPhD}.

Since $\kappa_{\rm C}\to 0$ as $r_{\rm C}\to r_{\rm e}$, Theorem~\ref{ThmI} describes QNMs in a shrinking neighborhood of $0$: they are approximately equal to $\kappa_{\rm C}$ times QNMs of the near-horizon geometry (see below). Our interest in QNMs $\sigma$ with small $-\Im\sigma$ stems from their importance in the context of Penrose's Strong Cosmic Censorship conjecture \cite{CardosoCostaDestounisHintzJansenSCC}: the regularity of solutions of the wave or Klein--Gordon equation at the future Cauchy horizon is $H^{\frac12+\beta-}$ \cite{HintzVasyCauchyHorizon,HintzKleinQuantumSCC} (which is expected to be sharp) where $\beta=\frac{1}{\kappa_{\rm C}}\min\{-\Im\sigma\}$ where $\sigma$ runs over all nonzero QNMs. If it holds that the QNMs identified in Theorem~\ref{ThmI} are those with smallest $-\Im\sigma$ (cf.\ Conjecture~\ref{ConjIShallow} below), we conclude that $\beta\to 1$ for $\mu=0$ in the extremal charge limit. A detailed analysis of $\beta$ in the full subextremal KNdS parameter space was performed in \cite{DaveyDiasGilSCCKNdS} following the earlier \cite{CasalsMarinhoSCCRotating}. Further results on the validity or failure of SCC based on QNM considerations are described in \cite{CardosoCostaDestounisHintzJansenSCC2,MoTianWangZhangZhongSCC,DafermosShlapentokhRothmanSCC,DiasEperonReallSantosSCC,DiasReallSantosSCCrough,DiasReallSantosSCCChargeddSBH}.

Other works in the physics literature on QNMs near extremality have mainly focused on near-extremal black hole spacetimes with vanishing cosmological constant $\Lambda=0$. Hod \cite{HodNearXRNQNM} studied the QNMs of massive scalar fields on near-extremal Reissner--Nordstr\"om (RN) black holes using a number of ad hoc approximations. His formula in \cite[equations~(13) and (39)]{HodNearXRNQNM} for $q=0$ is consistent with Theorem~\ref{ThmI} (with the identification $\bhm=r_{\rm e}\equiv r_+$, $\varkappa=1$ in the extremal RN limit $r_{\rm C}\equiv r_-=r_+-2\eps\to r_+$). The results of Kim--Myung--Park \cite{KimMyungParkNearXRNQNM} on the near-horizon geometry of extremal RN are consistent as well. See \cite{ZimmermanMarkZeroDampedQNM} for results in near-extremal Kerr--Newman geometries. Further references include \cite{HodQNMNearXKN,HodQNMNearXKerrKG,HodQNMChargedRN}. We also mention the work by Ficek--Warnick \cite{FicekWarnickXRNAdSQNM} presenting a numerical study of QNMs on near-extremal RN black holes with \emph{negative} cosmological constant $\Lambda<0$; the near-extremal modes analogous to those found in Theorem~\ref{ThmI} dominate in the extremal limit (cf.\ Conjecture~\ref{ConjIShallow} below, which however concerns $\Lambda>0$).

%%%%%%%%%%%%%%%%%%%%%%%%%%%%%%%%%%%%%%%%%%%%%%%%%%
\subsection{Near-extremality, extremality, near-horizon limit}

\begin{figure}[!ht]
\centering
\inclfig{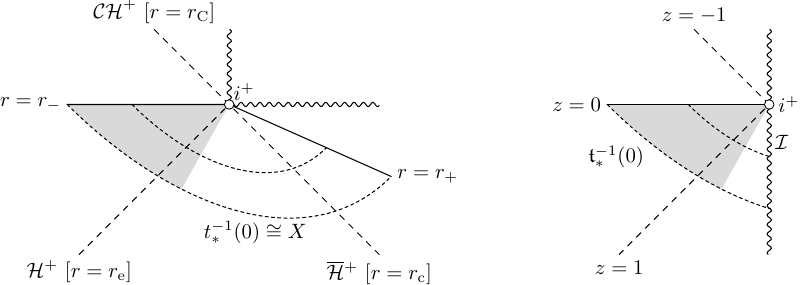}
\caption{\textit{On the left:} Penrose diagram of subextremal RNdS. \textit{On the right:} Penrose diagram of the near-horizon geometry $\AdS^2\times\Sph^2$. The level sets of the function $z$ meet at the point $i^+$ on the conformal boundary, and $t_*\to\infty$ as one approaches $i^+$. In the shaded regions on both sides, the metrics are close to being constant multiples of one another upon relating $t_*,r$ and $\ft_*,z$ as indicated after~\eqref{EqIz}.}
\label{FigINH}
\end{figure}

In order to prove Theorem~\ref{ThmI}, we recognize the extremal mass limit as being singular in the following sense. Write $g_\eps$ for the RNdS metric with parameters $r_{\rm e},r_{\rm c}$ (fixed) and $0<r_{\rm C}=r_{\rm e}-2\eps$. On the one hand, on every compact subset of $\{r>r_{\rm e}\}$ the metric $g_\eps$ converges to the extremal RNdS metric $g_0$. Near the event horizon on the other hand, let us pass to the rescaled radial coordinate
\begin{equation}
\label{EqIz}
  z := 2\frac{r-r_{\rm C}}{r_{\rm e}-r_{\rm C}} - 1 = \frac{r-r_{\rm C}}{\eps} - 1
\end{equation}
(so $z=-1$, resp.\ $z=+1$ defines the Cauchy, resp.\ event horizon); similarly introducing a rescaled time coordinate $\ft_*\sim\eps t_*$, the limit of $g_\eps$ as $\eps\to 0$ in the coordinates $\ft_*,z$ is isometric to $\AdS^2\times\Sph^2$ (for appropriate radii in the two factors), with $z=1$ being the past light cone based at a point $i^+$ on the conformal boundary. This space is (isometric to) the \emph{near-horizon geometry} of extremal RNdS \cite{CastroMarianiToldoNearXdS} (see the formula~\eqref{EqGNH} and Remark~\ref{RmkGNearHor} for more details). See Figure~\ref{FigINH}; the detailed computations are given in~\S\ref{SsGC}.

The QNMs observed in Theorem~\ref{ThmI} are then the rescalings of the QNMs of the near-horizon geometry; the corresponding resonant states are characterized as being functions in $z\geq 0$ that are smooth (in particular across the `horizon' $z=+1$) and decay as $z\to\infty$, i.e.\ towards the conformal boundary (\S\ref{SNH}). For the proof, we combine
\begin{itemize}
\item estimates at zero energy on extremal RNdS (\S\ref{S0}) and
\item estimates for the spectral family of the Klein--Gordon equation on the near-horizon geometry (\S\ref{SsNHFred})
\end{itemize}
in order to prove uniform estimates for the spectral family on near-extremal RNdS in the extremal charge limit (\S\ref{SPf}). The function spaces used for this uniform analysis are the weighted q-Sobolev spaces introduced in \cite{HintzKdSMS} which are equivalent to the function spaces (b-Sobolev spaces) appropriate for the analysis in the two asymptotic regimes. Theorem~\ref{ThmI} then follows from an application of Rouch\'e's theorem in the context of Schur's complement formula for suitably defined Grushin problems for the spectral family of the Klein--Gordon equation for the metric $g_\eps$.

\begin{rmk}[Comparison: zero mass limit]
\label{RmkI0Mass}
  The approach sketched above is, in spirit, related to \cite{HintzXieSdS,HintzKdSMS} where all shallow QNMs, now meaning $\Im\sigma\gtrsim-\sqrt\Lambda$, of Schwarzschild-- and Kerr--de~Sitter black holes are characterized in the zero-mass limit $\Lambda\bhm^2\searrow 0$: for fixed $\Lambda$, there are two geometries characterizing the zero-mass limit $\bhm\searrow 0$, namely the de~Sitter spacetime and, upon passing to $\hat r:=r/\bhm$, the unit mass Schwarzschild or Kerr spacetime. Unlike in those works, however, in the present setting we face an added difficulty: for the scalar field mass $\mu=0$, the zero energy operator on extremal RNdS fails to be invertible (since constants are mode solutions with frequency $0$). We surmount this using an idea from the low energy spectral analysis on asymptotically flat spaces as done in \cite{HintzGlueLocIII} by complementing the range of the non-surjective zero energy operator by the output of the spectral family on a singularly rescaled zero energy state; see Proposition~\ref{PropPf0Gr} and the discussion prior to it.
\end{rmk}

\begin{rmk}[QNMs of extremal RNdS]
\label{RmkIXQNM}
  Since the frequencies $\varsigma=\cO(\kappa_{\rm C})$ of interest in Theorem~\ref{ThmI} tend to $0$ in the extremal limit, we only need to study the zero frequency behavior of extremal RNdS here. In particular, we do not need to study nonzero QNMs on extremal RNdS. The analysis of QNMs on extremal RNdS with negative imaginary part is complicated by the vanishing surface gravity of the event horizon: on the level of analysis, the spectral family, near the event horizon, is akin to the spectral family of an asymptotically flat space near infinity, and thus delicate tools are necessary, such as the Gevrey analysis pioneered by Gajic--Warnick \cite{GajicWarnickXRNQNM,GajicWarnickKerrQNM} (extremal Reissner--Nordstr\"om, subextremal Kerr) or complex scaling methods as in S\'a Barreto--Zworski \cite{SaBarretoZworskiResonances} and Hitrik--Zworski \cite{HitrikZworskiQNM} (Schwarzschild) and Stucker \cite{StuckerKerrQNM} (Kerr). We conjecture that damped QNMs on extremal RNdS give rise to nearby damped QNMs of subextremal RNdS.
\end{rmk}

%%%%%%%%%%%%%%%%%%%%%%%%%%%%%%%%%%%%%%%%%%%%%%%%%%
\subsection{Related works on QNMs and resonance expansions}
\label{SsIExp}

Besset \cite{BessetRNdSDecay} adapted techniques of Bony--H\"afner \cite{BonyHaefnerDecay} and Georgescu--G\'erard--H\"afner \cite{GeorgescuGerardHafnerComplete} to obtain a complete resonance expansion for massive and weakly charged scalar waves (including across the horizons using ideas of \cite{DyatlovQNMExtended,DafermosInterior,DafermosRodnianskiRedShift}) propagating on subextremal RNdS spacetimes. Besset also developed a scattering theory and proved asymptotic completeness in \cite{BessetRNdSScattering}.

Allowing for the black hole to have nonzero angular momentum, Besset--H\"afner \cite{BessetHaefnerBomb} proved the existence of an unstable mode for weakly charged and weakly massive Klein--Gordon fields on slowly rotating KNdS spacetimes via a computation of the first order perturbation of the zero resonance in the massless and uncharged case. (See \cite{ShlapentokhRothmanBlackHoleBombs} for a related result for the Klein--Gordon equation on Kerr.) By contrast, for massless and uncharged scalar fields and in the slowly rotating setting, all QNMs except for $0$ (with resonant states being constants) have negative imaginary part (bounded away from $0$). The full nonlinear stability of slowly rotating KNdS black holes as solutions of the Einstein--Maxwell system was proved by the author in \cite{HintzKNdSStability} via an adaptation of the techniques introduced in joint work with Vasy \cite{HintzVasyKdSStability}. (Building on the earlier \cite{VasyMicroKerrdS,HintzVasySemilinear,HintzQuasilinearDS,HintzVasyQuasilinearKdS}, this work exploits information about QNMs, such as the absence of growing mode solutions, for the purpose of solving linear and nonlinear wave equations.) We also mention the work of Petersen--Vasy \cite{PetersenVasySubextremal} on partial expansions in the full subextremal range of Kerr--de~Sitter black holes, and \cite{GalkowskiZworskiHypo,PetersenVasyAnalytic} regarding the analyticity properties of resonant states.

Iantchenko \cite{IantchenkoRNdSDirac} studied QNMs for the massless charged Dirac equation on subextremal RNdS spacetimes, generalizing the influential earlier work by S\'a Barreto--Zworski \cite{SaBarretoZworskiResonances} on QNMs for the massless wave equation on Schwarzschild and Schwarzschild--de~Sitter spacetimes. The generalization to slowly rotating Kerr--Newman--de~Sitter (KNdS) backgrounds was done in \cite{IantchenkoKNdSDirac} following methods introduced by Dyatlov \cite{DyatlovQNM,DyatlovQNMExtended,DyatlovAsymptoticDistribution} in the Kerr--de~Sitter setting.

%%%%%%%%%%%%%%%%%%%%%%%%%%%%%%%%%%%%%%%%%%%%%%%%%%
\subsection{Outlook}

The aim of the present paper is to exhibit the mechanism through which near-horizon QNMs lift to QNMs of a near-extremal spacetime. We leave it to future work to study the following problems:

\begin{conj}[More precise asymptotics of QNMs]
\label{ConjIAsymp}
  If $\sigma_0$ is a simple QNM of the Klein--Gordon equation on the near-horizon geometry with parameters $r_{\rm e},r_{\rm c}$, then the unique nearby QNM on RNdS with $r_{\rm C}=r_{\rm e}-2\eps$ (for small $\eps>0$) depends on $\eps\in[0,r_{\rm e}/2)$ in a smooth or polyhomogeneous fashion.
\end{conj}

\begin{conj}[Shallow QNMs]
\label{ConjIShallow}
  Fix $0<r_{\rm e}<r_{\rm c}$. In the notation of Theorem~\usref{ThmI}, show that the set $\{\frac{\varsigma}{\kappa_{\rm C}}\colon\varsigma\in\QNM(r_{\rm C},r_{\rm e},r_{\rm c},\mu),\ \Im\varsigma>-C_0\kappa_{\rm C}\}$ converges to $\QNM_\NH(\mu)\cap\{\Im\sigma>-C_0\}$ as $r_{\rm C}\nearrow r_{\rm e}$.
\end{conj}

A proof of the latter conjecture would identify all QNMs in a half space including the real axis. It relates to Theorem~\ref{ThmI} in the same way that \cite{HintzKdSMS} relates to \cite{HintzXieSdS}. Finally, we mention:

\begin{prob}[Charged scalar waves]
\label{ProbICharged}
  Prove an analogue of Theorem~\usref{ThmI} for charged scalar waves and justify the numerical results of \cite{CardosoCostaDestounisHintzJansenSCC2} concerning near-extremal QNMs.
\end{prob}

\begin{prob}[Rotating black holes]
\label{PropIRotating}
  Prove analogues of Theorem~\usref{ThmI} for near-extremally charged (or near-extremally rotating) KNdS black holes.
\end{prob}

%%%%%%%%%%%%%%%%%%%%%%%%%%%%%%%%%%%%%%%%%%%%%%%%%%
\subsection{Outline}

The plan of the paper is as follows.
\begin{itemize}
\item \S\ref{SG}. We describe the geometry and the structure of the spectral family for the Klein--Gordon equation on RNdS spacetimes in the extremal charge limit: \S\ref{SsGX} for the exterior limit (extremal RNdS), \S\ref{SsGNH} for the near-horizon limit, and \S\ref{SsGC} for the combination and its relation to q-analysis \cite{HintzKdSMS}.
\item \S\ref{SNH}. We study the Klein--Gordon equation on the near-horizon geometry $\AdS^2\times\Sph^2$ and the notion of QNMs for it. In~\S\ref{SsNHSolv}, we develop the solvability and regularity theory for the Klein--Gordon equation, and in~\S\ref{SsNHFred} we prove Fredholm estimates for the spectral family.
\item \S\ref{S0}. We prove Fredholm estimates for the spectral family on extremal RNdS at zero frequency and identify resonant and co-resonant states.
\item \S\ref{SPf}. By combining the estimates from~\S\S\ref{SNH}--\ref{S0}, we prove uniform estimates for the spectral family on RNdS in the extremal charge limit: \S\S\ref{SsPfP} and \ref{SsPf0} treat the cases of massive and massless scalar waves, respectively.
\end{itemize}

%%%%%%%%%%%%%%%%%%%%%%%%%%%%%%%%%%%%%%%%%%%%%%%%%%%%%%%%%%%%%%%%%%%%%%
\section{Geometric singular analysis of the extremal charge limit}
\label{SG}

As a preparation for our analysis, we shall describe the uniform behavior of the RNdS metric and of the spectral family of the Klein--Gordon operator in the extremal charge limit.

Since the black hole charge $Q$ only enters the RNdS metric through the $Q^2$ term in~\eqref{EqIF}, we may restrict to the case $Q\geq 0$. Parameterizing subextremal RNdS parameters $\Lambda,\bhm,Q$ via the locations $0<r_{\rm C}<r_{\rm e}<r_{\rm c}$ of the horizons, the function $F$ in~\eqref{EqIF} takes the form
\begin{equation}
\label{EqGF}
  F(r) = -\frac{\Lambda}{3 r^2}(r-r_{\rm C})(r-r_{\rm e})(r-r_{\rm c})(r+r_{\rm C}+r_{\rm e}+r_{\rm c}),
\end{equation}
and comparison with~\eqref{EqIF} furthermore yields the following formulas for the RNdS parameters:
\begin{equation}
\label{EqGLambdaFromRad}
\begin{split}
  \frac{3}{\Lambda} &= (r_{\rm C}+r_{\rm e}+r_{\rm c})^2 - (r_{\rm C} r_{\rm e} + r_{\rm C} r_{\rm c} + r_{\rm e} r_{\rm c}), \\
  \frac{6\bhm}{\Lambda} &= (r_{\rm C}r_{\rm e}+r_{\rm C}r_{\rm c}+r_{\rm e}r_{\rm c})(r_{\rm C}+r_{\rm e}+r_{\rm c}) - r_{\rm C}r_{\rm e}r_{\rm c}, \\
  \frac{3 Q^2}{\Lambda} &= r_{\rm C}r_{\rm e}r_{\rm c}(r_{\rm C}+r_{\rm e}+r_{\rm c}).
\end{split}
\end{equation}
Fixing the locations
\begin{subequations}
\begin{equation}
\label{EqGrerc}
  0 < r_{\rm e} < r_{\rm c}
\end{equation}
of the event and cosmological horizons, we quantify the near-extremality using the parameter
\begin{equation}
\label{EqGeps}
  \eps := \frac{r_{\rm e} - r_{\rm C}}{2} \in [0,\eps_0),\quad \eps_0:=\frac{r_{\rm e}}{2};
\end{equation}
\end{subequations}
thus $r_{\rm C}=r_{\rm e}-2\eps$, and $\eps=0$ is the extremal case. We denote the function $F$ for these radii by $F_\eps$, so the RNdS metric is given by
\begin{equation}
\label{EqGMetric}
  g_\eps = -F_\eps(r)\,\dd t^2 + F_\eps(r)^{-1}\,\dd r^2 + r^2\slg.
\end{equation}
Since
\[
  \frac{3}{\Lambda} \equiv (r_{\rm c}+2 r_{\rm e})^2 - (r_{\rm e}^2 + 2 r_{\rm e} r_{\rm c}) \equiv r_{\rm c}^2 + 2 r_{\rm e} r_{\rm c} + 3 r_{\rm e}^2 \bmod \eps\CI([0,\eps_0)),
\]
the surface gravity of the Cauchy horizon is
\begin{equation}
\label{EqGSurfGrav}
\begin{split}
  \kappa_{\rm C,\eps} := \frac12|F_\eps'(r_{\rm C})| &= \frac12\frac{\Lambda}{3 r_{\rm C}^2}(r_{\rm e}-r_{\rm C})(r_{\rm c}-r_{\rm C})(2 r_{\rm C}+r_{\rm e}+r_{\rm c}) \\
    &\equiv \frac{\eps\Lambda}{3 r_{\rm e}^2}(r_{\rm c}-r_{\rm e})(r_{\rm c}+3 r_{\rm e}) \\
    &\equiv \frac{\eps}{r_{\rm e}^2}\varkappa \equiv \eps\varkappa_{\rm e} \bmod \eps^2\CI([0,\eps_0)),
\end{split}
\end{equation}
where we introduce
\begin{equation}
\label{EqGSurfGrav2}
  \varkappa := \frac{r_{\rm c}^2+2 r_{\rm e} r_{\rm c}-3 r_{\rm e}^2}{r_{\rm c}^2+2 r_{\rm e} r_{\rm c}+3 r_{\rm e}^2},\quad
  \varkappa_{\rm e} := \frac{\varkappa}{r_{\rm e}^2}.
\end{equation}
It is equal to the surface gravity $\frac12|F'_\eps(r_{\rm e})|$ of the event horizon up to $\eps^2\CI$ corrections; and it vanishes in the extremal limit $\eps\searrow 0$.

We proceed to describe the two limits of the RNdS metric $g_\eps$, given by~\eqref{EqGMetric} and \eqref{EqGF}, \eqref{EqGrerc}--\eqref{EqGeps}, as $\eps\searrow 0$: the extremal RNdS limit (when $r>r_{\rm e}$ is bounded away from $r_{\rm e}$) in~\S\ref{SsGX} and the near-horizon limit (when $r$ is $\eps$-close to $r_{\rm C},r_{\rm e}$) in~\S\ref{SsGNH}. A single perspective capturing both limits is described in~\S\ref{SsGC}.

%%%%%%%%%%%%%%%%%%%%%%%%%%%%%%%%%%%%%%%%%%%%%%%%%%
\subsection{Extremal RNdS}
\label{SsGX}

In compact subsets of $\{r>r_{\rm e}\}$, the metric $g_\eps$ converges, as $\eps\searrow 0$ and in the smooth topology, to the extremal RNdS metric
\begin{equation}
\label{EqGXg0}
  g_0 = -F_0(r)\,\dd t^2 + F_0(r)^{-1}\,\dd r^2 + r^2\slg,\quad F_0(r)=-\frac{\Lambda}{3 r^2}(r-r_{\rm e})^2(r-r_{\rm c})(r+2 r_{\rm e}+r_{\rm c}),
\end{equation}
where $\Lambda$ is given by~\eqref{EqGLambdaFromRad} with $r_{\rm C}=r_{\rm e}$. (This metric is sometimes called the \emph{cold RNdS solution} \cite{RomansEinsteinMaxwellCold}.)

%%%%%%%%%%%%%%%%%%%%%%%%%%%%%%%%%%%%%%%%%%%%%%%%%%
\subsection{Near-horizon geometry}
\label{SsGNH}

Between the Cauchy and event horizons, i.e.\ for $r_{\rm C}<r<r_{\rm e}$, we can use the form~\eqref{EqIMetric} of the RNdS metric. Recall the definition $z=2\frac{r-r_{\rm C}}{r_{\rm e}-r_{\rm C}}-1=\frac{r-r_{\rm C}}{\eps}-1=\frac{r-r_{\rm e}}{\eps}+1$ from~\eqref{EqIz}. We thus have
\begin{equation}
\label{EqGNHFeps}
\begin{split}
  F_\eps(r) &\equiv \frac{\Lambda}{3 r^2}\eps^2(z^2-1)(r_{\rm c}-r_{\rm e})(r_{\rm c}+3 r_{\rm e}) \\
    &\equiv \eps^2\varkappa_{\rm e}(z^2-1) \bmod \eps^3\CI([0,\eps_0)\times\R_z).
\end{split}
\end{equation}
Since $\dd r=\eps\,\dd z$, we have
\[
  F_\eps(r)^{-1}\,\dd r^2 \equiv \frac{1}{\varkappa_{\rm e}} (z^2-1)^{-1}\,\dd z^2
\]
modulo $\eps\CI([0,\eps_0)\times\R_z)$ (times $\dd z^2$). This suggests rescaling the time coordinate via
\begin{equation}
\label{EqGNHTime}
  \ft:=\kappa_{\rm C,\eps} t
\end{equation}
since then, by~\eqref{EqGSurfGrav}, $\dd t\equiv\frac{1}{\eps\varkappa_{\rm e}}\,\dd\ft\bmod\CI$ and thus
\[
  F_\eps(r)\,\dd t^2 \equiv \frac{1}{\varkappa_{\rm e}}(z^2-1)\,\dd\ft^2.
\]
In combination, we thus have, modulo tensors with coefficients (with respect to $\dd\ft$, $\dd z$, $\slg$) of class $\eps\CI([0,\eps_0)\times\R_z)$,
\begin{equation}
\label{EqGNH}
  g_\eps \equiv g_\NH := \frac{1}{\varkappa_{\rm e}}\bigl(-(z^2-1)\,\dd\ft^2 + (z^2-1)^{-1}\,\dd z^2 + \varkappa\slg\bigr).
\end{equation}
Note that the conformal class of $g_\NH$ depends on the ratio $r_{\rm e}/r_{\rm c}$ via $\varkappa$ in~\eqref{EqGSurfGrav2}, and hence is sensitive to the value of $\Lambda$.

\begin{rmk}[$g_\NH$ and the near-horizon geometry of extremal RNdS]
\label{RmkGNearHor}
  By definition, a near-horizon geometry is attached to an extremal horizon; in the case of the extremal RNdS metric $g_0$, with $r_{\rm C}=r_{\rm e}<r_{\rm c}$, it is obtained by introducing $r=r_{\rm e}+\eps\tilde z$ and $t=\frac{\tilde t}{\eps\varkappa_{\rm e}}$ and taking the limit $\eps\searrow 0$. Since $F_0\equiv\frac{\Lambda}{3 r_{\rm e}^2}\eps^2\tilde z^2(r_{\rm c}-r_{\rm e})(3 r_{\rm e}+r_{\rm c})\equiv\eps^2\varkappa_{\rm e}\tilde z^2\bmod\eps^3\CI$, this produces the metric
  \[
    \tilde g_\NH := \frac{1}{\varkappa_{\rm e}}\bigl( -\tilde z^2\,\dd\tilde t^2 + \tilde z^{-2}\,\dd\tilde z^2 + \varkappa\slg \bigr).
  \]
  This differs from~\eqref{EqGNH} in that $\tilde z^2$ (arising due to the extremality of the event horizon) replaces $z^2-1$ (arising from taking a limit along subextremal RNdS parameters such that the Cauchy and event horizon remain separated). Nonetheless, $\tilde g_\NH$ and $g_\NH$ are isometric: for $\tilde w:=\tilde z^{-1}$, we have $\tilde g_\NH=\frac{1}{\varkappa_{\rm e}}(\frac{-\dd\tilde t^2+\dd\tilde w^2}{\tilde w^2}+\varkappa\slg)$, which matches the expression~\eqref{EqNHCoord} for $g_\NH$ below upon identifying $(\tilde t,\tilde w)=(T,\rho)$.
\end{rmk}

\begin{rmk}[$g_\NH$ and the Einstein--Maxwell equations]
\label{RmkGEinMax}
  The RNdS metric $g_\eps$ solves the Einstein--Maxwell system
  \[
    \Ric(g_\eps) - \Lambda g_\eps = 2 T(g_\eps,\cF_\eps),\quad T(g,\cF)_{\mu\nu}:=\cF_{\mu\lambda}\cF_\nu{}^\lambda - \frac14 \cF_{\kappa\lambda}\cF^{\kappa\lambda}g_{\mu\nu},
  \]
  with electromagnetic 2-form $\cF_\eps=\dd\cA_\eps$, $\cA_\eps:=\frac{Q}{r}\,\dd t$, so $\cF_\eps=\frac{Q}{r^2}\,\dd t\wedge\dd r\equiv\frac{Q}{\varkappa}\,\dd\ft\wedge\dd z\bmod\eps\CI$. The $\eps\searrow 0$ limits $g_\NH$ and $\frac{Q}{\varkappa}\,\dd\ft\wedge\dd z$ (with $Q^2=\frac{r_{\rm e}^2 r_{\rm c}(2 r_{\rm e}+r_{\rm c})}{r_{\rm c}^2+2 r_{\rm e} r_{\rm c}+3 r_{\rm e}^2}$ being the square of the extremal charge) solve the Einstein--Maxwell system---as one can, of course, also verify by direct computation.
\end{rmk}

%%%%%%%%%%%%%%%%%%%%%%%%%%%%%%%%%%%%%%%%%%%%%%%%%%
\subsection{Combination via geometric singular analysis}
\label{SsGC}

In order to capture the uniform behavior of $g_\eps$ near the event horizon, we pass to regular coordinates there. We shall do this as in~\eqref{EqItstar}--\eqref{EqItstarMet} but now, for notational simplicity, using a function $\tilde T$ (independent of $\eps$) which equals $-1$ for $r<r_{\rm e}+\delta$ and $1$ for $r>r_{\rm c}-\delta$ where we fix $\delta:=\frac{r_{\rm c}-r_{\rm e}}{4}$; in particular,
\begin{equation}
\label{EqGCgeps}
  g_\eps = -F_\eps(r)\,\dd t_*^2 + 2\,\dd t_*\,\dd r + r^2\slg,\quad r<r_{\rm e}+\delta.
\end{equation}
We shall consider the metric $g_\eps$ in the region $r_{\rm e}-\eps\leq r\leq r_+$ for any fixed $r_+>r_{\rm c}$; the lower bound on $r$ corresponds to $z=\frac{r-r_{\rm C}}{\eps}-1\geq 0$.

We have
\[
  g^{-1} = -\frac{1-\tilde T(r)^2}{\tilde F_\eps(r)}\pa_{t_*}\otimes\pa_{t_*} - \tilde T(r)(\pa_{t_*}\otimes\pa_r + \pa_r\otimes\pa_{t_*}) + F_\eps(r)\pa_r\otimes\pa_r + r^{-2}\slg^{-1}.
\]
Therefore, writing $\slDelta=\Delta_\slg$ for the (non-negative) spherical Laplacian,
\begin{align*}
  P_\eps &:= \Box_{g_\eps} + \mu \\
    &= -\frac{1-\tilde T^2}{F_\eps}D_{t_*}^2 - D_{t_*}(r^{-2}D_r r^2\tilde T + \tilde T D_r) + r^{-2}D_r F_\eps r^2 D_r + r^{-2}\slDelta + \mu \\
    &= 2 r^{-1}D_r r D_{t_*} + r^{-2} D_r F_\eps r^2 D_r + r^{-2}\slDelta + \mu \qquad\text{for $r<r_{\rm e}+\delta$}.
\end{align*}
Being interested in resonances of size $\cO(\kappa_{\rm C,\eps})$ as $\eps\searrow 0$, we consider the spectral family $\wh{P_\eps}(\varsigma)$ of $P_\eps$ at frequencies $\varsigma=\kappa_{\rm C,\eps}\sigma$ where $\sigma=\cO(1)$; the operator $\wh{P_\eps}(\varsigma)$ is given by $e^{i\varsigma t_*}P_\eps e^{-i\varsigma t_*}$ acting on $t_*$-independent functions, so
\begin{equation}
\label{EqGCSpecFam}
\begin{split}
  \wh{P_\eps}(\varsigma) &= \wh{P_\eps}(\kappa_{\rm C,\eps}\sigma) \\
    &= -\frac{1-\tilde T^2}{F_\eps}(\kappa_{\rm C,\eps})^2\sigma^2 + \kappa_{\rm C,\eps}\sigma(r^{-2}D_r r^2\tilde T+\tilde T D_r) + r^{-2}D_r F_\eps r^2 D_r + r^{-2}\slDelta + \mu \\
    &= -2\kappa_{\rm C,\eps}\sigma r^{-1}D_r r + r^{-2}D_r F_\eps r^2 D_r + r^{-2}\slDelta + \mu\qquad\text{for $r<r_{\rm e}+\delta$}.
\end{split}
\end{equation}
Taking the limit $\eps\searrow 0$ (thus $\kappa_{\rm C,\eps}\searrow 0$) for $r>r_{\rm e}$ gives the spectral family
\begin{equation}
\label{EqGCSpecFam0}
  \wh{P_\ext}(0) = r^{-2}D_r F_0 r^2 D_r + r^{-2}\slDelta + \mu
\end{equation}
of the Klein--Gordon equation on extremal RNdS at frequency $0$. On the other hand, writing $r=r_{\rm e}+\eps(z-1)$ and recalling~\eqref{EqGSurfGrav} and \eqref{EqGNHFeps}, the limit $\eps\searrow 0$ for bounded $z$ yields the spectral family
\begin{equation}
\label{EqGCSpecFamNH}
  \wh{P_\NH}(\sigma) = \varkappa_{\rm e} \bigl( -2\sigma D_z + D_z(z^2-1)D_z + \varkappa^{-1}\slDelta \bigr) + \mu,
\end{equation}
of the Klein--Gordon operator $P_\NH=\Box_{g_\NH}+\mu$ on the near-horizon geometry
\begin{equation}
\label{EqGCgNH}
  g_\NH = \frac{1}{\varkappa_{\rm e}}\bigl( -(z^2-1)\,\dd\ft_*^2 + 2\,\dd\ft_*\,\dd z \bigr) + r_{\rm e}^2\slg
\end{equation}
at frequency $\sigma$ (relative to $\ft_*$); this metric is the $\eps\searrow 0$ limit of~\eqref{EqGCgeps} for bounded $z$ upon setting $\ft_*:=\kappa_{\rm C,\eps} t_*$ (and thus equal to~\eqref{EqGNH} via $\dd\ft_*=\dd\ft+\frac{\dd z}{z^2-1}$). The right panel of Figure~\ref{FigINH} illustrates $g_\NH$ (up to the minor inaccuracy that the level sets of $\ft_*$ as defined presently are null).

In order to combine the two scales, we now introduce:

\begin{definition}[Total space]
\label{DefGCTot}
  Fix $r_+>r_{\rm c}$. We then define $X:=[r_{\rm e},r_+]\times\Sph^2$ and the \emph{total space}
  \[
    \wt X = \bigl[ \, \{(\eps,r,\omega) \colon \eps\in[0,\eps_0),\ r_{\rm e}-\eps\leq r\leq r_+,\ \omega\in\Sph^2 \}; \{0\}\times\{r_{\rm e}\}\times\Sph^2\,\bigr]
  \]
  where $[M;N]$ denotes the real blow-up of the smooth submanifold $N\subset M$ \cite{MelroseDiffOnMwc}. We write $X_\NH$ for the front face and $X_\ext$ for the lift of $\{0\}\times X$. The manifold interior of $\wt X$ is denoted $\wt X^\circ$.
\end{definition}

Concretely, $\wt X$ is a manifold with corners which can be covered with the following three sets of coordinates (omitting the $\Sph^2$ factor and not making the ranges of the coordinate functions explicit):
\begin{alignat}{3}
\label{EqGCCoord1}
  \eps&\geq 0,&\ r&\in(r_{\rm e},r_+]; \\
\label{EqGCCoord2}
  \eps&\geq 0,&\ z&\geq 0,  && \text{related to~\eqref{EqGCCoord1} via $z=\frac{r-r_{\rm e}}{\eps}+1$}; \\
\label{EqGCCoord3}
  x&\geq 0,&\ \rho&\geq 0, && \text{related to~\eqref{EqGCCoord1} via $x=r-r_{\rm e}$, $\rho=\frac{\eps}{r-r_{\rm e}}$}, \\
    &&&&& \text{\hspace{1.05em} and to~\eqref{EqGCCoord2} via $x=\eps(z-1)$, $\rho=(z-1)^{-1}$}. \nonumber
\end{alignat}

\begin{figure}[!ht]
\centering
\inclfig{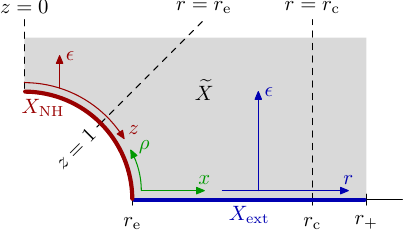}
\caption{The total space $\wt X$ for the spectral analysis of $\wh{P_\eps}(\varsigma)$ and its two boundary hypersurfaces $X_\NH$ (which carries $\wh{P_\NH}(\sigma)$ from~\eqref{EqGCSpecFamNH}) and $X_\ext$ (which carries $\wh{P_\ext}(0)$ from~\eqref{EqGCSpecFam0}). The local coordinates are defined in~\eqref{EqGCCoord1}--\eqref{EqGCCoord3}.}
\label{FigGCTotal}
\end{figure}

Thus, $X_\ext=[r_{\rm e},r_+]\times\Sph^2$ in the coordinates~\eqref{EqGCCoord1}, while $X_\NH$ is the compactification $[0,\infty]_z\times\Sph^2$ in the coordinates~\eqref{EqGCCoord2} where $[0,\infty]:=([0,\infty)_z\sqcup[0,\infty)_w)/\sim$, $z\sim w^{-1}$. See Figure~\ref{FigGCTotal}.

We recall from \cite[Definition~2.3]{HintzKdSMS} (with slightly different notation):

\begin{definition}[q-vector fields on the total space]
\label{DefGCVq}
  The space $\Vq(\wt X)$ of \emph{q-vector fields on $\wt X$} consists of all smooth vector fields $\wt V$ on $\wt X$ with $\wt V\eps=0$, i.e.\ $\wt V$ is tangent to the level sets of $\eps$ (and thus in particular to the boundary hypersurfaces $X_\NH$ and $X_\ext$ of $\wt X$).
\end{definition}

In the coordinates~\eqref{EqGCCoord1}, q-vector fields are thus linear combinations of $\pa_r$ and spherical vector fields with $\CI(\wt X)$-coefficients; in the coordinates~\eqref{EqGCCoord2} one uses $\pa_z=\eps\pa_r$, and in the coordinates~\eqref{EqGCCoord3} $x\pa_x-\rho\pa_\rho=(r-r_{\rm e})\pa_r=(z-1)\pa_z$. Globally on $\wt X$, we thus see that $\Vq(\wt X)$ is spanned, as a left $\CI(\wt X)$-module, by
\begin{equation}
\label{EqGCVqSpan}
  (r-r_{\rm C})\pa_r=(z+1)\pa_z
\end{equation}
and spherical vector fields.

Due to the tangency of q-vector fields to $X_\NH$ and $X_\ext$, one can restrict them to $X_\NH$ and $X_\ext$. Denote by $\Vb(X_\NH)$, resp.\ $\Vb(X_\ext)$ the space of smooth vector fields on $X_\NH$, resp.\ $X_\ext$ which are tangent to the boundary $z^{-1}=0$, resp.\ $r=r_{\rm e}$. (This space is spanned by $(z+1)\pa_z$, resp.\ $(r-r_{\rm e})\pa_r$ and spherical vector fields.) We thus obtain (surjective) restriction maps $\Vq(\wt X)\to\Vb(X_\NH),\Vb(X_\ext)$. We write $\Diffb^m(X_\NH)$ for the space of up to $m$-fold compositions of elements of $\Vb(X_\NH)$ (for $m=0$: multiplication by an element of $\CI(X_\NH)$), analogously for $\Diffb^m(X_\ext)$.

\begin{definition}[q-differential operators]
\label{DefGCDiffq}
  For $m\in\N_0$, we denote by $\Diffq^m(\wt X)$ the space of up to $m$-fold compositions of elements of $\Vq(\wt X)$ (for $m=0$: multiplication operators by elements of $\CI(\wt X)$). For $\wt P\in\Diffq^m(\wt X)$, we write $N_\NH(\wt P)\in\Diffb^m(X_\NH)$ and $N_\ext(\wt P)\in\Diffb^m(X_\ext)$ for its \emph{normal operators}, defined as the restrictions of $\wt P$ to $X_\NH$ and $X_\ext$, respectively.
\end{definition}

\begin{lemma}[Total spectral family]
\label{LemmaGCTot}
  Let $\sigma\in\C$ and define $\wt P\in\Diff^2(\wt X^\circ)$ by $\wh{P_\eps}(\kappa_{\rm C,\eps}\sigma)$ on the $\eps$-level sets of $\wt X$, $\eps\in(0,\eps_0)$. Then $\wt P$ extends to $\wt X$ as an element
  \[
    \wt P\in\Diffq^2(\wt X).
  \]
  The normal operators of $\wt P$ are
  \[
    N_\NH(\wt P) = \wh{P_\NH}(\sigma),\quad
    N_\ext(\wt P) = \wh{P_\ext}(0).
  \]
\end{lemma}
\begin{proof}
  We use the discussion around~\eqref{EqGCVqSpan}. The first term in~\eqref{EqGCSpecFam} equals
  \[
    -2\sigma\frac{\kappa_{\rm C,\eps}}{\eps}\frac{\eps}{r-r_{\rm C}}\Bigl((r-r_{\rm C})D_r - i\frac{r-r_{\rm C}}{r}\Bigr).
  \]
  Since $\frac{\kappa_{\rm C,\eps}}{\eps}$ and $\frac{\eps}{r-r_{\rm C}}=\frac{1}{z+1}=\frac{\rho}{1+2\rho}$ define elements of $\CI(\wt X)$, this lies in $\Diffq^1(\wt X)$. Similarly, using that $(z+1)^{-2}\eps^{-2}F_\eps\in\CI(\wt X)$ by~\eqref{EqGNHFeps}, one sees that the second term in~\eqref{EqGCSpecFam} lies in $\Diffq^2(\wt X)$. Lastly, $r^{-2}\slDelta\in\Diffq^2(\wt X)$ and $\mu\in\Diffq^0(\wt X)$. This shows $\wt P\in\Diffq^2(\wt X)$. The normal operators of $\wt P$ were already determined in~\eqref{EqGCSpecFam0}--\eqref{EqGCSpecFamNH}.
\end{proof}

The detailed analysis of $\wh{P_\NH}(\sigma)$ and $\wh{P_\ext}(0)$ is the subject of~\S\ref{SNH} and \S\ref{S0}, respectively.

%%%%%%%%%%%%%%%%%%%%%%%%%%%%%%%%%%%%%%%%%%%%%%%%%%%%%%%%%%%%%%%%%%%%%%
\section{Massive waves on the near-horizon geometry}
\label{SNH}

We study the operator $\wh{P_\NH}(\sigma)$, defined in~\eqref{EqGCSpecFamNH}, on the manifold $X_\NH=[0,\infty]_z\times\Sph^2$. Note that $\wh{P_\NH}(\sigma)$ is elliptic for $z>1$, hyperbolic for $z\in[0,1)$, and the transition between the two regimes at $z=1$ is qualitatively the same as for the spectral family of the Klein--Gordon operator on de~Sitter space near the cosmological horizon \cite{VasyMicroKerrdS,ZworskiRevisitVasy,HintzMicro}. A novel feature compared to the references is that we must analyze $\wh{P_\NH}(\sigma)$ also in the asymptotic regime $z\to\infty$. In terms of $w:=z^{-1}$, we have
\[
  \wh{P_\NH}(\sigma) = \varkappa_{\rm e} \bigl( 2 \sigma w\cdot w D_w + w^2 D_w(1-w^2)D_w + \varkappa^{-1}\slDelta\bigr) + \mu;
\]
this shows explicitly that $\wh{P_\NH}(\sigma)\in\Diffb^2(X_\NH)$, i.e.\ $\wh{P_\NH}(\sigma)$ is a b-differential operator on $X_\NH$ (cf.\ Lemma~\ref{LemmaGCTot}), and indeed it is elliptic as such for $w<1$. Its b-normal operator at $w=0$, obtained by freezing coefficients at $w=0$, is independent of $\sigma$ and given by
\begin{equation}
\label{EqNHNb}
  N_\bop(P_\NH) := \varkappa_{\rm e}(w^2 D_w^2+\varkappa^{-1}\slDelta) + \mu \in \Diffb^2([0,\infty)_w\times\Sph^2).
\end{equation}
The asymptotic behavior of elements in the nullspace of $\wh{P_\NH}(\sigma)$ at $w=0$ is governed by the indicial roots, i.e.\ those numbers $\lambda\in\R$ for which $N_\bop(P_\NH,\lambda):=w^{-\lambda}N_\bop(P_\NH)w^\lambda\in\Diff^2(\Sph^2)$ fails to be invertible.

\begin{lemma}[Indicial roots]
\label{LemmaNHInd}
  The indicial roots of $N_\bop(P_\NH)$ are given by
  \[
    \lambda_\ell^\pm(\mu) := \frac12\biggl(1 \pm \sqrt{1 + 4\frac{\ell(\ell+1)+r_{\rm e}^2\mu}{\varkappa}}\,\biggr),\quad \ell\in\N_0.
  \]
  The poles of $N_\bop(P_\NH,\lambda)^{-1}$ at these values of $\lambda$ have order $1$. A function $w^{\lambda_\ell^\pm(\mu)}Y(\omega)$ is an indicial solution, i.e.\ $N_\bop(P_\NH)(w^{\lambda_\ell^\pm(\mu)}Y(\omega))=0$, if and only if $Y$ is a spherical harmonic of degree $\ell$.
\end{lemma}
\begin{proof}
  This follows from $\frac{1}{\varkappa_{\rm e}}N_\bop(P_\NH,\lambda)=-\lambda(\lambda-1)+\varkappa^{-1}\slDelta+\varkappa^{-1}r_{\rm e}^2\mu$ (see~\eqref{EqGSurfGrav2}): acting on the eigenspace of $\slDelta$ with eigenvalue $\ell(\ell+1)$, $\ell\in\N_0$, this is multiplication by a constant which vanishes precisely for the stated values of $\lambda$.
\end{proof}

Since we only consider $\mu\geq 0$, we have $\lambda^-_\ell(\mu)\leq\lambda^-_0(\mu)\leq 0<1\leq\lambda^+_0(\mu)\leq\lambda^+_\ell(\mu)$ for all $\ell\in\N_0$, so $P_\NH$ has an \emph{indicial gap} $(\lambda^-_0(\mu),\lambda^+_0(\mu))\supseteq(0,1)$. We define quasinormal modes for $P_\NH$ by demanding Dirichlet boundary conditions at the conformal boundary, meaning that we demand resonant states to decay as $z\to\infty$. (This disallows for the presence of $w^{\lambda^-_\ell(\mu)}$ asymptotics.)

\begin{definition}[QNMs of the near-horizon geometry]
\label{DefNHQNM}
  We define $\QNM_\NH(\mu)\subset\C$ to consist of all $\sigma\in\C$ such that there exists a function (resonant state) $u\in\cA^1(X_\NH)$ such that $\wh{P_\NH}(\sigma)u=0$. Here, for $\beta\in\R$, we write
  \[
    \cA^\beta(X_\NH)\subset\CI([0,\infty)_z\times\Sph^2)
  \]
  for the space of all smooth functions on $[0,\infty)_z\times\Sph^2$ which are bounded by a constant times $(z+1)^{-\beta}$ together with derivatives (of any order) along $(z+1)\pa_z$ and spherical vector fields.
\end{definition}

The practical justification for this definition is that, as we shall see in~\S\ref{SPf}, estimates for $\wh{P_\eps}(\kappa_{\rm C,\eps}\sigma)$ (on function spaces adapted to its structure as a q-differential operator) will require estimates (proved in~\S\ref{SsNHFred}) for $\wh{P_\NH}(\sigma)$ on function spaces which encode decay as $z\to\infty$. The presence of a kernel on these spaces will be shown to be equivalent to $\sigma$ being a QNM for $P_\NH$.

The first main result of this section is the following.

\begin{thm}[QNMs of $P_\NH$]
\label{ThmNHQNM}
  We have $\QNM_\NH(\mu)=\{-i(\lambda_\ell^+(\mu)+n)\colon \ell,n\in\N_0\}$. Moreover, the space of resonant states, with spherical harmonic degree $\ell$, associated with the resonance $-i(\lambda_\ell^+(\mu)+n)$ has dimension $2\ell+1$. (An explicit basis is given by~\eqref{EqNHQNMResState}, with $Y_\ell$ there running over a basis of the space of degree $\ell$ spherical harmonics.)
\end{thm}

The proof of Theorem~\ref{ThmNHQNM} is given in~\S\ref{SsNHSolv}. Instead of relying on computations involving special functions, we use a conceptually cleaner argument in the spirit of \cite[\S{II}]{HintzXieSdS}. We pass from the coordinates $\ft_*,z$ used in~\eqref{EqGCgNH} to a coordinate system which highlights the $\AdS^2$ conformal boundary. To wit,\footnote{This coordinate change arises as follows. Let $h:=-(z^2-1)\,\dd\ft_*^2+2\,\dd\ft_*\,\dd z$. First, setting $\ft=\ft_*+\int\frac{\dd z}{1-z^2}$, we have $h=-(z^2-1)\,\dd\ft^2+(z^2-1)^{-1}\,\dd z^2$. Letting $\ft_0=\ft+\frac12\log(1-z^2)=\ft_*+\log(1+z)$ and then $T=-e^{-\ft_0}z$, $\rho=e^{-\ft_0}$ gives $h=\frac{-\dd T^2+\dd\rho^2}{\rho^2}$ and thus~\eqref{EqNHCoord}. Changing from $T,\rho$ to $\ft_0,z$ amounts to passing to coordinates on the blow-up of $\AdS^2$ at the point $(T,\rho)=(0,0)$ which are regular in the interior of the front face; changing from $\ft_0$ to $\ft$ amounts to passing to static coordinates; and changing from $\ft$ to $\ft_*$ amounts to passing to ingoing Eddington--Finkelstein type coordinates. See \cite[\S{4.3}]{VasyMicroKerrdS} for related computations on de~Sitter space.}
\begin{equation}
\label{EqNHCoord}
  T:=-e^{-\ft_*}\frac{z}{1+z},\ \rho:=e^{-\ft_*}\frac{1}{1+z} \implies g_\NH = \frac{1}{\varkappa_{\rm e}}\Bigl(\frac{-\dd T^2+\dd\rho^2}{\rho^2} + \varkappa\slg\Bigr),
\end{equation}
with $\rho=0$ defining the conformal boundary. For later use, we record the inverse transformation
\begin{equation}
\label{EqNHCoordInv}
  \ft_* = -\log(\rho-T),\ 
  z = -\frac{T}{\rho}.
\end{equation}
We will realize mode solutions $U:=e^{-i\sigma\ft_*}u(z,\omega)$ of $P_\NH$ as solutions of an initial boundary value problem on $M_\NH:=\R_T\times[0,\infty)_\rho\times\Sph^2$. After proving sharp regularity and polyhomogeneous asymptotics for solutions $U$ (lying in an appropriate space, in particular: satisfying Dirichlet boundary conditions at the conformal boundary) of $(\Box_{g_\NH}+\mu)U=0$ on a subset of $M_\NH$ containing $\{T=\rho=0\}=\{0\}\times\{0\}\times\Sph^2$, we deduce the possible values of $\sigma$ by comparison with the polyhomogeneous expansion of $U$ at $\{T=\rho=0\}$.

The second main result of this section gives Fredholm estimates for the operator $\wh{P_\NH}(\sigma)$ on appropriate b-Sobolev spaces on $X_\NH$, and its invertibility when $\sigma\notin\QNM_\NH(\mu)$; see Proposition~\ref{PropNHFred}.

%%%%%%%%%%%%%%%%%%%%%%%%%%%%%%%%%%%%%%%%%%%%%%%%%%
\subsection{Asymptotics of waves at the conformal boundary}
\label{SsNHSolv}

In light of~\eqref{EqNHCoord}, we have
\[
  P_\NH = \varkappa_{\rm e}(-\rho^2 D_T^2 + \rho^2 D_\rho^2 + \varkappa^{-1}\slDelta) + \mu.
\]
We rewrite the equation $(\varkappa_{\rm e}\rho^2)^{-1}P_\NH U=0$ on $\R_T\times(0,\infty)_\rho\times\Sph^2$ as
\begin{equation}
\label{EqNHSolvL}
  (-D_T^2 + L)U = 0,\quad L=D_\rho^2+\varkappa^{-1}\rho^{-2}\slDelta + \frac{\mu}{\varkappa_{\rm e}}\rho^{-2}.
\end{equation}

The operator $L$ is qualitatively similar to the Laplacian on a manifold with a conic singularity at $\rho=0$. We shall analyze~\eqref{EqNHSolvL} using the spectral theory of $L$. To this end, it is convenient to first remove the noncompact end $\rho\to\infty$. Concretely, let $Y:=\Sph^3$ and $\fp\in Y$, and let $\rho>0$, $\omega\in\Sph^2$ be polar coordinates on the stereographic projection of $Y\setminus\{-\fp\}$ (so $\rho=0$ at $\fp$ and $\rho\to\infty$ as one approaches $-\fp$). Fix a cutoff function
\begin{equation}
\label{EqNHSolvCutoff}
  \chi\in\CIc([0,\infty)_\rho),\quad \chi(\rho)=1\ \text{for}\ \rho\in[0,4].
\end{equation}
Fix a Riemannian metric $g_Y$ on $Y$ and set
\[
  g := \chi(\dd\rho^2 + \varkappa\rho^2\slg) + (1-\chi)g_Y.
\]
Let $\varrho\in\CI(Y\setminus\{\fp\})$ be equal to $\rho$ for $\rho\leq 4$ and positive for $\rho\geq 4$. Then the operator
\[
  \cL := \varrho\Delta_g\varrho^{-1} + \frac{\mu}{\varkappa_{\rm e}}\varrho^{-2} \in \Diff^2(Y\setminus\{\fp\})
\]
is elliptic on $Y\setminus\{\fp\}$ and equal to $L$ for $\rho=\varrho\leq 4$. Moreover, on $L^2(Y\setminus\{\fp\})$ with volume density $\dd\mu:=c\varrho^{-2}|\dd g|$, $c>0$, it is symmetric with domain $\CIc(Y\setminus\{\fp\})$. We fix $c=\varkappa^{-1}$, so
\[
  \dd\mu = c\varrho^{-2}|\dd g|=|\dd\rho\,\dd\slg|\quad \text{for}\ \rho\leq 4.
\]
For $u\in\CIc(Y\setminus\{\fp\})$, we compute
\begin{equation}
\label{EqNHSolvInner}
  \la\cL u,u\ra_{L^2(Y,\dd\mu)} = \|\varrho\nabla^g(\varrho^{-1}u)\|_{L^2(Y,\dd\mu)}^2 + \frac{\mu}{\varkappa_{\rm e}}\|\varrho^{-1}u\|_{L^2(Y,\dd\mu)}^2.
\end{equation}
We wish to find a self-adjoint extension of $\cL$. Let $Y':=[Y;\{\fp\}]$, so $Y'$ is the smooth manifold with boundary that is covered by the two charts $[0,\infty)_\rho\times\Sph^2$ and $Y\setminus\{\fp\}$. For $s\in\N_0$, $\alpha\in\R$, we define the function space
\[
  \Hb^{s,\alpha}(Y')
\]
to consist of all $u$ with $(1-\chi)u\in H^s(Y)$ and $\|\chi u\|_{\Hb^{s,\alpha}([0,\infty)\times\Sph^2)}<\infty$ where
\[
  \|v\|_{\Hb^{s,\alpha}([0,\infty)\times\Sph^2)}^2=\sum_{i+|\beta|\leq s}\int_{\Sph^2}\int_0^\infty|\rho^{-\alpha}(\rho\pa_\rho)^i\Omega^\beta v|^2\,\frac{\dd\rho}{\rho}\,\dd\slg<\infty;
\]
here $\Omega=\{\Omega_1,\Omega_2,\Omega_3\}\subset\cV(\Sph^2)$ is the set of rotation vector fields around coordinate axes.

\begin{lemma}[Completion]
\label{LemmaNHSolvComp}
  The completion of $\CIc(Y\setminus\{\fp\})$ with respect to the squared norm given by the right hand side of~\eqref{EqNHSolvInner} is equal to the space $\Hb^{1,\frac12}(Y')$.
\end{lemma}
\begin{proof}
  Working with $u$ supported in $\rho<4$, we note that the right hand side of~\eqref{EqNHSolvInner} is equivalent (i.e.\ bounded from above and below by a constant times)
  \begin{equation}
  \label{EqNHSolvCompPf}
    \int_{\Sph^2}\int_0^\infty \rho\Bigl( |\rho\pa_\rho(\rho^{-1}u)|^2 + |\slnabla(\rho^{-1}u)|^2 + \frac{\mu}{\varkappa_{\rm e}}|\rho^{-1}u|^2\Bigr)\,\frac{\dd\rho}{\rho}\,\dd\slg.
  \end{equation}
  The Hardy inequality gives, for $v:=\rho^{-1}u$,
  \[
    \int_0^\infty |\rho^{\frac12}v|^2\,\frac{\dd\rho}{\rho} \leq 4\int_0^\infty |\rho^{\frac12}(\rho\pa_\rho v)|^2\,\frac{\dd\rho}{\rho}.
  \]
  Therefore,~\eqref{EqNHSolvCompPf} is equivalent to $\|\rho^{\frac12}\rho^{-1}u\|_{\Hb^1([0,\infty)\times\Sph^2)}^2$, and hence to $\|u\|_{\Hb^{1,\frac12}(Y')}^2$. Conversely, every element of $\Hb^{1,\frac12}(Y')$ can be approximated in this norm by an element of $\CIc(Y\setminus\{\fp\})$ by first cutting it off to the complement of a sufficiently small neighborhood of $\pa Y'$ and then using a standard mollifier.
\end{proof}

We now take as the self-adjoint extension of $\cL$ the Friedrichs extension; we denote this by $\cL$ still, and the domain by $\cD(\cL)$. Note that $\cL\geq 0$.

\begin{prop}[Domains of powers of $\cL$]
\label{PropNHSolvDom}
  Let $k\in\N$ and recall~\eqref{EqNHSolvCutoff}.
  \begin{enumerate}
  \item If $u\in\cD(\cL^k)$, then $(1-\chi)u\in H^{2 k}(Y)$ and there exist spherical harmonics $Y_\ell^n\in\CI(\Sph^2)$, $n\in\N_0$, of degree $\ell\in\N_0$, such that
    \begin{equation}
    \label{EqNHSolvDom}
      \chi(\rho)u(\rho,\omega) = \chi\sum_{\ell,n}\rho^{\lambda_\ell^+(\mu)+2 n}Y_\ell^n(\omega) + \tilde u(\rho,\omega)
    \end{equation}
    where the sum is over all $\ell,n$ with $\lambda_\ell^+(\mu)+2 n<2 k-\frac12$, and $\tilde u\in\bigcap_{\eta>0}\Hb^{2 k,2 k-\frac12-\eta}(Y')$. (One can take $\eta=0$ if $\lambda_\ell^+(\mu)+2 n\neq 2 k-\frac12$ for all $\ell,n$.)
  \item\label{ItNHSolvDomConv} Conversely, if $(1-\chi)u\in H^{2 k}(Y)$ and $\chi u$ is of the form~\eqref{EqNHSolvDom} with $\tilde u\in\Hb^{2 k,2 k-\frac12}(Y')$, then $u\in\cD(\cL^k)$.
  \end{enumerate}
\end{prop}
\begin{proof}
  Consider first the case $k=1$. If $u\in\cD(\cL)$, then $u\in\Hb^{1,\frac12}(Y')$ and $\cL u\in L^2(Y,\dd\mu)=\Hb^{0,-\frac12}(Y')$. Elliptic regularity gives $u\in H^2_\loc(Y'\setminus\pa Y')$. Near $\rho=0$, we use $[\rho^2\cL,\chi]\in\rho\Diffb^1(Y')$ to compute
  \[
    \rho^2\cL(\chi u) = \chi\rho^2\cL u+[\rho^2\cL,\chi]u \in \Hb^{0,\frac32}.
  \]
  Now, in $\rho<4$, the operator $\rho^2\cL=\rho^2 D_\rho^2+\varkappa^{-1}\slDelta+\frac{\mu}{\varkappa_{\rm e}}$ is dilation-invariant and, upon identifying $w$ and $\rho$, equal to $(\varkappa_{\rm e})^{-1}N_\bop(P_\NH)$ in~\eqref{EqNHNb}. Passing to the Mellin transform in $\rho$ and using Lemma~\ref{LemmaNHInd} and the meromorphicity of $N_\bop(P_\NH,\lambda)^{-1}$, one can thus extract a partial asymptotic expansion of $\chi u$, namely
  \[
    \chi u(\rho,\omega) \equiv Y_0\rho \bmod \Hb^{2,\frac32}
  \]
  where $Y_0$ is a constant. (Note that $\lambda_\ell^+(\mu)\geq 2$ for $\ell\geq 1$ since $\mu\geq 0$ and $0<\varkappa<1$.)

  Consider now $k\geq 2$. Fix $\chi^\flat\in\CIc([0,4))$ with $\chi^\flat=1$ on $[0,3]$. If $u\in\cD(\cL^k)$, then $u\in\Hb^{1,\frac12}(Y')$ and $\cL u\in\cD(\cL^{k-1})$, so
  \[
    \rho^2\cL(\chi^\flat u) \equiv \chi\sum_{\ell,n} \rho^{\lambda_\ell^+(\mu)+2+2 n}Y_\ell^n(\omega) \bmod \bigcap_{\eta>0}\Hb^{2(k-1),2(k-1)+2-\frac12-\eta}(Y')
  \]
  Solving this using the Mellin transform and noting that $\lambda_\ell^+(\mu)+2+2 n$ is not an indicial root admitting degree $\ell$ spherical harmonics as indicial solutions, one obtains the expansion~\eqref{EqNHSolvDom} for $\chi^\flat u$. Since $u\in H^{2 k}_\loc(Y'\setminus\pa Y')$, this implies~\eqref{EqNHSolvDom} as stated.

  For the converse, consider $u$ for which $(1-\chi)u\in H^{2 k}(Y)$ and which admit an expansion~\eqref{EqNHSolvDom}. Then $(1-\chi)\cL u\in H^{2 k-2}(Y)$. Moreover, in view of $\cL(\rho^{\lambda_\ell^+(\mu)}Y_\ell^n(\omega))=0$, we have
  \[
    \chi^\flat\cL u = \chi^\flat\sum_\ell\sum_{n\geq 1} \rho^{(\lambda_\ell^+(\mu)+2 n)-2}\tilde Y_\ell^n(\omega) + \tilde u^\flat
  \]
  where $\tilde Y_\ell^n$ is a degree $\ell$ spherical harmonic and $\tilde u^\flat\in\Hb^{2 k-2,2 k-2-\frac12}(Y')$. Thus, $\cL u$ satisfies the same conditions as $u$ but with $k$ reduced by $1$. Proceeding in this fashion shows that $\cL^k u\in\Hb^{0,-\frac12}(Y')=L^2(Y,\dd\mu)$, which completes the proof of $u\in\cD(\cL^k)$.
\end{proof}

Let $I\subset\R$ be an interval. Consider
\[
  U \in \CI\bigl(I;\Hb^{1,\frac12}(Y')\bigr)
\]
which is a (distributional) solution of $(-D_T^2+\cL)U=0$ on $I\times(Y'\setminus\pa Y')$. Since $\cL U=D_T^2 U\in\CI(I;L^2(Y,\dd\mu))$, we have $U\in\CI(I;\cD(\cL))$. Iterating this argument gives
\begin{equation}
\label{EqNHSolvReg}
  U \in \bigcap_{k\in\N} \CI(I;\cD(\cL^k)).
\end{equation}
By Proposition~\ref{PropNHSolvDom}, this implies that $U$ has a full asymptotic expansion at $\rho=0$.

\begin{proof}[Proof of Theorem~\usref{ThmNHQNM}]
  \pfstep{Upper bound on $\QNM_\NH(\mu)$.} Suppose that $\sigma\in\QNM_\NH(\mu)$, and let $0\neq u\in\cA^1(X_\NH)\cap\ker\wh{P_\NH}(\sigma)$ be a resonant state. Using a normal operator argument at $z^{-1}=0$ and Lemma~\ref{LemmaNHInd}, we find that
  \begin{equation}
  \label{EqNHQNM}
    u(z,\omega)\equiv (z+1)^{-1}Y_0\bmod\cA^2(X_\NH).
  \end{equation}
  Express $U^\flat:=e^{-i\sigma\ft_*}u(z,\omega)\in\ker P_\NH$ in the coordinates~\eqref{EqNHCoordInv}; then
  \[
    U^\flat(T,\rho) := (\rho-T)^{i\sigma}u\Bigl(-\frac{T}{\rho},\omega\Bigr),\quad T<0,\ \rho\in(0,\infty),\ \omega\in\Sph^2,
  \]
  is a solution of $P_\NH(U^\flat)=0$.

  In order to relate $U^\flat$ to the operator $-D_T^2+\cL$, we shall first extend $U^\flat|_{\{\rho<4\}}$ to $[-1,0)_T\times Y'$. To this end, define $U_0,U_1\in\CI(Y\setminus\{\fp\})$ such that $U_0(\rho)=U^\flat(-1,\rho)$ and $U_1(\rho)=\pa_T U^\flat(-1,\rho)$. Proposition~\ref{PropNHSolvDom}\eqref{ItNHSolvDomConv} and~\eqref{EqNHQNM} imply $U_0,U_1\in\cD(\cL)$, and therefore
  \begin{equation}
  \label{EqNHQNMSolv}
    U(T) := \cos\bigl((T+1)\sqrt\cL\bigr)U_0 + \frac{\sin\bigl((T+1)\sqrt\cL\bigr)}{\sqrt\cL}U_1
  \end{equation}
  defines a solution of $(-D_T^2+\cL)U=0$ of class $\cC^0(\R;\cD(\cL))\cap\cC^2(\R;L^2(Y,\dd\mu))$. By the finite speed of propagation for (distributional) solutions of wave equations, we must have $U=U^\flat$ for $T\in(-1,0)$ and $\rho\leq 3-T$. See Figure~\ref{FigNHQNMProp}. In particular, the restriction of $U$ to $\{-1\leq T<0\}$ is of class $\CI([-1,0);\Hb^{1,\frac12}(Y'))$, which in view of~\eqref{EqNHSolvReg} shows that $U_0,U_1\in\cD(\cL^k)$ for all $k\in\N$. The formula~\eqref{EqNHQNMSolv} then shows that, in fact, $U\in\CI(\R;\cD(\cL^k))$ for all $k$.

  \begin{figure}[!ht]
  \centering
  \inclfig{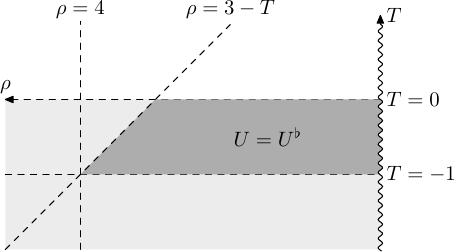}
  \caption{Illustration of the passage from the mode solution $U^\flat$ (defined in the light gray region) to a global solution $U$ of a wave-type equation on $\R_T\times Y'$ which agrees with $U^\flat$ where $\rho\leq 3-T$ and $-1<T<0$ (dark gray). The operators $(\varkappa_{\rm e}\rho^2)^{-1}P_\NH$ and $-D_T^2+\cL$ agree $\rho\leq 4$.}
  \label{FigNHQNMProp}
  \end{figure}

  We now take advantage of the expansion~\eqref{EqNHSolvDom} which shows that $U$ is an asymptotic sum (as $\rho\to 0$) of terms of the form
  \begin{equation}
  \label{EqNHQNMTerm}
    \rho^{\lambda_\ell^+(\mu)+2 n}Y_\ell(T,\omega)
  \end{equation}
  where $Y_\ell$ is smooth in $T$, with $Y_\ell(T,\cdot)$ valued in the space of degree $\ell$ spherical harmonics. Expanding $Y_\ell$ in Taylor series around $T=0$, so $Y_\ell(T,\omega)\sim\sum_{j\geq 0} T^j Y_{\ell,j}(\omega)$, we find that the expression for~\eqref{EqNHQNMTerm} in terms of the coordinates $\ft_*$ and $z\geq 0$ (see~\eqref{EqNHCoord}) is an asymptotic sum (as $\ft_*\to\infty$) of terms
  \[
    \Bigl(-\frac{z}{1+z}\Bigr)^j(1+z)^{-(\lambda_\ell^+(\mu)+2 n)}Y_{\ell,j}(\omega) e^{-(\lambda_\ell^+(\mu)+2 n+j)\ft_*}
  \]
  On the other hand, we have $U=e^{-i\sigma\ft_*}u(z,\omega)$ for $z\geq 0$, and therefore we must have
  \[
    \sigma = -i(\lambda_\ell^+(\mu)+2 n+j)
  \]
  for some $\ell,n,j\in\N_0$.

  %%%%%%%%%%
  \pfstep{Lower bound on $\QNM_\NH(\mu)$.} Fix $n\in\N_0$. For a suitable polynomial $a=a(T)$, we will produce a solution of $P_\NH U=0$ with leading order behavior $\rho^{\lambda_\ell^+(\mu)}a(T)Y_\ell(\omega)$ at $\rho=0$, where $Y_\ell\neq 0$ is any fixed degree $\ell$ spherical harmonic; expressing this in terms of~\eqref{EqNHCoord} will furnish a resonant state of $P_\NH$ with frequency $-i(\lambda_\ell^+(\mu)+n)$. In more detail, recall that $\rho^2 L=(\varkappa_{\rm e})^{-1}N_\bop(P_\NH)$ acts on $\rho^\lambda Y_\ell$ via multiplication with $-p_\ell(\lambda)$ where
  \begin{equation}
  \label{EqNHQNMPoly}
    p_\ell(\lambda) := \lambda(\lambda-1) - \frac{\ell(\ell+1)+r_{\rm e}^2\mu}{\varkappa}.
  \end{equation}
  (This polynomial has roots $\lambda_\ell^\pm(\mu)$.) Therefore, 
  \[
    \rho^2(-D_T^2+L)\bigl(\rho^{\lambda_\ell^+(\mu)}a(T)Y_\ell\bigr) = \rho^{\lambda_\ell^+(\mu)+2}a''(T)Y_\ell.
  \]
  The right hand side equals
  \[
    -\rho^2 L\Bigl(\frac{1}{p_\ell(\lambda_\ell^+(\mu)+2)}\rho^{\lambda_\ell^+(\mu)+2}a''(T)Y_\ell\Bigr),
  \]
  and we thus find
  \[
    \rho^2(-D_T^2+L)\Bigl(\rho^{\lambda_\ell^+(\mu)}a(T)Y_\ell+\frac{1}{p_\ell(\lambda_\ell^+(\mu)+2)}\rho^{\lambda_\ell^+(\mu)+2}a''(T)Y_\ell\Bigr) = \frac{1}{p_\ell(\lambda_\ell^+(\mu)+2)}\rho^{\lambda_\ell^+(\mu)+4}a^{(4)}(T)Y_\ell.
  \]
  We continue in this fashion; if $\deg(a)=:n$, we find for $k\in\N$ with $2 k\geq n$ that
  \[
    U(T,\rho,\omega) := \sum_{j=0}^k \frac{1}{\prod_{m=1}^j p_\ell(\lambda_\ell^+(\mu)+2 m)}\rho^{\lambda_\ell^+(\mu)+2 j}a^{(2 j)}(T)Y_\ell(\omega)
  \]
  solves $P_\NH U=0$. Consider the special case $a(T)=(-T)^n$ and insert~\eqref{EqNHCoord}; we then conclude that upon setting
  \begin{equation}
  \label{EqNHQNMResState}
    u_{\ell,n}(z,\omega):=\sum_{j=0}^{\lfloor n/2\rfloor} \frac{n!}{(n-2 j)!\prod_{m=1}^j p_\ell(\lambda_\ell^+(\mu)+2 m)}(1+z)^{-(\lambda_\ell^+(\mu)+2 j)}\Bigl(\frac{z}{1+z}\Bigr)^{n-2 j}Y_\ell(\omega),
  \end{equation}
  the function $e^{-(\lambda_\ell^+(\mu)+n)\ft_*}u_{\ell,n}(z)$ is a mode solution. Therefore, $-i(\lambda_\ell^+(\mu)+n)\in\QNM_\NH(\mu)$, and $u$ is a corresponding resonant state. Our computations imply that, in fact, $u$ spans the space of mode solutions with spherical harmonic degree $\ell$.
\end{proof}

As a simple example for the formula~\eqref{EqNHQNMResState}, the resonant state corresponding to $-i\lambda_\ell^+(\mu)$ is thus given by $(1+z)^{-\lambda_\ell^+(\mu)}Y_\ell(\omega)$.

We remark that the analysis of the equation $-D_T^2+\cL$ could also be done by applying more general black box results such as \cite[Theorem~3.22]{HintzConicWave}. Alternatively, one could also analyze the asymptotic boundary value problem by adapting the methods introduced in the AdS setting by Holzegel~\cite{HolzegelAdS}.

%%%%%%%%%%%%%%%%%%%%%%%%%%%%%%%%%%%%%%%%%%%%%%%%%%
\subsection{Fredholm theory for the spectral family}
\label{SsNHFred}

Recall that $\wh{P_\NH}(\sigma)$ acts on functions on $[0,\infty)_z\times\Sph^2$. We shall state quantitative estimates for $\wh{P_\NH}(\sigma)$ using the following function spaces capturing b-behavior at $z=\infty$:

\begin{definition}[b-Sobolev spaces]
\label{DefNHFredb}
  Let $\Omega=\{\Omega_1,\Omega_2,\Omega_3\}\subset\cV(\Sph^2)$ be the set of rotation vector fields around coordinate axes. Let $s\in\N_0$, $\alpha\in\R$. Let $I\subseteq[0,\infty]$. We then define the space $\Hbext^{s,\alpha}(I\times\Sph^2)$ to consist of all $u\in L^2_\loc(I^\circ\times\Sph^2)$ such that
  \begin{equation}
  \label{EqNHFredb}
    \|u\|_{\Hbext^{s,\alpha}(I\times\Sph^2)}^2 := \sum_{i+|\beta|\leq s} \int_{\Sph^2}\int_I | (z+1)^\alpha((z+1)\pa_z)^i\Omega^\beta u(z,\omega) |^2\,\dd z\,\dd\slg.
  \end{equation}
  For $I=[0,\infty]$, we denote this space by $\Hbext^{s,\alpha}(X_\NH)$.
\end{definition}

An equivalent norm on $\Hbext^{s,\alpha}$ is given by $\|(z+1)^{-\alpha}u\|_{\Hbext^{s,0}}$. The spaces $\Hb^{s,\alpha}([0,\infty]\times\Sph^2)$ can be defined more generally for real $s\in\R$ via duality and interpolation. A hands-on definition, using a partition of unity, is as follows: the squared norm of $u$ supported in $z\geq 4$ is defined as the sum of squares of $H^s$-norms of $[0,3]\times\Sph^2\ni (Z,\omega)\mapsto 2^{\alpha j}\chi(Z)u(2^j 2^Z,\omega)$ for $j\in\N_0$, where $\chi\in\CIc((0,3))$ equals $1$ on $[1,2]$ (note here that writing $z=2^j 2^Z$, we have $z\pa_z=\frac{1}{\log 2}\pa_Z$), whereas the squared norm of $u$ supported in $z\leq 8$ is defined as the minimal $H^s(\R\times\Sph^2)$-norm of all extensions of $u$ to distributions supported in $[-1,9]\times\Sph^2$. The $L^2$-dual space of $\Hbext^{s,\alpha}(X_\NH)$ is equal to $\Hbsupp^{-s,-\alpha}(X_\NH)$, the space of all elements of $\Hbext^{-s,-\alpha}([-\frac12,\infty]\times\Sph^2)$ with support in $z\geq 0$. (See also \cite[Appendix~B]{HormanderAnalysisPDE3} and \cite[Chapter~10.3]{HintzMicro}.) We finally recall that the inclusion map $\Hbext^{s,\alpha}(X_\NH)\to\Hbext^{s_0,\alpha_0}(X_\NH)$ is compact for $s>s_0$, $\alpha>\alpha_0$; this is a simple consequence of the usual Rellich compactness theorem.

\begin{prop}[Fredholm estimates and index $0$]
\label{PropNHFred}
  Let $\alpha\in(-\frac12,\frac12)$, $C_0\in\R$, and $s>\frac12+C_0$.
  \begin{enumerate}
  \item\label{ItNHFred} For all $\sigma\in\C$ with $\Im\sigma>-C_0$, the operator\footnote{Since $\wh{P_\NH}(\sigma)-\wh{P_\NH}(0)\in\Diffb^1$, one can equally well use $\wh{P_\NH}(\sigma)$ in the definition of the space $\cX^{s,\alpha}$.}
    \begin{equation}
    \label{EqNHFred}
      \wh{P_\NH}(\sigma) \colon \{ u\in\Hbext^{s,\alpha}(X_\NH) \colon \wh{P_\NH}(0)u\in\Hbext^{s-1,\alpha}(X_\NH) \} \to \Hbext^{s-1,\alpha}(X_\NH)
    \end{equation}
    is Fredholm of index $0$.
  \item\label{ItNHInv} The operator~\eqref{EqNHFred} is invertible if and only if $\sigma\notin\QNM_\NH(\mu)$. In this case, there exists a constant $C$ such that
    \begin{equation}
    \label{EqNHFred2}
      \|u\|_{\Hbext^{s,\alpha}(X_\NH)} \leq C\|\wh{P_\NH}(\sigma)u\|_{\Hbext^{s-1,\alpha}(X_\NH)}.
    \end{equation}
  \end{enumerate}
\end{prop}
\begin{proof}
  \pfstep{Fredholm estimate.} As hinted at at the beginning of the section, we can, for $z\in[0,5]$, analyze the operator $\wh{P_\NH}(\sigma)$, given by~\eqref{EqGCSpecFamNH}, using standard microlocal and energy arguments (see \cite[\S4]{VasyMicroKerrdS}, \cite[\S2]{ZworskiRevisitVasy}, \cite[Chapter~12]{HintzMicro}). The radial point estimate at $N^*\{z=1\}\setminus o$ uses the threshold regularity assumption $s>\frac12+C_0$. Thus,
  \begin{equation}
  \label{EqNHFredRad}
    \|u\|_{H^s([0,4]\times\Sph^2)} \leq C\bigl( \|\wh{P_\NH}(\sigma)u \|_{H^{s-1}([0,5]\times\Sph^2)} + \|u\|_{H^{s_0}([0,5]\times\Sph^2)} \bigr),
  \end{equation}
  where we fix $s_0$ with $s>s_0>\frac12+C_0$. (For a self-contained proof of this estimate for separated $u$, we refer the reader to \cite[\S{II.A}]{HintzXieSdS}.) For $z\in[3,\infty)$ on the other hand, the operator $\wh{P_\NH}(\sigma)$ is elliptic, including at $z=\infty$ as a b-operator (equivalently, it is uniformly elliptic when expressed in terms of $\log z$). Therefore, for any fixed $\alpha$,
  \[
    \|u\|_{\Hbext^{s,\alpha}([3,\infty]\times\Sph^2)} \leq C\bigl( \|\wh{P_\NH}(\sigma)u\|_{\Hbext^{s-2,\alpha}([2,\infty]\times\Sph^2)} + \|u\|_{\Hbext^{s_0,\alpha}([2,\infty]\times\Sph^2)} \bigr).
  \]
  Combining the two estimates gives
  \[
    \|u\|_{\Hbext^{s,\alpha}(X_\NH)} \leq C\bigl( \|\wh{P_\NH}(\sigma)u\|_{\Hbext^{s-1,\alpha}(X_\NH)} + \|u\|_{\Hbext^{s_0,\alpha}(X_\NH)} \bigr).
  \]

  We proceed to improve the weight of the weak norm on the right using standard elliptic b-theory. Fix $\chi\in\CIc([0,2))$ with $\chi=1$ on $[0,1]$. Set $w=z^{-1}$. We have $\|u\|_{\Hbext^{s_0,\alpha}(X_\NH)}\leq\|\chi(w)u\|_{\Hbext^{s_0,\alpha}(X_\NH)}+\|(1-\chi(w))u\|_{\Hbext^{s_0,\alpha-1}(X_\NH)}$, the weight in the second summand being irrelevant since $z$ is bounded on $\supp(1-\chi)$. We estimate the first summand by passing to the Mellin transform in $w$ and inverting $N_\bop(P_\NH,\lambda)$ for $\Re\lambda=\alpha+\frac12$, which can be done for weights $\alpha$ satisfying $\alpha+\frac12\in(0,1)$ (which is contained in the indicial gap). (The shift by $\frac12$ arises from the fact that the Plancherel theorem gives an isomorphism of $w^\alpha L^2([0,\infty)_w\times\Sph^2,|\dd(w^{-1})\,\dd\slg|)=w^{\alpha+\frac12}L^2([0,\infty)\times\Sph^2,|\frac{\dd w}{w}\,\dd\slg|)$ with $L^2(\{\Re\lambda=\alpha+\frac12\};L^2(\Sph^2))$ via $(\cM u)(\lambda)=\int_0^\infty w^{-\lambda}u(w,\omega)\,\frac{\dd w}{w}$.) This gives
  \begin{equation}
  \label{EqNHFredNb}
    \|\chi u\|_{\Hbext^{s_0,\alpha}(X_\NH)} \leq C\|N_\bop(P_\NH)(\chi u)\|_{\Hbext^{s_0-2,\alpha}(X_\NH)}.
  \end{equation}
  Replacing $N_\bop(P_\NH)$ by the operator $\wh{P_\NH}(\sigma)$ differing from it by an element of $w\Diffb^2$ produces an error term $\|u\|_{\Hbext^{s_0,\alpha-1}(X_\NH)}$; similarly for the error term produced subsequently by commuting $\wh{P_\NH}(\sigma)$ through $\chi$. Altogether, we get
  \begin{equation}
  \label{EqNHFredPf}
    \|u\|_{\Hbext^{s,\alpha}(X_\NH)} \leq C\bigl(\|\wh{P_\NH}(\sigma)u\|_{\Hbext^{s-1,\alpha}(X_\NH)} + \|u\|_{\Hbext^{s_0,\alpha-1}(X_\NH)}\bigr).
  \end{equation}
  Since $\Hbext^{s,\alpha}\hra\Hbext^{s_0,\alpha-1}$ is compact, this implies that $\wh{P_\NH}(\sigma)$ has finite-dimensional nullspace and closed range.

  Similar arguments prove the estimate
  \begin{equation}
  \label{EqNHFredAdj}
    \|u^*\|_{\Hbsupp^{-s+1,-\alpha}(X_\NH)} \leq C\bigl(\|\wh{P_\NH}(\sigma)^*u^*\|_{\Hbsupp^{-s,-\alpha}(X_\NH)} + \|u^*\|_{\Hbsupp^{s_0^*,-\alpha-1}(X_\NH)}\bigr)
  \end{equation}
  for the adjoint of $\wh{P_\NH}(\sigma)$; here we fix any $s_0^*<-s+1$. This implies the finite-dimensionality of the cokernel of $\wh{P_\NH}(\sigma)$ and thus implies the Fredholm statement of part~\eqref{ItNHFred}. (See \cite[Chapter~12.3]{HintzMicro} for details in a closely related setting.)

  \pfstep{Nullspace of $\wh{P_\NH}(\sigma)$ and resonances.} Since $\cA^1(X_\NH)\subset\Hbext^{s,\alpha}(X_\NH)$ for all $s\in\R$ and $\alpha<\frac12$, the nullspace of $\wh{P_\NH}(\sigma)$ is nontrivial when $\sigma\in\QNM_\NH(\mu)$. For the converse, we need to show that $u\in\Hbext^{s,\alpha}(X_\NH)$, $\wh{P_\NH}(\sigma)u=0$ implies $u\in\cA^1$. The weaker statement $u\in\bigcap_N\Hbext^{N,\alpha}(X_\NH)$ follows from the fact that the estimate~\eqref{EqNHFredPf} (per its proof) holds in the strong sense for all $s>\frac12-\Im\sigma$: if the right hand side is finite, then so is the left hand side. Sobolev embedding now gives $u\in\cA^{\alpha+\frac12}(X_\NH)$. Since the smallest indicial root $\geq\alpha+\frac12$ is $\lambda^+_0(\mu)\geq 1$, we in fact have $u\in\cA^1(X_\NH)$ by a Mellin transform/normal operator argument.

  \pfstep{Index $0$.} It suffices to show that $\wh{P_\NH}(\sigma)$ is invertible for sufficiently large $\Im\sigma$; we shall show this here for $\Im\sigma>\frac32$. Injectivity holds for such $\sigma$ by Theorem~\ref{ThmNHQNM}. Consider $u^*\in\Hbsupp^{-s+1,-\alpha}(X_\NH)$ with $\wh{P_\NH}(\sigma)^*u^*=\wh{P_\NH}(\bar\sigma)u^*=0$. Since~\eqref{EqNHFredAdj} holds in the strong sense for $s>\frac12-\Im\sigma$, we have
  \begin{equation}
  \label{EqNHFredustarReg}
    u^*\in\bigcap_{\eta>0}\Hbsupp^{\frac12+\Im\sigma-\eta,-\alpha}(X_\NH),
  \end{equation}
  so a fortiori $u^*\in\Hbsupp^{1,-\alpha}$; and a normal operator argument shows that in fact
  \[
    u^*\in\bigcap_{\beta<\frac12}\Hbsupp^{1,\beta}(X_\NH).
  \]
  Finally, $u^*=0$ for $z<1$ since $u^*=0$ for $z<0$ and $\wh{P_\NH}(\bar\sigma)u^*=0$ is a wave equation in $z<1$, with $z$ a time function. The function $u^*=u^*(z,\omega)$ gives rise to a mode solution $e^{-i\bar\sigma\ft_*}u^*$ which in the coordinates~\eqref{EqNHCoord} is given by
  \begin{equation}
  \label{EqNHFredUstar}
    U^*(\rho,T,\omega) = (\rho-T)^{i\bar\sigma}u^*\Bigl(-\frac{T}{\rho},\omega\Bigr),\quad T<0;
  \end{equation}
  it vanishes for $\rho>-T$. We extend $U^*$ by $0$ to $T\in(-1,1)$, $\rho+T>0$. Recalling the notation $Y',\cL$ from~\S\ref{SsNHSolv}, we can regard $U^*$ as a function on $(-1,1)\times Y'$ (defined by $0$ on $Y'\setminus\{\rho\leq 1\}$) that is a distributional solution of $P_\NH^*U^*=P_\NH U^*=0$ on $(-1,1)\times(Y'\setminus\pa Y')$. We claim that
  \begin{equation}
  \label{EqNHFredC0}
    U^* \in \cC^0\bigl((-1,1);\Hb^{1,\frac12}(Y')\bigr).
  \end{equation}
  To verify this, consider $v=z^{-\beta}v_0$, $v_0\in L^2(X_\NH,|\dd z\,\dd\slg|)$, vanishing for $z=-\frac{T}{\rho}<1$; for $T<0$, we then have
  \begin{align*}
    &\int_{\Sph^2}\int_0^{-T} \rho^{-1}\Bigl|(\rho-T)^{i\bar\sigma} v\Bigl(-\frac{T}{\rho},\omega\Bigr)\Bigr|^2\,\frac{\dd\rho}{\rho}\,\dd\slg \\
    &\qquad = |T|^{-1+2\Im\sigma}\int_{\Sph^2}\int_1^\infty z^{-2\beta}\Bigl(\frac{z+1}{z}\Bigr)^{2\Im\sigma}|v_0(z,\omega)|^2\,\dd z\,\dd\slg \\
    &\qquad \leq C|T|^{-1+2\Im\sigma}
  \end{align*}
  provided $\beta\geq 0$. If $\Im\sigma>\frac12$, this tends to $0$ as $T\nearrow 0$. Applying this with $v=u^*$, we conclude that $U^*\in\cC^0((-1,1);\Hb^{0,\frac12}(Y'))$. Note moreover that
  \[
    \rho\pa_\rho U^*(\rho,T,\omega) = -(\rho-T)^{i\bar\sigma} (z\pa_z u^*)\Bigl(-\frac{T}{\rho},\omega\Bigr) + i\bar\sigma(\rho-T)^{i\cdot\ol{(\sigma-i)}}u^*\Bigl(-\frac{T}{\rho},\omega\Bigr),
  \]
  so applying the above estimate with $v=z\pa_z u^*$ as well as with $v=u^*$ and $\sigma-i$ in place of $\sigma$, and to $v=\Omega_a u^*$ implies~\eqref{EqNHFredC0} for $\Im(\sigma-i)=\Im\sigma-1>\frac12$.

  \begin{figure}[!ht]
  \centering
  \inclfig{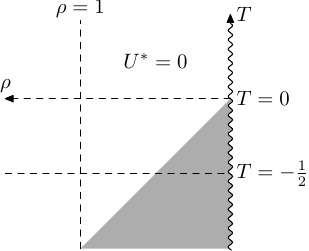}
  \caption{Illustration of the argument for the absence of cokernel for $\Im\sigma>\frac32$. The extension by $0$ of a putative mode solution $U^*$ for $P_\NH^*$ (which has support contained in the shaded region) solves the wave equation $(-D_T^2+\cL)U^*=0$ (after mollification in $T$, cf.\ \eqref{EqNHFredMolly}) for $T\in(-\frac12,\frac12)$, and hence vanishes identically.}
  \label{FigNHFred}
  \end{figure}

  We claim that~\eqref{EqNHFredC0} and $P_\NH U^*=0$ imply $U^*=0$. (See Figure~\ref{FigNHFred}.) To this end, fix $\phi\in\CIc((-1,1))$ with $\int_{-1}^1\phi(T)\,\dd T=1$. For $\eta>0$, set $\phi_\eta(T):=\eta^{-1}\phi(\frac{T}{\eta})$. Using convolution in $T$, define then
  \begin{equation}
  \label{EqNHFredMolly}
    U^*_\eta := \phi_\eta * U^*.
  \end{equation}
  Note that $U^*_\eta\in\CI((-\frac12,\frac12),\Hb^{1,\frac12}(Y'))$ solves $P_\NH U^*=0$ still since $P_\NH$ commutes with $T$-translations. Therefore,
  \[
    (-D_T^2 + \cL)U^*_\eta = 0.
  \]
  The arguments leading to~\eqref{EqNHSolvReg} give $U^*_\eta\in\CI((-\frac12,\frac12),\cD(\cL^k))$ for all $k$. In particular, since $U^*_\eta=0$ for $T\geq\frac14$ when $\eta<\frac14$, the formula $U^*_\eta(T)=\cos((T-\frac14)\sqrt\cL)U^*_\eta(\frac14)+\frac{\sin((T-\frac14)\sqrt\cL)}{\sqrt\cL}(\pa_T U^*_\eta)(\frac14)$ implies that $U^*_\eta=0$ for $-\frac12<T<\frac12$. Taking the limit $\eta\searrow 0$ yields the same conclusion for $U^*$. From the vanishing of~\eqref{EqNHFredUstar} for $T=-\frac14$, say, we conclude that $u^*=0$.
\end{proof}

%%%%%%%%%%%%%%%%%%%%%%%%%%%%%%%%%%%%%%%%%%%%%%%%%%%%%%%%%%%%%%%%%%%%%%
\section{Zero energy estimate on extremal RNdS}
\label{S0}

We shall prove an estimate for the zero operator $\wh{P_\ext}(0)=r^{-2}D_r F_0 r^2 D_r+r^{-2}\slDelta+\mu$ on extremal RNdS from~\eqref{EqGCSpecFam0} on b-Sobolev spaces on $X_\ext=[r_{\rm e},r_+]\times\Sph^2$ (see Definition~\ref{DefGCTot}) defined analogously to Definition~\ref{DefNHFredb}. Concretely, for $s\in\N_0$ and $\gamma\in\R$, we set
\begin{equation}
\label{Eq0Hb}
  \|u\|_{\Hbext^{s,\gamma}(X_\ext)}^2 := \sum_{i+|\beta|\leq s} \int_{\Sph^2}\int_{r_{\rm e}}^{r_+} \bigl| (r-r_{\rm e})^{-\gamma} \bigl((r-r_{\rm e})\pa_r\bigr)^i \Omega^\beta u(r,\omega)\bigr|^2\,r^2\dd r\,\dd\slg.
\end{equation}
The $L^2$-dual space $\Hbsupp^{-s,-\gamma}(X_\ext)$ is equal to the space of elements of $\Hbext^{-s,-\gamma}([r_{\rm e},r_{\rm c}+1]\times\Sph^2)$ with support in $r\leq r_{\rm c}$.

\begin{prop}[Zero energy estimate]
\label{Prop0}
  Let $\gamma\in(-\frac12,\frac12)$, $s>\frac12$. Then the operator
  \begin{equation}
  \label{Eq0Op}
    \wh{P_\ext}(0) \colon \cX^{s,\gamma} := \{ u\in\Hbext^{s,\gamma}(X_\ext) \colon \wh{P_\ext}(0)u\in\Hbext^{s-1,\gamma}(X_\ext) \} \to \Hbext^{s-1,\gamma}(X_\ext)
  \end{equation}
  is Fredholm of index $0$. Moreover:
  \begin{enumerate}
  \item if the scalar field mass $\mu$ is strictly positive, then the map~\eqref{Eq0Op} is invertible;
  \item\label{It0Ker} in the case $\mu=0$, define $u_{(0)}:=1$ and $u_{(0)}^*(r,\omega)=1_{[r_{\rm e},r_{\rm c}]}(r)$. Then
    \begin{equation}
    \label{Eq0OpKer}
      \ker_{\Hbext^{s,\gamma}(X_\ext)}\wh{P_\ext}(0)=\mathspan\{u_{(0)}\}, \quad
      \ker_{\Hbsupp^{-s+1,-\gamma}(X_\ext)}\wh{P_\ext}(0)^*=\mathspan\{u_{(0)}^*\}.
    \end{equation}
  \end{enumerate}
\end{prop}
\begin{proof}
  \pfstep{Fredholm property.} The proof is very similar to that of Proposition~\ref{PropNHFred}. Indeed, since $F_0=(r-r_{\rm e})^2(r-r_{\rm c})\cdot(-\frac{\Lambda}{3 r^2}(r+2 r_{\rm e}+r_{\rm c}))$, we first observe that the operator $\wh{P_\ext}(0)$ is an elliptic b-operator near $r-r_{\rm e}=0$. Its b-normal operator $N_\bop(\wh{P_\ext}(0))=\varkappa_{\rm e}D_r(r-r_{\rm e})^2 D_r+r_{\rm e}^{-2}\slDelta+\mu$, and thus its indicial roots are equal to $-\lambda^\pm_\ell(\mu)$, $\ell\in\N_0$ in the notation of Lemma~\ref{LemmaNHInd}. Since $(r-r_{\rm e})^\gamma L^2([r_{\rm e},r_+]\times\Sph^2;|\dd r\,\dd\slg|)=(r-r_{\rm e})^{\gamma-\frac12}L^2([r_{\rm e},r_+]\times\Sph^2;|\frac{\dd r}{r-r_{\rm e}}\,\dd\slg|)$, this means that we need $\gamma-\frac12\neq-\lambda^\pm_\ell(\mu)$ for all $\ell\in\N_0$---which is in particular satisfied for $\gamma\in(-\frac12,\frac12)$---in order to obtain
  \[
    \|\chi u\|_{\Hbext^{s_0,\gamma}(X_\ext)} \leq C\|N_\bop(\wh{P_\ext}(0))(\chi u)\|_{\Hbext^{s_0-2,\gamma}(X_\ext)}
  \]
  where $\chi\in\CIc([r_{\rm e},r_+))$ equals $1$ near $r_{\rm e}$, and $s_0$ is arbitrary but fixed. (This is the analogue of~\eqref{EqNHFredNb}.)

  Moreover, the analysis of $\wh{P_\ext}(0)$ near the non-degenerate horizon $r=r_{\rm c}$ is again standard; for $s>s_0>\frac12$, and recalling $\delta=\frac{r_{\rm c}-r_{\rm e}}{4}$, we can thus estimate
  \begin{equation}
  \label{Eq0PfRad}
    \|u\|_{\bar H^s([r_{\rm c}-\delta,r_+]\times\Sph^2)} \leq C\bigl( \|\wh{P_\ext}(0)u\|_{\bar H^{s-1}([r_{\rm c}-2\delta,r_+]\times\Sph^2)} + \|u\|_{\bar H^{s_0}([r_{\rm c}-2\delta,r_+]\times\Sph^2)}\bigr).
  \end{equation}
  The combined estimate, analogous to~\eqref{EqNHFredPf}, reads
  \begin{equation}
  \label{Eq0Pf}
    \|u\|_{\Hbext^{s,\gamma}(X_\ext)} \leq C\bigl( \|\wh{P_\ext}(0)u\|_{\Hbext^{s-1,\gamma}(X_\ext)} + \|u\|_{\Hbext^{s_0,\gamma-1}(X_\ext)}\bigr).
  \end{equation}
  From an analogous estimate on the dual spaces, we then deduce the Fredholm property of the map~\eqref{Eq0Op}.

  \pfstep{Kernel.} Suppose now $u\in\Hbext^{s,\gamma}(X_\ext)$ lies in the kernel of $\wh{P_\ext}(0)$. Then $u\in\bigcap_{N\in\R}\Hbext^{N,\gamma}(X_\ext)$ since~\eqref{Eq0Pf} holds in the strong sense: the finiteness of the right implies that of the left hand side. A normal operator argument implies that, in fact, $u\in\cA^0([r_{\rm e},r_+]\times\Sph^2)$, i.e.\ $u$ is bounded together with all of its b-derivatives (i.e.\ derivatives along $(r-r_{\rm e})\pa_r$ and spherical derivatives). We can thus integrate by parts to find
  \begin{equation}
  \label{Eq0PfIBP}
    0 = \int_{\Sph^2}\int_{r_{\rm e}}^{r_{\rm c}} \wh{P_\ext}(0)u\,\bar u\,r^2\,\dd r\,\dd\slg = \int_{\Sph^2}\int_{r_{\rm e}}^{r_{\rm c}} F_0 r^2|D_r u|^2 + |\slnabla u|^2 + \mu r^2|u|^2\,\dd r\,\dd\slg.
  \end{equation}
  The boundary term at $r=r_{\rm c}$ vanishes since $F_0(r_{\rm c})=0$. In the case $\mu>0$, the vanishing of~\eqref{Eq0PfIBP} implies $u=0$ for $r_{\rm e}\leq r\leq r_{\rm c}$. Since $u$ thus vanishes to infinite order at $r=r_{\rm c}$, a simple energy estimate in $r>r_{\rm c}$, where $\wh{P_\ext}(0)$ is hyperbolic (with $r$ a time function) implies the vanishing of $u$ also for $r>r_{\rm c}$ (cf.\ \cite[Lemma~1]{ZworskiRevisitVasy}); therefore, $u=0$. In the case $\mu=0$, we deduce from~\eqref{Eq0PfIBP} that $u$ equals a constant $c$ for $r_{\rm e}\leq r\leq r_{\rm c}$. Since constants lie in the kernel of $\wh{P_\ext}(0)$, also $u-c$ lies in $\ker\wh{P_\ext}(0)$, and since $u-c$ is smooth and vanishes to infinite order at $r=r_{\rm c}$, energy estimates in $r>r_{\rm c}$ imply $u-c=0$ on $X_\ext$.

  \pfstep{Cokernel.} We next show that the cokernel of $\wh{P_\ext}(0)$ is trivial when $\mu>0$. We adapt the arguments from \cite[Lemma~3.4]{HintzXieSdS}. Consider thus $u^*\in\ker\wh{P_\ext}(0)^*$; by b-ellipticity and a normal operator argument near $r=r_{\rm e}$, we have $u^*\in\cA^0([r_{\rm e},r_{\rm e}+\delta)\times\Sph^2)$, further $u^*$ is smooth for $r\neq r_{\rm c}$, vanishes for $r>r_{\rm c}$, and lies in $H^{\frac12-}$ near $r=r_{\rm c}$. Projecting $u^*$ in the angular variables to the space of spherical harmonics of degree $\ell$, we furthermore have
  \begin{equation}
  \label{Eq0PfSep}
    0 = (r^{-2}D_r F_0 r^2 D_r + r^{-2}\ell(\ell+1) + \mu)u^* = 0.
  \end{equation}
  Upon multiplication by $r-r_{\rm c}$, this is a regular-singular ODE at $r=r_{\rm c}$ with double indicial root $0$, and hence $u^*=c_1\log(r_{\rm c}-r)+c_0+\tilde u^*$ in $r<r_{\rm c}$ for some $c_1,c_2\in\C$ where $\tilde u^*\in\cA^{1-}((r_{\rm c}-\delta,r_{\rm c}])$ is conormal at $r=r_{\rm c}$ and bounded by $(r_{\rm c}-r)^{1-\eta}$ for all $\eta>0$ (together with all derivatives along $(r_{\rm c}-r)\pa_r$ and spherical derivatives). Letting $H$ denote the Heaviside function, one now computes that $(r^{-2}D_r F_0 r^2 D_r+r^{-2}\ell(\ell+1)+\mu)(c_1\log(r_{\rm c}-r)_++c_0 H(r_{\rm c}-r)+\tilde u^*)$ is equal to a nonzero multiple of $c_1\delta(r_{\rm c}-r)$ plus a distribution in $L^1_\loc$; thus we must have $c_1=0$, so $u^*=(c_0+\tilde u^*)H(r_{\rm c}-r)$ near $r=r_{\rm c}$. We may now multiply~\eqref{Eq0PfSep} by $r^2\ol{u^*}$, integrate over $[r_{\rm e},r_{\rm c}]$, and integrate by parts to obtain $\int_{r_{\rm e}}^{r_{\rm c}} \mu r^2|u^*|^2\,\dd r=0$, so $u^*=0$ in $(r_{\rm e},r_{\rm c})$. Since also $u^*=0$ on $(r_{\rm c},\infty)$, we have $\supp u^*\subset\{r_{\rm c}\}$; but $u^*\in H^{\frac12-}$ then implies that $u^*=0$ on $(r_{\rm e},\infty)$.

  We have also shown now that the Fredholm index of~\eqref{Eq0Op} is zero for $\mu>0$. Since $\wh{P_\ext}(0)$ is Fredholm between the $\mu$-independent spaces in~\eqref{Eq0Op}, its index is $\mu$-independent as well, and hence it is $0$ also for $\mu=0$. A direct computation shows that $\wh{P_\ext}(0)^*u_{(0)}^*=0$, which gives~\eqref{Eq0OpKer}.
\end{proof}

%%%%%%%%%%%%%%%%%%%%%%%%%%%%%%%%%%%%%%%%%%%%%%%%%%%%%%%%%%%%%%%%%%%%%%
\section{QNMs on near-extremal RNdS: proof of Theorem~\usref{ThmI}}
\label{SPf}

We now return to the study of the spectral family
\[
  \wh{P_\eps}(\varsigma) = \wh{P_\eps}(\kappa_{\rm C,\eps}\sigma)
\]
of $\Box_{g_\eps}+\mu$; see~\eqref{EqGCSpecFam}. First, we note that for every fixed $\eps\in(0,\eps_0)$, we have $\wt P_\eps=\wh{P_\eps}(\kappa_{\rm C,\eps}\sigma)\in\Diff^2(X_\eps)$ where
\[
  X_\eps := [r_{\rm e}-\eps,r_+]\times\Sph^2.
\]
Since $X_\eps$ contains the subextremal event horizon $r=r_{\rm e}$ and the subextremal cosmological horizon $r=r_{\rm c}$, while its hypersurfaces at $r=r_{\rm e}-\eps$ and $r=r_+>r_{\rm c}$ are spacelike, standard arguments \cite{VasyMicroKerrdS} imply that for $s>\max(\frac12-\Im\sigma,\frac12)$, there exists $\eps_1\in(0,\eps_0)$ such that the map
\begin{equation}
\label{EqPfPeps}
  \wh{P_\eps}(\kappa_{\rm C,\eps}\sigma) \colon \cX^s(X_\eps) := \{ u\in H^s(X_\eps) \colon \wh{P_\eps}(0)u\in H^{s-1}(X_\eps) \} \to H^{s-1}(X_\eps)
\end{equation}
is Fredholm when $\eps\in(0,\eps_1]$.\footnote{The threshold regularity is the maximum of the threshold $\frac12$ at the cosmological horizon for frequency $0$ and the threshold $\frac12-\Im\frac{\kappa_{\rm C,\eps}\sigma}{\kappa_{\rm e,\eps}}=\frac12-\Im\sigma+\cO(\eps)$ at the event horizon; here $\kappa_{\rm e,\eps}$ is the surface gravity of the event horizon of $g_\eps$. One can also directly quote the semi-Fredholm estimate~\eqref{EqPfHi} below.} Elements in its nullspace are automatically smooth on $X_\eps$, and hence nonzero such elements are resonant states as defined in~\eqref{EqIBoxSol}. Furthermore, the map~\eqref{EqPfPeps} has index $0$, as follows for sufficiently large $\Im\sigma$ from an energy estimate (cf.\ \cite[Proposition~12.18]{HintzMicro}). Thus, its inverse is finite-meromorphic for $\Im\sigma>\frac12-s$.

The technical heart of our argument is the proof of appropriate uniform estimates for $\wh{P_\eps}(\kappa_{\rm C,\eps}\sigma)$ as $\eps\searrow 0$ on function spaces adapted to the nature of the family
\begin{equation}
\label{EqPftildeP}
  \wt P=(\wt P_\eps)_{\eps\in(0,\eps_0)},\quad \wt P_\eps:=\wh{P_\eps}(\kappa_{\rm C,\eps}\sigma),
\end{equation}
as a q-differential operator on $\wt X$ (Lemma~\ref{LemmaGCTot}). Fix the smooth defining functions
\[
  \rho_\NH = r-r_{\rm C},\quad
  \rho_\ext = \frac{\eps}{r-r_{\rm C}} \in \CI(\wt X)
\]
of $X_\NH$, $X_\ext$. We will localize to neighborhoods of $X_\NH$ and $X_\ext$ using cutoff functions
\begin{subequations}
\begin{equation}
\label{EqPfCutoffs}
  \chi_\NH,\ \chi_\ext\in\CI(\wt X);
\end{equation}
concretely, fixing $\chi_0\in\CIc([0,\min(\frac12,r_{\rm c}-r_{\rm e}-2\delta)))$ with $\chi_0=1$ near $0$, we may take
\begin{equation}
\label{EqPfCutoffs2}
  \chi_\NH = \chi_0(\rho_\NH),\quad
  \chi_\ext = \chi_0(\rho_\ext).
\end{equation}
\end{subequations}
Let $\Omega\subset\cV(\Sph^2)$ be as in Definition~\ref{DefNHFredb}, and recall~\eqref{EqGCVqSpan}.

\begin{definition}[Weighted q-Sobolev spaces]
\label{DefPfQ}
  Let $s\in\N_0$, $\alpha_\NH,\alpha_\ext\in\R$. Then $\bar H_{\qop,\eps}^{s,\alpha_\NH,\alpha_\ext}(X_\eps)$ is the vector space $H^s(X_\eps)$ equipped with the $\eps$-dependent squared norm
  \begin{equation}
  \label{EqPfQNorm}
    \|u\|_{\bar H_{\qop,\eps}^{s,\alpha_\NH,\alpha_\ext}(X_\eps)}^2 := \sum_{i+|\beta|\leq s} \int_{\Sph^2}\int_{r_{\rm e}-\eps}^{r_+} \bigl| \rho_\NH^{-\alpha_\NH}\rho_\ext^{-\alpha_\ext} \bigl((r-r_{\rm C})\pa_r\bigr)^i \Omega^\beta u(r,\omega)\bigr|^2\,r^2\dd r\,\dd\slg.
  \end{equation}
\end{definition}

This is analogous to \cite[Definition~2.5]{HintzKdSMS}. Given $\wt L\in\rho_\NH^{-\beta_\NH}\rho_\ext^{-\beta_\ext}\Diffq^m(\wt X)$ (i.e.\ $\rho_\NH^{\beta_\NH}\rho_\ext^{\beta_\ext}\wt L\in\Diffq^m(\wt X)$), given on the $\eps$-level set $X_\eps$ of $\wt X$ by $\wt L_\eps\in\Diff^m(X_\eps)$, and given $\eps_1\in(0,\eps_0)$, there exists a constant $C$ such that for all $\eps\in(0,\eps_1]$,
\begin{equation}
\label{EqPfQUnif}
  \|\wt L_\eps u\|_{\bar H_{\qop,\eps}^{s-m,\alpha_\NH-\beta_\NH,\alpha_\ext-\beta_\ext}(X_\eps)} \leq C\|u\|_{\bar H_{\qop,\eps}^{s,\alpha_\NH,\alpha_\ext}(X_\eps)}.
\end{equation}
That is, $\wt L_\eps$ is \emph{uniformly bounded} as a map between q-Sobolev spaces.

Near $X_\NH$ and $X_\ext$, we can relate~\eqref{EqPfQNorm} to simpler, uniformly (in $\eps$) equivalent, norms. To wit,
\begin{subequations}
\begin{align}
\label{EqPfqNH}
  \|\chi_\NH u\|_{\bar H_{\qop,\eps}^{s,\alpha_\NH,\alpha_\ext}(X_\eps)} &\sim \eps^{-\alpha_\NH+\frac12}\|\chi_\NH u\|_{\bar H_\bop^{s,\alpha_\ext-\alpha_\NH}(X_\NH)}, \\
\label{EqPfqExt}
  \|\chi_\ext u\|_{\bar H_{\qop,\eps}^{s,\alpha_\NH,\alpha_\ext}(X_\eps)} &\sim \eps^{-\alpha_\ext}\|\chi_\ext u\|_{\bar H_\bop^{s,\alpha_\NH-\alpha_\ext}(X_\ext)}.
\end{align}
\end{subequations}
Here `$\sim$' means that, for all $u$, the left hand side is bounded by a \emph{uniform} constant times the right hand side and vice versa. Regarding the first norm equivalence, we can reduce to the case $\alpha_\NH=0$ by multiplying both sides by $\eps^{\alpha_\NH}$ and relabeling $\alpha_\ext-\alpha_\NH$ as $\alpha_\ext$. We change variables via $z=\frac{r-r_{\rm C}}{\eps}-1$, so $(r-r_{\rm C})\pa_r=(z+1)\pa_z$ and $\rho_\ext=(z+1)^{-1}$. Comparison with~\eqref{EqNHFredb} gives~\eqref{EqPfqNH}, the extra power of $\eps^{\frac12}$ being due to $\dd r=\eps\,\dd z$. To prove~\eqref{EqPfqExt}, we may reduce to $\alpha_\ext=0$; comparison with~\eqref{Eq0Hb} and recalling $r_{\rm C}=r_{\rm e}-2\eps$ then gives~\eqref{EqPfqExt}.

As a consequence of~\eqref{EqPfqNH}--\eqref{EqPfqExt}, we have
\begin{equation}
\label{EqPfq}
  \|u\|_{\bar H_{\qop,\eps}^{s,\alpha_\NH,\alpha_\ext}(X_\eps)} \sim \eps^{-\alpha_\NH+\frac12}\|\chi_\NH u\|_{\Hbext^{s,\alpha_\ext-\alpha_\NH}(X_\NH)} + \eps^{-\alpha_\ext}\|\chi_\ext u\|_{\Hbext^{s,\alpha_\NH-\alpha_\ext}(X_\ext)}\,.
\end{equation}
We can use the right hand side to define weighted q-Sobolev norms also for $s\in\R$.

The starting point of our analysis of $\wt P$ is the following uniform high frequency estimate.

\begin{prop}[q-regularity estimate]
\label{PropPfHi}
  Let $\alpha_\NH,\alpha_\ext\in\R$ and $s>s_0>\max(\frac12-\Im\sigma,\frac12)$. Then there exist $\eps_1\in(0,\eps_0)$ and a constant $C$ such that for all $\eps\in(0,\eps_1]$,
  \begin{equation}
  \label{EqPfHi}
    \| u \|_{\bar H_{\qop,\eps}^{s,\alpha_\NH,\alpha_\ext}(X_\eps)} \leq C\Bigl( \| \wt P_\eps u \|_{\bar H_{\qop,\eps}^{s-1,\alpha_\NH,\alpha_\ext}(X_\eps)} + \| u \|_{\bar H_{\qop,\eps}^{s_0,\alpha_\NH,\alpha_\ext}(X_\eps)}\Bigr).
  \end{equation}
\end{prop}
\begin{proof}
  Starting with~\eqref{EqPfq}, we can estimate $\chi_\NH u$ for $z\leq 4$ (where $\chi_\NH u=u$ for small $\eps$) as in~\eqref{EqNHFredRad}, except with $\wt P_\eps=\wh{P_\eps}(\kappa_{\rm C,\eps}\sigma)$ on the right. This estimate holds uniformly for all sufficiently small $\eps>0$ by the stability of the radial point and propagation estimates underlying~\eqref{EqNHFredRad}; see \cite[Remark~2.5 and \S{2.7}]{VasyMicroKerrdS}. Similarly, we can estimate $\chi_\ext u$ for $r\geq r_{\rm c}-\delta$ (where $\chi_\ext u=u$ for small $\eps$) as in~\eqref{Eq0PfRad}, except with $\wt P_\eps$ on the right.

  Define now $\psi(\eps,r)=\psi_0(z)\psi_1(r)$, $z=\frac{r-r_{\rm C}}{\eps}-1$, where $\psi_0\in\CI(\R)$ equals $0$ for $z\leq 2$ and $1$ for $z\geq 3$, and $\psi_1\in\CI(\R)$ equals $0$ for $r\geq r_{\rm c}-\frac{\delta}{3}$ and $1$ for $r\leq r_{\rm c}-\frac{2\delta}{3}$. With $(1-\psi)u$ already controlled, it remains to prove for $\psi u$ the uniform elliptic estimate
  \begin{equation}
  \label{EqPfHiEll}
    \|\psi u\|_{\bar H_{\qop,\eps}^{s,\alpha_\NH,\alpha_\ext}(X_\eps)} \leq C\Bigl(\|\wt P_\eps u\|_{\bar H_{\qop,\eps}^{s-2,\alpha_\NH,\alpha_\ext}(X_\eps)} + \|u\|_{\bar H_{\qop,\eps}^{s_0,\alpha_\NH,\alpha_\ext}(X_\eps)}\Bigr).
  \end{equation}
  Now, for $z\geq\frac32$, the operator $\wh{P_\NH}(\sigma)$ is elliptic as a b-operator, i.e.\ its leading order part is a positive definite quadratic form in $(z+1)\pa_z$ and $\slnabla$; similarly, for $r\leq r_{\rm c}-\frac{\delta}{6}$, the operator $\wh{P_\ext}(0)$ is b-elliptic, i.e.\ its leading order part is a positive definite quadratic form in $(r-r_{\rm e})\pa_r=(1-2\rho_\ext)(r-r_{\rm C})\pa_r$ and $\slnabla$. By Lemma~\ref{LemmaGCTot} and the discussion around~\eqref{EqGCVqSpan}, the leading order part of $\wt P_\eps$ is therefore a positive definite quadratic form in $(r-r_{\rm C})\pa_r$ and $\slnabla$ in the region $z\geq 2$, $r\leq r_{\rm c}-\frac{\delta}{3}$ and for all sufficiently small $\eps$. This implies~\eqref{EqPfHiEll}. (In more detail, one can reduce the proof of~\eqref{EqPfHiEll} to $\alpha_\ext=0$, and then to $\alpha_\NH=0$ by conjugating $\wt P_\eps$ by $(r-r_{\rm C})^{-\alpha_\NH}$, which does not affect its ellipticity properties. Passing from $r$ to $\tilde r:=-\log(r-r_{\rm C})$ turns $\wt P_\eps$ into a uniformly bounded family of uniformly elliptic operators on appropriate subsets of $\R_{\tilde r}\times\Sph^2$, and~\eqref{EqPfHiEll} is the corresponding elliptic estimate.)
\end{proof}

Below, we shall use the fact that the estimate~\eqref{EqPfHi} holds uniformly for all $\sigma$ (entering via~\eqref{EqPftildeP}) in a fixed compact subset of $\C$.

Now, $\varsigma=\kappa_{\rm C,\eps}\sigma$ is not a QNM of $\Box_{g_\eps}+\mu$ if and only if $\wt P_\eps=\wh{P_\eps}(\kappa_{\rm C,\eps}\sigma)$ is injective on $H^s(X_\eps)$ or, equivalently, surjective onto $H^{s-1}(X_\eps)$ with domain $\cX^s(X_\eps)$. Our strategy for proving the injectivity/surjectivity of $\wt P_\eps$ for appropriate values of $\sigma$ is to estimate the second term in~\eqref{EqPfHi} using the estimates for the two normal operators. The details differ depending on the mapping properties of $\wh{P_\ext}(0)$, which is determined by the value of the scalar field mass $\mu$ (see Proposition~\ref{Prop0}\eqref{It0Ker}):
\begin{enumerate}
\item The simpler setting is when $\wh{P_\ext}(0)$ is injective (i.e.\ $\mu>0$). QNMs of $\Box_{g_\eps}+\mu$ near $\kappa_{\rm C,\eps}$ times those of $P_\NH$ can be detected using a Grushin problem and Rouch\'e's theorem.
\item When $\wh{P_\ext}(0)$ is not injective (i.e.\ $\mu=0$) but $\wh{P_\NH}(\sigma)$ is, then $\wt P_\eps$ can, using a carefully chosen Grushin problem, be shown to be surjective unless $\sigma=0$ (\S\ref{SssPf0N}). We detect QNMs of $\Box_{g_\eps}$ using a Grushin problem featuring two augmentations (\S\ref{SssPf0Y}).
\end{enumerate}

\textit{Henceforth, we shall write `$A\lesssim B$' for $\eps$-dependent quantities $A,B$ when there exists a constant $C$ such that $A\leq C B$ for all $\eps\in(0,\eps_1]$  for some $\eps_1\in(0,\eps_0)$.}

%%%%%%%%%%%%%%%%%%%%%%%%%%%%%%%%%%%%%%%%%%%%%%%%%%
\subsection{Massive scalar waves}
\label{SsPfP}

We consider scalar field masses
\[
  \mu>0.
\]

%%%%%%%%%%%%%%%%%%%%%%%%%%%%%%
\subsubsection{Absence of QNMs}

By Proposition~\ref{Prop0}, we have an estimate
\begin{equation}
\label{EqPfP0}
  \|u\|_{\Hbext^{s,\gamma}(X_\ext)} \leq C\|\wh{P_\ext}(0)u\|_{\Hbext^{s-1,\gamma}(X_\ext)}
\end{equation}
for any fixed $s>\frac12$ and $\gamma\in(-\frac12,\frac12)$.

\begin{prop}[Absence of QNMs]
\label{PropPfPY}
  Let $\sigma\in\C$, $\sigma\notin\QNM_\NH(\mu)$. Then there exists $\eps_1\in(0,\eps_0)$ such that for all $\eps\in(0,\eps_1]$, we have $\kappa_{\rm C,\eps}\sigma\notin\QNM(r_{\rm C},r_{\rm e},r_{\rm c},\mu)$ where $r_{\rm C}=r_{\rm e}-2\eps$.
\end{prop}
\begin{proof}
  Consider the estimate~\eqref{EqPfHi} for $s\geq s_0+2$ where we fix $s_0$ with $s_0>\max(\frac12-\Im\sigma,\frac12)$, and for $\alpha_\NH,\alpha_\ext\in\R$ with $\gamma:=\alpha_\NH-\alpha_\ext\in(-\frac12,\frac12)$.

  \pfstep{Estimate near $X_\ext$ via inversion of $\wh{P_\ext}(0)$.} We use the zero energy estimate~\eqref{EqPfP0} to bound the second term on the right in~\eqref{EqPfHi} using~\eqref{EqPfqExt} by a uniform constant times
  \begin{align*}
    &\eps^{-\alpha_\ext}\|\chi_\ext u\|_{\Hbext^{s_0,\gamma}(X_\ext)} + \|(1-\chi_\ext)u\|_{\bar H_{\qop,\eps}^{s_0,\alpha_\NH,\alpha_\ext}(X_\eps)} \\
    &\qquad \lesssim \eps^{-\alpha_\ext}\|\wh{P_\ext}(0)(\chi_\ext u)\|_{\Hbext^{s_0-1,\gamma}(X_\ext)} + \|u\|_{\bar H_{\qop,\eps}^{s_0,\alpha_\NH,\alpha_\ext-1}(X_\eps)}.
  \end{align*}
  We proceed to estimate the first term on the right by
  \begin{align}
    &\eps^{-\alpha_\ext}\|\wt P_\eps(\chi_\ext u)\|_{\Hbext^{s_0-1,\gamma}(X_\ext)} + \eps^{-\alpha_\ext}\|(\wt P_\eps-\wh{P_\ext}(0))(\chi_\ext u)\|_{\Hbext^{s_0-1,\gamma}(X_\ext)} \nonumber\\
    &\qquad \lesssim \|\wt P_\eps u\|_{\bar H_{\qop,\eps}^{s_0-1,\alpha_\NH,\alpha_\ext}(X_\eps)} + \|[\wt P_\eps,\chi_\ext]u\|_{\bar H_{\qop,\eps}^{s_0-1,\alpha_\NH,\alpha_\ext}(X_\eps)} \nonumber\\
    &\qquad \hspace{12em} + \|(\wt P_\eps-\wh{P_\ext}(0))(\chi_\ext u)\|_{\bar H_{\qop,\eps}^{s_0-1,\alpha_\NH,\alpha_\ext}(X_\eps)} \nonumber\\
  \label{EqPfPYExt}
    &\qquad \lesssim \|\wt P_\eps u\|_{\bar H_{\qop,\eps}^{s_0-1,\alpha_\NH,\alpha_\ext}(X_\eps)} + \|u\|_{\bar H_{\qop,\eps}^{s_0+1,\alpha_\NH,\alpha_\ext-1}(X_\eps)};
  \end{align}
  in the passage to the final line we used $[\wt P_\eps,\chi_\ext]\in\rho_\ext^N\Diffq^1(\wt X)$ (for all $N$) and $(\wt P_\eps-\wh{P_\ext}(0))\circ\chi_\ext\in\rho_\ext\Diffq^2(\wt X)$ (see Lemma~\ref{LemmaGCTot}) together with~\eqref{EqPfQUnif}. Strengthening the $X_\ext$-weight from $\alpha_\ext-1$ to $\alpha_\ext-\eta$ for $\eta\in(0,1]$ increases the norm; hence, we have now proved
  \begin{equation}
  \label{EqPfPY1}
    \| u \|_{\bar H_{\qop,\eps}^{s,\alpha_\NH,\alpha_\ext}(X_\eps)} \lesssim \| \wt P_\eps u \|_{\bar H_{\qop,\eps}^{s-1,\alpha_\NH,\alpha_\ext}(X_\eps)} + \| u \|_{\bar H_{\qop,\eps}^{s_0+1,\alpha_\NH,\alpha_\ext-\eta}(X_\eps)}.
  \end{equation}
  This improves on~\eqref{EqPfHi} in the $X_\ext$-weight, at an acceptable loss in the q-regularity order. We shall use this estimate for a value $\eta>0$ for which $\gamma+\eta=\alpha_\NH-(\alpha_\ext-\eta)\in(-\frac12,\frac12)$ still.

  \pfstep{Estimate near $X_\NH$ via inversion of $\wh{P_\NH}(\sigma)$.} We next exploit $\sigma\notin\QNM_\NH(\mu)$ by using the estimate~\eqref{EqNHFred2}, with $s_0+1$ in place of $s$ and for $\alpha:=(\alpha_\ext-\eta)-\alpha_\NH=-\gamma-\eta\in(-\frac12,\frac12)$, in a similar fashion. Thus,
  \begin{equation}
  \label{EqPfPY2}
    \|u\|_{\bar H_{\qop,\eps}^{s_0+1,\alpha_\NH,\alpha_\ext-\eta}(X_\eps)} \lesssim \eps^{-\alpha_\NH+\frac12}\|\chi_\NH u\|_{\Hbext^{s_0+1,\alpha}(X_\NH)} + \|u\|_{\bar H_{\qop,\eps}^{s_0+1,\alpha_\NH-1,\alpha_\ext-\eta}(X_\eps)},
  \end{equation}
  with the first summand further bounded by
  \begin{align}
    &\eps^{-\alpha_\NH+\frac12}\|\wh{P_\NH}(\sigma)(\chi_\NH u)\|_{\Hbext^{s_0,\alpha}(X_\NH)} \nonumber\\
    &\qquad \lesssim \eps^{-\alpha_\NH+\frac12}\|\wt P_\eps(\chi_\NH u)\|_{\Hbext^{s_0,\alpha}(X_\NH)} + \eps^{-\alpha_\NH+\frac12}\|(\wt P_\eps-\wh{P_\NH}(\sigma))(\chi_\NH u)\|_{\Hbext^{s_0,\alpha}(X_\NH)} \nonumber\\
    &\qquad \lesssim \|\wt P_\eps u\|_{\bar H_{\qop,\eps}^{s_0,\alpha_\NH,\alpha_\ext-\eta}(X_\eps)} + \|[\wt P_\eps,\chi_\NH]u\|_{\bar H_{\qop,\eps}^{s_0,\alpha_\NH,\alpha_\ext-\eta}(X_\eps)} \nonumber\\
    &\qquad \hspace{12em} + \|(\wt P_\eps-\wh{P_\NH}(\sigma))(\chi_\NH u)\|_{\bar H_{\qop,\eps}^{s_0,\alpha_\NH,\alpha_\ext-\eta}(X_\eps)} \nonumber\\
  \label{EqPfPY3}
    &\qquad \lesssim \|\wt P_\eps u\|_{\bar H_{\qop,\eps}^{s_0,\alpha_\NH,\alpha_\ext-\eta}(X_\eps)} + \|u\|_{\bar H_{\qop,\eps}^{s_0+2,\alpha_\NH-1,\alpha_\ext-\eta}(X_\eps)};
  \end{align}
  here we used $[\wt P_\eps,\chi_\NH]$, $(\wt P_\eps-\wh{P_\NH}(\sigma))\circ\chi_\NH\in\rho_\NH\Diffq^2(\wt X)$. Plugging this into~\eqref{EqPfPY1} yields
  \begin{equation}
  \label{EqPfPY4}
    \|u\|_{\bar H_{\qop,\eps}^{s,\alpha_\NH,\alpha_\ext}(X_\eps)} \lesssim \| \wt P_\eps u \|_{\bar H_{\qop,\eps}^{s-1,\alpha_\NH,\alpha_\ext}(X_\eps)} + \| u \|_{\bar H_{\qop,\eps}^{s_0+2,\alpha_\NH-1,\alpha_\ext-\eta}(X_\eps)}.
  \end{equation}
  Since $s_0+2\leq s$, the second term on the right is $\lesssim\eps^\eta\|u\|_{\bar H_{\qop,\eps}^{s,\alpha_\NH,\alpha_\ext}(X_\eps)}$; for sufficiently small $\eps>0$, this can be absorbed into the left hand side. This, finally, yields the existence of $\eps_1\in(0,\eps_0)$ such that
  \[
    \|u\|_{\bar H_{\qop,\eps}^{s,\alpha_\NH,\alpha_\ext}(X_\eps)} \lesssim \| \wt P_\eps u \|_{\bar H_{\qop,\eps}^{s-1,\alpha_\NH,\alpha_\ext}(X_\eps)},\quad \eps\leq\eps_1.
  \]
  In particular, $\wt P_\eps=\wh{P_\eps}(\kappa_{\rm C,\eps}\sigma)$ is injective on $H^s(X_\eps)$ for such $\eps$.
\end{proof}

By the local uniformity of the estimate~\eqref{EqPfHi}, the above proof in fact yields the following stronger statement:

\begin{prop}[Absence of QNMs: uniform statement]
\label{PropPfPYUnif}
  Let $K\subset\C$ be a compact set disjoint from $\QNM_\NH(\mu)$. Then there exists $\eps_1\in(0,\eps_0)$ such that for all $\eps\in(0,\eps_1]$, the set $\{\kappa_{\rm C,\eps}\sigma\colon\sigma\in K\}$ is disjoint from $\QNM(r_{\rm C},r_{\rm e},r_{\rm c},\mu)$ (with $r_{\rm C}=r_{\rm e}-2\eps$).
\end{prop}

%%%%%%%%%%%%%%%%%%%%%%%%%%%%%%
\subsubsection{Existence of QNMs}
\label{SssPfPN}

We now turn to the existence of QNMs for $\Box_{g_\eps}+\mu$ near points $\kappa_{\rm C,\eps}\sigma$ where $\sigma$ is a near-horizon QNM.

\begin{thm}[Existence of QNMs]
\label{ThmPfPN}
  Let $\sigma_0\in\QNM_\NH(\mu)$, and write $m(\mu;\sigma_0)$ for the multiplicity of $\sigma_0$. Let $r_0>0$ be so small that for all $\sigma\in\QNM_\NH(\mu)\setminus\{\sigma_0\}$ we have $|\sigma-\sigma_0|\geq 2 r_0$. Then there exists $\eps_1\in(0,\eps_0)$ such that for all $\eps\in(0,\eps_0]$, there are $m(\mu;\sigma_0)$ many QNMs $\varsigma\in\QNM(r_{\rm C},r_{\rm e},r_{\rm c},\mu)$, $r_{\rm C}=r_{\rm e}-2\eps$, of $\Box_{g_\eps}+\mu$ (counted with multiplicity) with
  \[
    \Bigl|\frac{\varsigma}{\kappa_{\rm C,\eps}}-\sigma_0\Bigr| < r_0.
  \]
  Denote by $\Sigma_\eps$ the set of these QNMs $\varsigma$. Then:
  \begin{enumerate}
  \item\label{ItPfPNConv} $\Sigma_\eps\subset i\R$, and $\{\frac{\varsigma}{\kappa_{\rm C,\eps}}\colon\varsigma\in\Sigma_\eps\}\to\{\sigma_0\}$ in the Hausdorff distance sense as $\eps\to 0$;
  \item\label{ItPfPNPole} $\wh{P_\eps}(\zeta)^{-1}$ has a pole of order $1$ at $\zeta=\varsigma$ for every such $\varsigma$.
  \end{enumerate}
  Finally:
  \begin{enumerate}
  \setcounter{enumi}{2}
  \item\label{ItPfPNConvState} Let $\ell\in\N_0$ be such that $\Sigma_\eps$ contains a (necessarily unique) element $\varsigma_\eps=\kappa_{\rm C,\eps}(\sigma_0+o(1))$ for which a resonant state with angular dependence $Y_\ell$ (a fixed degree $\ell$ spherical harmonic) exists.\footnote{Due to the spherical symmetry of the RNdS metric, there exists such $\ell$ for every QNM; and unless there are coincidences among the QNMs in Theorem~\ref{ThmNHQNM}, $\ell$ is uniquely determined by the QNM.} Then we can normalize such a resonant state $u_\eps\in\CI(X_\eps)$ of $\wh{P_\eps}(\varsigma_\eps)$ in such a way that
    \begin{equation}
    \label{EqPfNConvState}
      \Bigl\|u_\eps(r,\omega) - u_0\Bigl(\frac{r-r_{\rm e}}{\eps}+1,\omega\Bigr)\Bigr\|_{\cC_{\bop,\eps}^{k,\theta}(X_\eps)} \xra{\eps\to 0} 0
    \end{equation}
    for all $\theta<1$, where $u_0$ is a resonant state of $\wh{P_\NH}(\sigma_0)$ (i.e.\ of the form~\eqref{EqNHQNMResState} for a suitable value of $n$). Here,
    \begin{equation}
    \label{EqPfNConvNorm}
      \|v\|_{\cC_{\bop,\eps}^{k,\theta}(X_\eps)} = \sum_{i+|\beta|\leq k}\sup_{r_{\rm C}\leq r\leq r_+} \Bigl(\frac{r-r_{\rm C}}{\eps}\Bigr)^\theta\,\bigl|\bigl((r-r_{\rm C})\pa_r\bigr)^i\Omega^\beta v(r,\omega)\bigr|,\quad r_{\rm C}:=r_{\rm e}-2\eps.
    \end{equation}
  \end{enumerate}
\end{thm}

Due to the weight $\frac{r-r_{\rm C}}{\eps}$ in~\eqref{EqPfNConvNorm}, the convergence~\eqref{EqPfNConvState} implies in particular the localization of $u_\eps$ to $r-r_{\rm C}\lesssim\eps$.

\begin{rmk}[Spherical harmonics]
\label{RmkPfPNYlm}
  Separation into spherical harmonics plays no role in the proof. We only use it in part~\eqref{ItPfPNConvState} for the clarity of the statement. In the (non-generic) case that $k\geq 2$ of the numbers $\lambda^+_\ell(\mu)+n$ for $\ell,n\in\N_0$ coincide, this QNM may split into up to $k$ different QNMs for $0<\eps\ll 1$.
\end{rmk}

\begin{rmk}[Co-resonant states]
\label{RmkPfPNCoRes}
  Repeating the arguments below regarding $u_\eps$ for the adjoint $\wh{P_\eps}(\kappa_{\rm C,\eps}\sigma)^*$, one can show that also the co-resonant state for the QNM $\varsigma_\eps\in\QNM(r_{\rm C},r_{\rm e},r_{\rm c},\mu)$ is well-approximated by the co-resonant state for the limiting near-horizon QNM in a space capturing $\frac12+\Im\sigma-\eta$ degrees of Sobolev regularity near $z=1$ and $\frac12-\eta$ degrees of Sobolev regularity near $r=r_{\rm c}$ (and arbitrary regularity in between), and almost $(z+1)^{-1}$ decay as $z\to\infty$. The latter localization property means that the contributions of the QNMs described by Theorem~\ref{ThmPfPN} in the late-time asymptotics of solutions of the Klein--Gordon equation are very small if the initial data are localized away from the event horizon.
\end{rmk}

Given the order $1$ property of the poles of $\wh{P_\eps}(\zeta)^{-1}$ asserted in Theorem~\ref{ThmPfPN}\eqref{ItPfPNPole}, the multiplicity of a QNM $\varsigma$ is equal to the dimension of the nullspace of $\wh{P_\eps}(\varsigma)$ on $\CI(X_\eps)$, and thus equal to the sum of $\ell(\ell+1)$ where $\ell$ ranges over all spherical harmonic degrees represented by resonant states associated with $\varsigma$. (For generic scalar field masses $\mu$, there is only ever one such $\ell$.)

We also note that if $\zeta$ is a QNM with resonant state $u$, then so is $-\bar\zeta$ with resonant state $\bar u$. Since upon restriction to fixed spherical harmonic dependence $Y_\ell(\omega)$ the space of resonant states is 1-dimensional, the near-horizon QNMs located on the negative imaginary axis cannot split; this proves the first half of part~\eqref{ItPfPNConv}. We begin the proof of the rest of Theorem~\ref{ThmPfPN}. In view of Proposition~\ref{PropPfPY}, we may shrink the value of $r_0$ throughout the proof, as long as it remains independent of $\eps$.

\pfstep{Step 1.~Grushin problem for the near-horizon operator.} For notational simplicity, we consider only the case $m(\mu;\sigma_0)=1$. Let $0\neq v_0\in\cA^1(X_\NH)$ be a resonant state, i.e.\ $\wh{P_\NH}(\sigma_0)v_0=0$, and let $0\neq v_0^*\in\bigcap_{\eta>0}\Hbsupp^{\frac12+\Im\sigma_0-\eta,-\alpha}(X_\NH)$, $\alpha\in(-\frac12,\frac12)$, be a co-resonant state, i.e.\ $\wh{P_\NH}(\sigma_0)^*v_0^*=0$ (cf.\ the arguments leading to~\eqref{EqNHFredustarReg}). Pick\footnote{Here $X_\NH^\circ=(0,\infty)_z\times\Sph^2_\omega$.} $w_0^\sharp,w_0^\flat\in\CIc(X_\NH^\circ)$ such that
\[
  \la v_0,w_0^\sharp\ra_{L^2(X_\NH)} \neq 0,\quad
  \la w_0^\flat,v_0^*\ra_{L^2(X_\NH)} \neq 0.
\]
(Here we write $\la f,g\ra_{L^2(X_\NH)}=\int_{\Sph^2}\int_0^\infty f(z)\ol{g(z)}\,\dd z\,\dd\slg.$) Thus $w_0^\flat$ spans the complement of the range of $\wh{P_\NH}(\sigma_0)$ as a map on the spaces in~\eqref{EqNHFred}. The augmented operator
\begin{equation}
\label{EqPfPNNHaug}
  P_\NH^\aug(\sigma) := \begin{pmatrix} \wh{P_\NH}(\sigma) & w_0^\flat \\ \la\cdot,w_0^\sharp\ra_{L^2(X_\NH)} & 0 \end{pmatrix}
\end{equation}
is then Fredholm of index $0$ between the direct sum of the spaces in~\eqref{EqNHFred} with $\C$. Since it is invertible for $\sigma=\sigma_0$, it moreover satisfies uniform bounds
\begin{equation}
\label{EqPfPNNHEst}
  \|(u,c)\|_{\Hbext^{s,\alpha}(X_\NH)\oplus\C} \leq C\|P_\NH^\aug(\sigma)(u,c)\|_{\Hbext^{s-1,\alpha}(X_\NH)\oplus\C}
\end{equation}
for $|\sigma-\sigma_0|<2 r_0$ for sufficiently small $r_0>0$. Writing the inverse as
\begin{equation}
\label{EqPfPNNHInv}
  P_\NH^\aug(\sigma)^{-1} = \begin{pmatrix} A(\sigma) & B(\sigma) \\ C(\sigma) & D(\sigma) \end{pmatrix},
\end{equation}
we have $D(\sigma_0)=0$. Here $D(\sigma)$ is a $1\times 1$ matrix, where $1=m(\mu;\sigma_0)$; i.e.\ it is a complex number. The Schur complement formula expresses $\wh{P_\NH}(\sigma)^{-1}$ in terms of $D(\sigma)^{-1}$ and implies that $m(\mu;\sigma_0)$ is equal to the order of vanishing of $D(\sigma)$ at $\sigma=\sigma_0$. (In the case $m(\mu;\sigma_0)>1$, one instead works with $\tilde m:=\dim\ker\wh{P_\NH}(\sigma_0)\leq m(\mu;\sigma_0)$ many $w_0^\sharp,w_0^\flat$ such that the span of the $w_0^\flat$ complements the range of $\wh{P_\NH}(\sigma_0)$, while the linear functionals given by the $\la\cdot,w_0^\sharp\ra_{L^2(X_\NH)}$ are linearly independent on the kernel of $\wh{P_\NH}(\sigma_0)$. Then $D(\sigma)$ is an $\tilde m\times\tilde m$ matrix, and $\det D(\sigma)$ has a zero of order $m(\mu;\sigma_0)$ at $\sigma=\sigma_0$.)

%%%%%%%%%%
\pfstep{Step 2.\ Grushin problem for the spectral family.} Consider now the augmentation
\begin{equation}
\label{EqPfPNAugOp}
  \wt P_\eps^\aug(\sigma) := \begin{pmatrix} \wh{P_\eps}(\kappa_{\rm C,\eps}\sigma) & w_0^\flat \\ \la\cdot,w_0^\sharp\ra_{L^2(X_\NH)}  & 0 \end{pmatrix}
\end{equation}
of $\wt P_\eps=\wh{P_\eps}(\kappa_{\rm C,\eps}\sigma)$. Taking into account the $\eps$-scaling in~\eqref{EqPfqNH}, we introduce
\begin{equation}
\label{EqPfPNCNorm}
  |c|_{\eps^q\C} := \eps^{-q}|c|,\quad c\in\C,
\end{equation}
and claim:

\begin{lemma}[Uniform estimates for the augmented operator]
\label{LemmaPfPNAug}
  Let $s\geq s_0+2$ where $s_0>\max(\frac12-\Im\sigma_0,\frac12)$ and $\alpha_\NH,\alpha_\ext\in\R$ with $\alpha_\NH-\alpha_\ext\in(-\frac12,\frac12)$. Then there exist $r_0>0$ and $\eps_1\in(0,\eps_0)$ such that for all $\sigma$ with $|\sigma-\sigma_0|<2 r_0$, we have a uniform estimate
  \begin{equation}
  \label{EqPfPNAug}
    \|(u,c)\|_{\bar H_{\qop,\eps}^{s,\alpha_\NH,\alpha_\ext}(X_\eps)\oplus\eps^{\alpha_\NH-\frac12}\C} \leq C\|\wt P_\eps^\aug(\sigma)(u,c)\|_{\bar H_{\qop,\eps}^{s-1,\alpha_\NH,\alpha_\ext}(X_\eps)\oplus\eps^{\alpha_\NH-\frac12}\C}
  \end{equation}
  for all $\eps\in(0,\eps_1]$.
\end{lemma}

We remark that the weights in~\eqref{EqPfPNAug} are consistent with the mapping properties of $\wt P_\eps^\aug(\sigma)$: we can use~\eqref{EqPfqNH} to see that the off-diagonal terms in~\eqref{EqPfPNAugOp} obey uniform bounds
\begin{equation}
\label{EqPfPNAug2}
\begin{split}
  \|c w_0^\flat\|_{\bar H_{\qop,\eps}^{s-1,\alpha_\NH,\alpha_\ext}(X_\eps)} &\lesssim \eps^{-\alpha_\NH+\frac12}|c| = |c|_{\eps^{\alpha_\NH-\frac12}\C}, \\
  |\la u,w_0^\sharp\ra_{L^2(X_\NH)}|_{\eps^{\alpha_\NH-\frac12}\C} &\lesssim \eps^{-\alpha_\NH+\frac12}\|\chi_\NH u\|_{\Hbext^{s,\alpha_\ext-\alpha_\NH}(X_\NH)} \lesssim \|u\|_{\bar H_{\qop,\eps}^{s,\alpha_\ext,\alpha_\NH}(X_\eps)}.
\end{split}
\end{equation}
(In the second estimate, we use that $\chi_\NH w_0^\sharp=w_0^\sharp$ for small $\eps$. This estimate in fact holds for every $\alpha_\ext\in\R$ due to the compact support property of $w_0^\sharp$.)

\begin{proof}[Proof of Lemma~\usref{LemmaPfPNAug}]
  We first combine~\eqref{EqPfPY1} with~\eqref{EqPfPNAug2} to obtain the uniform (for $\sigma$ near $\sigma_0$ and $\eps$ near $0$) estimate
  \begin{equation}
  \label{EqPfPN1}
  \begin{split}
    &\| (u,c) \|_{\bar H_{\qop,\eps}^{s,\alpha_\NH,\alpha_\ext}\oplus\eps^{\alpha_\NH-\frac12}\C} \\
    &\qquad \lesssim \| \wt P_\eps^\aug(\sigma)(u,c) \|_{\bar H_{\qop,\eps}^{s-1,\alpha_\NH,\alpha_\ext}\oplus\eps^{\alpha_\NH-\frac12}\C} + \| (u,c) \|_{\bar H_{\qop,\eps}^{s_0+1,\alpha_\NH,\alpha_\ext-\eta}\oplus\eps^{\alpha_\NH-\frac12}\C}\,.
  \end{split}
  \end{equation}
  We estimate the second term on the right similarly to the arguments starting with~\eqref{EqPfPY2}, now using~\eqref{EqPfPNNHEst}; thus, it is bounded by $\|u\|_{\bar H_{\qop,\eps}^{s_0+1,\alpha_\NH-1,\alpha_\ext-\eta}}$ plus
  \begin{align}
    &\eps^{-\alpha_\NH+\frac12}\|P_\NH^\aug(\sigma)(\chi_\NH u,c)\|_{\Hbext^{s_0,\alpha}(X_\NH)\oplus\C} \nonumber\\
    &\qquad \lesssim \|\wt P_\eps^\aug(\sigma)(u,c)\|_{\bar H_{\qop,\eps}^{s_0,\alpha_\NH,\alpha_\ext-\eta}\oplus\eps^{\alpha_\NH-\frac12}\C} + \| [\wt P_\eps^\aug(\sigma),\chi_\NH\oplus I](u,c) \|_{\bar H_{\qop,\eps}^{s_0,\alpha_\NH,\alpha_\ext-\eta}\oplus\eps^{\alpha_\NH-\frac12}\C} \nonumber\\
    &\qquad \hspace{15.4em} + \|(\wt P_\eps^\aug(\sigma)-P_\NH^\aug(\sigma))(\chi_\NH u,c)\|_{\bar H_{\qop,\eps}^{s_0,\alpha_\NH,\alpha_\ext-\eta}\oplus\eps^{\alpha_\NH-\frac12}\C} \nonumber\\
  \label{EqPfPN2}
    &\qquad \lesssim \|\wt P_\eps^\aug(\sigma)(u,c)\|_{\bar H_{\qop,\eps}^{s_0,\alpha_\NH,\alpha_\ext-\eta}\oplus\eps^{\alpha_\NH-\frac12}\C} + \|u\|_{\bar H_{\qop,\eps}^{s_0+2,\alpha_\NH-1,\alpha_\ext-\eta}}.
  \end{align}
  Here we use that
  \[
    [\wt P_\eps^\aug(\sigma),\chi_\NH\oplus I] = \begin{pmatrix} [\wt P_\eps,\chi_\NH] & (1-\chi_\NH)w_0^\flat \\ \la(\chi_\NH-1)\cdot,w_0^\sharp\ra_{L^2(X_\NH)} & 0 \end{pmatrix}
  \]
  has vanishing off-diagonal entries for sufficiently small $\eps>0$, similarly for $\wt P_\eps^\aug(\sigma)-P_\NH^\aug(\sigma)$ (by definition of $\wt P_\eps^\aug(\sigma)$), and thus the commutator and difference terms can be estimated as in~\eqref{EqPfPY3}. Absorbing the second term in~\eqref{EqPfPN2} into the left hand side of~\eqref{EqPfPN1} yields~\eqref{EqPfPNAug}.
\end{proof}

%%%%%%%%%%
\pfstep{Step 3.\ Inverse of the augmented spectral family.} In view of~\eqref{EqPfPNAug} and the index $0$ property of $\wh{P_\eps}(\kappa_{\rm C,\eps}\sigma)$ and thus of $\wt P_\eps^\aug(\sigma)$, we have
\[
  \wt P_\eps^\aug(\sigma)^{-1} = \begin{pmatrix} A_\eps(\sigma) & B_\eps(\sigma) \\ C_\eps(\sigma) & D_\eps(\sigma) \end{pmatrix}
\]
where $D_\eps$ is holomorphic for $|\sigma-\sigma_0|<2 r_0$ and uniformly bounded as $\eps\searrow 0$ (as a linear map $\eps^{\alpha_\NH-\frac12}\C\to\eps^{\alpha_\NH-\frac12}\C$, i.e.\ as a complex number). We claim:

\begin{lemma}[Continuity of $D_\eps(\sigma)$]
\label{LemmaPfPNCont}
  $D_\eps(\sigma)$ converges uniformly to $D(\sigma)$ in the disk $\{|\sigma-\sigma_0|\leq r_0\}$.
\end{lemma}
\begin{proof}
  In view of the uniform boundedness and holomorphicity of $D_\eps(\sigma)$ for $|\sigma-\sigma_0|<2 r_0$, it suffices to prove pointwise convergence. For fixed $\sigma$, consider thus
  \begin{equation}
  \label{EqPfPNInv}
    (u_\eps,c_\eps) := \wt P_\eps^\aug(\sigma)^{-1}(0,1) \implies \wt P_\eps u_\eps + c_\eps w_0^\flat=0,\quad \la u_\eps,w_0^\sharp\ra_{L^2(X_\NH)}=1.
  \end{equation}
  We apply~\eqref{EqPfPNAugOp} with $\alpha_\NH=\frac12$ (and, correspondingly, $\alpha_\ext\in(0,1)$) and deduce uniform bounds
  \begin{equation}
  \label{EqPfPNInvBd}
    \|u_\eps\|_{\bar H_{\qop,\eps}^{s,\frac12,\alpha_\ext}(X_\eps)} + |c_\eps| \lesssim 1.
  \end{equation}
  By~\eqref{EqPfqNH}, this implies $\|\chi_\NH u_\eps\|_{\Hbext^{s,\alpha}(X_\NH)}\lesssim 1$ where $\alpha:=\alpha_\ext-\frac12\in(-\frac12,\frac12)$. Consider a subsequence $\chi_\NH u_{\eps_j}$ converging weakly to some $u_{\NH,0}$ in $\Hbext^{s,\alpha}(X_\NH)$, and thus strongly in $\Hbext^{s',\alpha'}(X_\NH)$ for $s'<s$, $\alpha'<\alpha$; here $\eps_j\searrow 0$ is such that, moreover, $c_{\eps_j}\to c_0\in\C$. We claim that
  \begin{equation}
  \label{EqPfPNLim}
     \wh{P_\NH}(\sigma)u_{\NH,0} + c_0 w_0^\flat = 0,\quad \la u_{\NH,0},w_0^\sharp\ra_{L^2(X_\NH)}=1.
  \end{equation}
  Only the first equation requires an argument. Let $\psi,\tilde\psi\in\CIc((0,\infty)_z)$, with $\tilde\psi=1$ near $\supp\psi$. For small $\eps=\eps_j$, we analyze
  \begin{equation}
  \label{EqPfPNLimComp}
    \psi\wt P_\eps u_\eps = \psi\wh{P_\NH}(\sigma)(\chi_\NH u_\eps) + \psi(\wt P_\eps-\wh{P_\NH}(\sigma))(\tilde\psi\chi_\NH u_\eps) - \psi[\wt P_\eps,\chi_\NH]\tilde\psi u_\eps
  \end{equation}
  in the coordinates $z\in(0,\infty)$, $\omega\in\Sph^2$. The first term converges in distributions to $\psi\wh{P_\NH}(\sigma)u_{\NH,0}$. By Lemma~\ref{LemmaGCTot} and using~\eqref{EqPfQUnif}, the second term is bounded by
  \[
    \|\tilde\psi \chi_\NH u_\eps\|_{\bar H_{\qop,\eps}^{s-2,-\frac12,\alpha_\ext}(X_\eps)} \sim \eps\|\chi_\NH u_\eps\|_{\Hbext^{s-2,\alpha}(X_\NH)} \lesssim \eps.
  \]
  (Here we use that $z$ is bounded on $\tilde\psi$, and hence weights at $z=\infty$ are arbitrary.) The third term likewise converges to $0$ as $\eps\to 0$. Therefore, $\psi\wt P_\eps u_\eps$ converges in distributions to $\psi\wh{P_\NH}(\sigma)u_{\NH,0}$.
  
  The system~\eqref{EqPfPNLim} is equivalent to $(u_{\NH,0},c_0)=P_\NH^\aug(\sigma)^{-1}(0,1)$; therefore, $c_0=D(\sigma)$ in the notation of~\eqref{EqPfPNNHInv}. This proves that $c_\eps=D_\eps(\sigma)\to c_0$ as $\eps\searrow 0$.
\end{proof}

%%%%%%%%%
\pfstep{Step 4.\ QNMs and resonant states.} If $r_0>0$ is so small that $\sigma_0$ is the unique zero of $D(\sigma)$ in the disk $\{|\sigma-\sigma_0|\leq r_0\}$, then also $D_\eps(\sigma)$ has a unique zero, $\sigma_\eps$, in this disk for all sufficiently small $\eps>0$ by Rouch\'e's theorem; and $\sigma_\eps$ depends continuously on $\sigma_0$. By the Schur complement formula, $\wh{P_\eps}(\kappa_{\rm C,\eps}\sigma)^{-1}$ has a unique pole in this disk at $\sigma=\sigma_\eps$.

Finally, if $u^\flat_\eps\in H^s(X_\eps)$ is a resonant state of $\wh{P_\eps}(\kappa_{\rm C,\eps}\sigma)$, then $\wt P_\eps^\aug(\sigma_\eps)(u^\flat_\eps,0)=(0,c)$ for some $c\neq 0$. Inverting $\wt P_\eps^\aug(\sigma_\eps)$ shows that we can obtain a resonant state via the formula
\[
  u_\eps^\res = \pi_1\bigl(\wt P_\eps^\aug(\sigma_\eps)^{-1}(0,1)\bigr)
\]
where $\pi_1\colon H^s(X_\eps)\oplus\C\to H^s(X_\eps)$ is the projection on the first summand. Since we have the uniform bounds~\eqref{EqPfPNInvBd} for all $s\in\R$ and $\alpha_\ext\in(0,1)$, we conclude that
\[
  1 \gtrsim \|(1-\chi_\NH)u_\eps^\res\|_{\bar H_{\qop,\eps}^{s,\frac12,\alpha_\ext}(X_\eps)} \sim \eps^{-\alpha_\ext}\|(1-\chi_\NH)u_\eps^\res\|_{\Hbext^{s,\frac12-\alpha_\ext}(X_\ext)}\,.
\]
Since $\inf_{\supp(1-\chi_\NH)}(r-r_{\rm e})>0$, Sobolev embedding implies $(1-\chi_\NH)u_\eps^\res=\cO(\eps^{1-})$ in $\CI(X_\ext)$. On the other hand, the arguments following~\eqref{EqPfPNLim} show that $\chi_\NH u_\eps^\res$ converges to the resonant state $u_\NH^\res:=\pi_1 P_\NH^\aug(\sigma_0)^{-1}(0,1)$ of the near-horizon geometry in $\Hbext^{s,\alpha}(X_\NH)$ for all $s\in\R$ and $\alpha<\frac12$. Since $u_\NH^\res\in\cA^1(X_\NH)$, regarded as a function on $\wt X$, vanishes simply at $X_\ext$ (which is consistent with the order of vanishing of $(1-\chi_\NH)u_\eps^\res$ recorded above), we obtain~\eqref{EqPfNConvState} by Sobolev embedding. This completes the proof of Theorem~\ref{ThmPfPN}.\hfill\textsquare

Proposition~\ref{PropPfPY} and Theorem~\ref{ThmPfPN} prove (a strengthening of) Theorem~\ref{ThmI} in the case $\mu>0$.

\begin{rmk}[The case $\mu=0$, $\ell\geq 1$]
\label{RmkPfP0ell}
  In the case $\mu=0$, the operator $\wh{P_\ext}(0)$ is not invertible. However, if we work on spaces of functions with vanishing spherical averages (i.e.\ their projections to degree $0$ spherical harmonics vanish), then $\wh{P_\ext}(0)$ \emph{is} invertible by Proposition~\ref{Prop0}, and thus (the proofs of) Proposition~\ref{PropPfPY} and Theorem~\ref{ThmPfPN} apply \emph{mutatis mutandis}.
\end{rmk}

%%%%%%%%%%%%%%%%%%%%%%%%%%%%%%%%%%%%%%%%%%%%%%%%%%
\subsection{Massless scalar waves}
\label{SsPf0}

We now turn to the case
\[
  \mu=0,\quad \text{therefore}\ P_\eps=\Box_{g_\eps},\ \wt P_\eps=\wh{P_\eps}(\kappa_{\rm C,\eps}\sigma),
\]
which is more delicate since $\wh{P_\ext}(0)$ in Proposition~\ref{Prop0} fails to be invertible then. We recall the notation $u_{(0)}=1$, $u_{(0)}^*=H(r_{\rm c}-r)$ from Proposition~\ref{Prop0}\eqref{It0Ker}.

%%%%%%%%%%%%%%%%%%%%%%%%%%%%%%
\subsubsection{Absence of QNMs}
\label{SssPf0N}

We first consider $\sigma\notin\QNM_\NH(0)$ and aim to prove an analogue of Proposition~\ref{PropPfPY}. We first sketch the setup of a Grushin problem for the zero energy operator on extremal RNdS. Using~\eqref{EqGCSpecFam}, we compute the derivative of the spectral family of $P_\eps=\Box_{g_\eps}$ at $0$ to be independent of $\eps$:
\[
  \pa_\sigma\wh{P_\eps}(0)=r^{-2}D_r r^2\tilde T + \tilde T D_r =: \pa_\sigma\wh{P_\ext}(0).
\]
This is formally self-adjoint with respect to the $L^2(X_\ext,r^2\,\dd r\,\dd\slg)$-inner product. Fix\footnote{Here $X_\ext^\circ=(r_{\rm e},r_+)_r\times\Sph^2_\omega$.}
\begin{equation}
\label{EqPf0u0fs}
  u_0^\flat:=\pa_\sigma\wh{P_\ext}(0)u_{(0)};\qquad
  u_0^\sharp\in\CIc(X_\ext^\circ),\quad \la u_{(0)},u_0^\sharp\ra_{L^2(X_\ext)}=1.
\end{equation}
Note that $u_0^\flat=-i r^{-2}\pa_r(r^2\tilde T)\in\CI(X_\ext)$. In particular,
\begin{equation}
\label{EqPf0SketchMem}
  u_0^\flat\in\Hbext^{s,\gamma}(X_\ext),\quad s>\tfrac12,\ \gamma<\tfrac12.
\end{equation}
Recalling the space $\cX^{s,\gamma}$ from~\eqref{Eq0Op}, we can thus consider
\begin{equation}
\label{EqPf0Pextaug}
  P_\ext^\aug
  := \begin{pmatrix}
       \wh{P_\ext}(0) & u_0^\flat \\
       \la\cdot,u_0^\sharp\ra_{L^2(X_\ext)} & 0
     \end{pmatrix}
\end{equation}
as an index $0$ operator $\cX^{s,\gamma}\oplus\C\to\Hbext^{s-1,\gamma}(X_\ext)\oplus\C$. This operator is, in fact, invertible, as follows from the following computation:

\begin{lemma}[Nondegenerate pairing]
\label{LemmaPf0}
  $\la u_0^\flat,u_{(0)}^*\ra_{L^2(X_\ext)}=-4\pi i(r_{\rm e}^2+r_{\rm c}^2)\neq 0$.
\end{lemma}
\begin{proof}
  The pairing equals $-4\pi\cdot i$ times
  \[
    \int_{r_{\rm e}}^{r_{\rm c}} r^{-2}\pa_r(r^2\tilde T)\,r^2\,\dd r = r_{\rm c}^2\tilde T(r_{\rm c}) - r_{\rm e}^2\tilde T(r_{\rm e}).
  \]
  Since $\tilde T(r_{\rm e})=-1$ and $\tilde T(r_{\rm c})=1$, the claim follows.
\end{proof}

In order to set up a Grushin problem for $\wt P_\eps$, it is then particular natural to use $\eps^{-1}\wh{P_\eps}(\kappa_{\rm C,\eps}\sigma)u_{(0)}\approx\frac{\kappa_{\rm C,\eps}}{\eps}\sigma\wh{P_\ext}(0)u_{(0)}$, which is a multiple of $u_0^\flat$ and thus, by Lemma~\ref{LemmaPf0}, spans a complement to the range of the $X_\ext$-model $\wh{P_\ext}(0)$.\footnote{The resolvent analysis near $0$ energy on asymptotically flat spaces for spectral families $\hat P(\sigma)$ admitting a zero energy state $u_{(0)}$ as done in a concrete setting \cite[\S{3.3}]{HintzGlueLocIII} follows a similar route. To be more concrete, if $\pa_\sigma\hat P(0)u_{(0)}\notin\ran\hat P(0)$ (and the space of zero energy states is spanned by $u_{(0)}$), one can set up a Grushin problem for $\hat P(\sigma)$ by using $\sigma^{-1}\hat P(\sigma)u_{(0)}$ (or refinements thereof) as the $(1,2)$ entry of an augmented operator $\hat P^{\rm aug}(\sigma)$, and the uniform invertibility of $\hat P^{\rm aug}(\sigma)$ near $\sigma=0$ then gives the invertibility of $\hat P(\sigma)$ for $\sigma\neq 0$ with a first order pole at $\sigma=0$. Cf.\ \cite[(3.30)]{HintzGlueLocIII}.} We divide this further by $\sigma$ to avoid the degeneracy as $\sigma\to 0$, and we normalize it for consistency with~\eqref{EqPf0Pextaug}.

\begin{prop}[Grushin problem for $\wt P_\eps$]
\label{PropPf0Gr}
  Suppose that $\sigma\notin\QNM_\NH(0)$. Set\footnote{Recall from~\eqref{EqGSurfGrav} that $\frac{\kappa_{\rm C,\eps}}{\eps}\equiv\varkappa_{\rm e}\bmod\eps\CI([0,\eps_0))$.}
  \begin{equation}
  \label{EqPf0Grueps}
    u_\eps^\flat := \Bigl(\frac{\kappa_{\rm C,\eps}}{\eps}\Bigr)^{-1}\wh{P_\eps}(\kappa_{\rm C,\eps}\sigma)\bigl((\eps\sigma)^{-1}u_{(0)}\bigr) = u_0^\flat - \frac{1-\tilde T^2}{F_\eps}\kappa_{\rm C,\eps}\sigma.
  \end{equation}
  (This is defined through the final expression on the right for $\sigma=0$.) Define the operator
  \begin{equation}
  \label{EqPf0GrOp}
    \wt P_\eps^\aug(\sigma) :=
    \begin{pmatrix}
      \wh{P_\eps}(\kappa_{\rm C,\eps}\sigma) & u_\eps^\flat \\
      \la\cdot,u_0^\sharp\ra_{L^2(X_\ext)} & 0
    \end{pmatrix}.
  \end{equation}
  Let $s\geq s_0+2$ where $s_0>\max(\frac12-\Im\sigma,\frac12)$, and let $\alpha_\NH,\alpha_\ext\in\R$ with $\alpha_\NH-\alpha_\ext\in(-\frac12,\frac12)$. Then for sufficiently small $\eps$ we have a uniform estimate
  \begin{equation}
  \label{EqPf0GrEst}
    \|(u,c)\|_{\bar H_{\qop,\eps}^{s,\alpha_\NH,\alpha_\ext}(X_\eps)\oplus\eps^{\alpha_\ext}\C} \leq C\|\wt P_\eps^\aug(\sigma)(u,c)\|_{\bar H_{\qop,\eps}^{s-1,\alpha_\NH,\alpha_\ext}(X_\eps)\oplus\eps^{\alpha_\ext}\C}\,,
  \end{equation}
  where we use the notation~\eqref{EqPfPNCNorm}.
\end{prop}

Before giving the proof of Proposition~\usref{PropPf0Gr}, note that since $u_\eps^\flat\in\CI(\wt X)$, we have
\begin{equation}
\label{EqPf0uflatEst}
  \|c u_\eps^\flat\|_{\bar H_{\qop,\eps}^{s-1,\alpha_\NH,\alpha_\ext}(X_\eps)} \leq |c|_{\eps^{\alpha_\ext}\C} \|u_\eps^\flat\|_{\bar H_{\qop,\eps}^{s-1,\gamma,0}(X_\eps)} \lesssim |c|_{\eps^{\alpha_\ext}\C}
\end{equation}
where
\[
  \gamma:=\alpha_\NH-\alpha_\ext\in(-\tfrac12,\tfrac12);
\]
the second bound is due to $\gamma<\frac12$ and $\int_0^{r_+-r_{\rm e}}x^{-2\gamma}\,\dd x<\infty$. Since $u_0^\sharp\in\CIc(X_\ext^\circ)$, we moreover have, for all sufficiently small $\eps>0$, $u_0^\sharp=\chi_\ext u_0^\sharp$ and thus
\[
  |\la u,u_0^\sharp\ra_{L^2(X_\ext)}|_{\eps^{\alpha_\ext}\C} \lesssim \eps^{-\alpha_\ext} \|\chi_\ext u\|_{\Hbext^{s,\gamma}(X_\ext)} \lesssim \|u\|_{\bar H_{\qop,\eps}^{s,\alpha_\NH,\alpha_\ext}(X_\eps)}\,.
\]
(This is analogous to~\eqref{EqPfPNAug2}.)

\begin{proof}[Proof of Proposition~\usref{PropPf0Gr}]
  We argue as in the proof of Proposition~\ref{PropPfPY}, except we now use the invertibility of $P_\ext^\aug$ in the first step. Thus, we start with~\eqref{EqPfHi}, write $u=\chi_\ext u+(1-\chi_\ext)u$ and obtain
  \begin{align*}
    &\| (u,c) \|_{\bar H_{\qop,\eps}^{s,\alpha_\NH,\alpha_\ext}(X_\eps)\oplus\eps^{\alpha_\ext}\C} \\
    &\qquad \lesssim \| \wt P_\eps^\aug(u,c) \|_{\bar H_{\qop,\eps}^{s-1,\alpha_\NH,\alpha_\ext}(X_\eps)\oplus\eps^{\alpha_\ext}\C} + \| (\chi_\ext u,c) \|_{\bar H_{\qop,\eps}^{s_0,\alpha_\NH,\alpha_\ext}(X_\eps)\oplus\eps^{\alpha_\ext}\C} \\
    &\qquad \hspace{16.75em} + \| u \|_{\bar H_{\qop,\eps}^{s_0,\alpha_\NH,\alpha_\ext-\eta}(X_\eps)}
  \end{align*}
  where we fix $\eta>0$ such that $\gamma+\eta\in(-\frac12,\frac12)$. We next use~\eqref{EqPfqExt} and estimate the second term on the right using the invertibility of~\eqref{EqPf0Pextaug} by
  \begin{align*}
    \eps^{-\alpha_\ext}\|(\chi_\ext u,c)\|_{\Hbext^{s_0,\gamma}(X_\ext)\oplus\C} &\lesssim \eps^{-\alpha_\ext} \| P_\ext^\aug(\chi_\ext u,c) \|_{\Hbext^{s_0-1,\gamma}(X_\ext)\oplus\C} \\
      &\lesssim \| P_\ext^\aug(\chi_\ext u,c) \|_{\bar H_{\qop,\eps}^{s_0-1,\alpha_\NH,\alpha_\ext}(X_\eps)\oplus\eps^{\alpha_\ext}\C}\,.
  \end{align*}
  We replace $P_\ext^\aug$ by $\wt P_\eps^\aug(\sigma)$. We can bound the action of the difference
  \[
    \wt P_\eps^\aug(\sigma)\circ(\chi_\ext\oplus 1) - P_\ext^\aug\circ(\chi_\ext\oplus I)
    = \begin{pmatrix}
        \bigl(\wh{P_\eps}(\kappa_{\rm C,\eps}\sigma)-\wh{P_\ext}(0)\bigr)\chi_\ext & u_\eps^\flat-u_0^\flat \\ 0 & 0
      \end{pmatrix}
  \]
  on $(u,c)$ in $\bar H_{\qop,\eps}^{s_0-1,\alpha_\NH,\alpha_\ext}(X_\eps)\oplus\eps^{\alpha_\ext}\C$ by $\|u\|_{\bar H_{\qop,\eps}^{s_0+1,\alpha_\NH,\alpha_\ext-1}(X_\eps)} + |c|_{\eps^{\alpha_\ext-1}\C}$ since the fact that $u_\eps^\flat-u_0^\flat=:\eps\tilde u$ with $\tilde u\in\CI(\wt X)$ implies
  \[
    \|c\cdot(u_\eps^\flat-u_0^\flat)\|_{\bar H_{\qop,\eps}^{s_0-1,\alpha_\NH,\alpha_\ext}(X_\eps)} \leq \eps^{-\alpha_\ext}|c| \|\eps\tilde u\|_{\bar H_{\qop,\eps}^{s_0-1,\gamma,0}(X_\eps)} \lesssim \eps|c|_{\eps^{\alpha_\ext}\C}
  \]
  (cf.\ the justification of~\eqref{EqPf0uflatEst}).

  Next, we commute $\wt P_\eps^\aug(\sigma)$ through $\chi_\ext\oplus I$. We can bound the norm of the output of the commutator
  \[
    [\wt P_\eps^\aug(\sigma),\chi_\ext\oplus 1] = \begin{pmatrix} [\wt P_\eps,\chi_\ext] & (1-\chi_\ext)u_\eps^\flat \\ \la(\chi_\ext-1)\cdot,u_0^\sharp\ra_{L^2(X_\ext)} & 0 \end{pmatrix}
  \]
  acting on $(u,c)$ as follows. Since $\chi_\ext-1=0$ on $\supp u_0^\sharp$ for sufficiently small $\eps>0$, only the first row is nonzero. We can estimate the contribution of $[\wt P_\eps,\chi_\ext]\in\rho_\ext^N\Diffq^1(\wt X)$ as in~\eqref{EqPfPYExt}. Furthermore, since $(1-\chi_\ext)u_\eps^\flat$ is smooth on $\wt X$ and vanishes near $X_\ext$, we have
  \[
    \|c\cdot (1-\chi_\ext)u_\eps^\flat\|_{\bar H_{\qop,\eps}^{s_0-1,\alpha_\NH,\alpha_\ext}(X_\eps)} = \eps^\eta|c|_{\eps^{\alpha_\ext}\C} \|(1-\chi_\ext)u_\eps^\flat\|_{\bar H_{\qop,\eps}^{s_0-1,\gamma+\eta,\eta}(X_\eps)} \lesssim \eps^\eta|c|_{\eps^{\alpha_\ext}\C}
  \]
  since $\gamma+\eta<\frac12$.

  In summary, we have now established
  \begin{equation}
  \label{EqPf0NEstExt}
  \begin{split}
    &\| (u,c) \|_{\bar H_{\qop,\eps}^{s,\alpha_\NH,\alpha_\ext}(X_\eps)\oplus\eps^{\alpha_\ext}\C} \\
    &\qquad \lesssim \| \wt P_\eps^\aug(u,c) \|_{\bar H_{\qop,\eps}^{s-1,\alpha_\NH,\alpha_\ext}(X_\eps)\oplus\eps^{\alpha_\ext}\C} + \| (u,c) \|_{\bar H_{\qop,\eps}^{s_0+1,\alpha_\NH,\alpha_\ext-\eta}(X_\eps)\oplus\eps^{\alpha_\ext-\eta}\C}\,.
  \end{split}
  \end{equation}
  We then write $u=\chi_\NH u+(1-\chi_\NH)u$ and estimate $\|\chi_\NH u\|_{\bar H_{\qop,\eps}^{s_0+1,\alpha_\NH,\alpha_\ext-\eta}(X_\eps)}$ as around~\eqref{EqPfPY2}. This leads to the following analogue of~\eqref{EqPfPY4}:
  \begin{align*}
    &\|(u,c)\|_{\bar H_{\qop,\eps}^{s,\alpha_\NH,\alpha_\ext}(X_\eps)\oplus\eps^{\alpha_\ext}\C} \\
    &\qquad \lesssim \| \wt P_\eps^\aug(u,c) \|_{\bar H_{\qop,\eps}^{s-1,\alpha_\NH,\alpha_\ext}(X_\eps)\oplus\eps^{\alpha_\ext}\C} + \| (u,c) \|_{\bar H_{\qop,\eps}^{s_0+2,\alpha_\NH-1,\alpha_\ext-\eta}(X_\eps)\oplus\eps^{\alpha_\ext-\eta}\C}\,.
  \end{align*}
  For sufficiently small $\eps>0$, the second term on the right can be absorbed into the left hand side.
\end{proof}

The estimate~\eqref{EqPf0GrEst} is, in fact, locally uniform in $\sigma$, as follows from its proof. This allows us to conclude:

\begin{prop}[Absence of QNMs except $0$]
\label{PropPf0N}
  Recall the relationship $r_{\rm C}=r_{\rm e}-2\eps$.
  \begin{enumerate}
  \item\label{ItPf0N0} For all $\eps>0$, we have $0\in\QNM(r_{\rm C},r_{\rm e},r_{\rm c})$.
  \item Let $K\subset\C$ be a compact set disjoint from $\QNM_\NH(0)$. Then there exists $\eps_1\in(0,\eps_0)$ such that for all $\eps\in(0,\eps_1]$, we have
    \[
      \{ \kappa_{\rm C,\eps}\sigma \colon \sigma\in K \} \cap \bigl( \QNM(r_{\rm C},r_{\rm e},r_{\rm c})\setminus\{0\} \bigr) = \emptyset.
    \]
  \end{enumerate}
\end{prop}

Thus, unlike in the setting of Proposition~\ref{PropPfPYUnif} where $\wh{P_\ext}(0)$ was invertible, the presence of the zero mode $u_{(0)}$ for massless scalar waves on extremal RNdS leads to the existence of the QNM $0$ for nearly extremal RNdS.

\begin{proof}[Proof of Proposition~\usref{PropPf0N}]
  The first part follows from the fact that $\Box_{g_\eps}u_{(0)}=0$ (constants solve the wave equation) for all $\eps>0$. For the second part, the estimate~\eqref{EqPf0GrEst} holds uniformly for all $\sigma\in K$ and $\eps\in(0,\eps_1]$ when $\eps_1\in(0,\eps_0)$ is sufficiently small. For $\sigma\in K$ and $\eps\in(0,\eps_1]$, and given any $f\in H^{s-1}(X_\eps)$, define then
  \[
    (u,c) := \wt P_\eps^\aug(\sigma)^{-1}(f,0).
  \]
  By definition of $u_\eps^\flat$ in~\eqref{EqPf0Grueps}, we then have
  \[
    \wh{P_\eps}(\kappa_{\rm C,\eps}\sigma)u' = f,\quad u':=u + \Bigl(\frac{\kappa_{\rm C,\eps}}{\eps}\Bigr)^{-1}(\eps\sigma)^{-1}u_{(0)}
  \]
  provided $\sigma\neq 0$ (so that $u'$ is well-defined), with $u'\in H^s(X_\eps)$. Therefore, $\wh{P_\eps}(\kappa_{\rm C,\eps}\sigma)$ is surjective as a map~\eqref{EqPfPeps}, thus injective since it has index $0$, and hence $\kappa_{\rm C,\eps}\sigma\notin\QNM(r_{\rm C},r_{\rm e},r_{\rm c})$.
\end{proof}

%%%%%%%%%%%%%%%%%%%%%%%%%%%%%%
\subsubsection{Existence of QNMs}
\label{SssPf0Y}

Note that $0\notin\QNM_\NH(0)$. Besides the QNM $0$ observed in Proposition~\ref{PropPf0N}\eqref{ItPf0N0}, we next find the QNMs arising from the near-horizon QNMs.

\begin{thm}[Existence of QNMs]
\label{ThmPf0Y}
  Let $\sigma_0\in\QNM_\NH(0)$, and write $m(\sigma_0)$ for the multiplicity of $\sigma_0$. Let $r_0>0$ be so small that for all $\sigma\in\{0\}\cup\QNM_\NH(0)\setminus\{\sigma_0\}$ we have $|\sigma-\sigma_0|\geq 2 r_0$. Then there exists $\eps_1\in(0,\eps_0)$ such that for all $\eps\in(0,\eps_0]$, there are $m(\sigma_0)$ many QNMs $\varsigma\in\QNM(r_{\rm C},r_{\rm e},r_{\rm c})$, $r_{\rm C}=r_{\rm e}-2\eps$, of $\Box_{g_\eps}$ (counted with multiplicity) with
  \[
    \Bigl| \frac{\varsigma}{\kappa_{\rm C,\eps}} - \sigma_0\Bigr| < r_0.
  \]
  Denote by $\Sigma_\eps$ the set of these QNMs $\varsigma$. Then:
  \begin{enumerate}
  \item\label{ItPf0YConv} $\Sigma_\eps\subset i\R$, and $\{\frac{\varsigma}{\kappa_{\rm C,\eps}}\colon\varsigma\in\Sigma_\eps\}\to\{\sigma_0\}$ in the Hausdorff distance sense as $\eps\to 0$;
  \item\label{ItPf0YPole} $\wh{P_\eps}(\zeta)^{-1}$ has a pole of order $1$ at $\zeta=\varsigma$ for every such $\varsigma$.
  \end{enumerate}
  Finally:
  \begin{enumerate}
  \setcounter{enumi}{2}
  \item\label{ItPf0YState} suppose $\Sigma_\eps$ contains a (necessarily unique) element $\varsigma_\eps=\kappa_{\rm C,\eps}(\sigma_0+o(1))$ for which a spherically symmetric resonant state exists.\footnote{Resonant states with angular dependence given by a degree $\ell\geq 1$ spherical harmonic were already described before; see Remark~\ref{RmkPfP0ell}. See also Remark~\ref{RmkPfPNYlm} regarding separation into spherical harmonics.} Then we can normalize such a resonant state $u_\eps\in\CI(X_\eps)$ of $\wh{P_\eps}(\varsigma_\eps)$ in such a way that, for some constant $c_\eps\in\C$ which is uniformly bounded as $\eps\to 0$,
    \[
      \Bigl\| u_\eps(r,\omega) - \Bigl[c_\eps + u_0\Bigl(\frac{r-r_{\rm e}}{\eps}+1\Bigr)\Bigr] \Bigr\|_{\cC_{\bop,\eps}^{k,\theta}(X_\eps)} \xra{\eps\to 0} 0
    \]
    for all $\theta<1$, where $u_0$ is a resonant state of $\wh{P_\NH}(\sigma_0)$ (i.e.\ of the form~\eqref{EqNHQNMResState} for $\ell=0$, thus without $\omega$-dependence, and a suitable value of $n$); the norm here is defined in~\eqref{EqPfNConvNorm}.
  \end{enumerate}
\end{thm}

We shall prove this theorem by means of a Grushin problem similar to~\eqref{EqPfPNAugOp}, except that now, due to the failure of invertibility of the $X_\ext$-model problem $\wh{P_\ext}(0)$, we use the augmented operator~\eqref{EqPf0GrOp} in place of $\wh{P_\eps}(\kappa_{\rm C,\eps}\sigma)$ in~\eqref{EqPfPNAugOp}. We use different notation for the latter operator now and write for $\sigma\neq 0$
\[
  \wt P_\eps^{\rm aug,1}(\sigma) := \begin{pmatrix} \wh{P_\eps}(\kappa_{\rm C,\eps}\sigma) & c_{\eps,\sigma}\wh{P_\eps}(\kappa_{\rm C,\eps}\sigma)u_{(0)} \\ \la\cdot,u_0^\sharp\ra_{L^2(X_\ext)} & 0 \end{pmatrix},\quad c_{\eps,\sigma}:=\Bigl(\frac{\kappa_{\rm C,\eps}}{\eps}\Bigr)^{-1}(\eps\sigma)^{-1}.
\]
This operator detects QNMs in the following sense:

\begin{lemma}[First augmentation]
\label{LemmaPf0YAug1}
  Let $\sigma\neq 0$ and $s>\frac12-\Im\sigma$. Then $\wh{P_\eps}(\kappa_{\rm C,\eps}\sigma)\colon\cX^s(X_\eps)\to H^{s-1}(X_\eps)$ is invertible if and only if $\wt P_\eps^{\rm aug,1}(\sigma)\colon\cX^s(X_\eps)\oplus\C\to H^{s-1}(X_\eps)\oplus\C$ is.
\end{lemma}
\begin{proof}
  Given $f\in H^{s-1}(X_\eps)$, consider $\wt P_\eps^{\rm aug,1}(\sigma)^{-1}(f,0)=:(u,c)$; then $\wh{P_\eps}(\kappa_{\rm C,\eps}\sigma)(u+c c_{\eps,\sigma}u_{(0)})=f$. Conversely, given $(f,c)\in H^{s-1}(X_\eps)\oplus\C$, let $u':=\wh{P_\eps}(\kappa_{\rm C,\eps}\sigma)^{-1}f$. Since also
  \[
    \wh{P_\eps}(\kappa_{\rm C,\eps}\sigma) \bigl( u'-a c_{\eps,\sigma}u_{(0)}\bigr) + a c_{\eps,\sigma}\wh{P_\eps}(\kappa_{\rm C,\eps}\sigma)u_{(0)} = f
  \]
  for all $a\in\C$, we note that $\la u'-a c_{\eps,\sigma}u_{(0)},u_0^\sharp\ra_{L^2(X_\ext)}=c$ for $a=(\la u',u_0^\sharp\ra-c)/(c_{\eps,\sigma}\la u_{(0)},u_0^\sharp\ra)$. The denominator is nonzero by~\eqref{EqPf0u0fs}.
\end{proof}

We assume (for notational simplicity as in~\S\ref{SssPfPN}) that $m(\sigma_0)=1$, and we write $v_0\in\cA^1(X_\NH)$ for a resonant state and $v_0^*\in\bigcap_{\eta>0}\Hbsupp^{\frac12+\Im\sigma_0-\eta,-\alpha}(X_\NH)$ (where $\alpha\in(-\frac12,\frac12)$) for a co-resonant state. We pick $w_0^\sharp,w_0^\flat\in\CIc(X_\NH^\circ)$ with $\la v_0,w_0^\sharp\ra_{L^2(X_\NH)}$, $\la w_0^\flat,v_0^*\ra_{L^2(X_\NH)}\neq 0$. The augmented operator for the near-horizon analysis is then denoted
\[
  P^{\rm aug}_\NH(\sigma) := \begin{pmatrix} \wh{P_\NH}(\sigma) & w_0^\flat \\ \la\cdot,w_0^\sharp\ra_{L^2(X_\NH)} & 0 \end{pmatrix};
\]
it was already analyzed in Step~1 of the proof of Theorem~\ref{ThmPfPN} following~\eqref{EqPfPNNHaug}; in particular, we have
\begin{equation}
\label{EqPf0NH}
  P^{\rm aug}_\NH(\sigma)^{-1} = \begin{pmatrix} A(\sigma) & B(\sigma) \\ C(\sigma) & D(\sigma) \end{pmatrix};\quad
  D(\sigma)\ \text{has a simple zero at $\sigma=\sigma_0$.}
\end{equation}

Recalling $u_\eps^\flat:=c_{\eps,\sigma}\wh{P_\eps}(\kappa_{\rm C,\eps}\sigma)u_{(0)}$, the full augmented operator is
\[
  \wt P_\eps^{\rm aug}(\sigma) := \begin{pmatrix} \wh{P_\eps}(\kappa_{\rm C,\eps}\sigma) & u_\eps^\flat & w_0^\flat \\ \la\cdot,u_0^\sharp\ra_{L^2(X_\ext)} & 0 & 0 \\ \la\cdot,w_0^\sharp\ra_{L^2(X_\NH)} & 0 & 0 \end{pmatrix}.
\]

\begin{prop}[Grushin problem for $\wt P_\eps$]
\label{PropPf0GrY}
  Let $\sigma_0\in\QNM_\NH(0)$. Let $s\geq s_0+2$ where $s_0>\max(\frac12-\Im\sigma_0,\frac12)$, and let $\alpha_\NH,\alpha_\ext\in\R$ with $\gamma:=\alpha_\NH-\alpha_\ext\in(-\frac12,\frac12)$. Then there exist $r_0>0$ and $\eps_1\in(0,\eps_0)$ such that for all $\sigma\in\C$ with $|\sigma-\sigma_0|<2 r_0$ and for all $\eps\in(0,\eps_1]$,
  \begin{equation}
  \label{EqPf0GrY}
    \|(u,c_1,c_2)\|_{\bar H_{\qop,\eps}^{s,\alpha_\NH,\alpha_\ext} \oplus \eps^{\alpha_\ext}\C \oplus \eps^{\alpha_\NH-\frac12}\C}
    \lesssim \|\wt P_\eps^{\rm aug}(\sigma)(u,c_1,c_2)\|_{\bar H_{\qop,\eps}^{s-1,\alpha_\NH,\alpha_\ext} \oplus \eps^{\alpha_\ext}\C \oplus \eps^{\alpha_\NH-\frac12}\C}\,.
  \end{equation}
\end{prop}
\begin{proof}
  The invertibility of $\wt P_\eps^{\aug,1}(\sigma)$ allows us to estimate $u,c_1$ as in~\eqref{EqPf0NEstExt}, so using also the triangle inequality to split up the second term on the right,
  \begin{align*}
    &\|(u,c_1,c_2)\|_{\bar H_{\qop,\eps}^{s,\alpha_\NH,\alpha_\ext} \oplus \eps^{\alpha_\ext}\C \oplus \eps^{\alpha_\NH-\frac12}\C} \\
    &\qquad \lesssim \|\wt P_\eps^{\rm aug}(\sigma)(u,c_1,c_2)\|_{\bar H_{\qop,\eps}^{s-1,\alpha_\NH,\alpha_\ext} \oplus \eps^{\alpha_\ext}\C \oplus \eps^{\alpha_\NH-\frac12}\C} \\
    &\qquad \qquad + \|(\chi_\NH u,0,c_2)\|_{\bar H_{\qop,\eps}^{s_0+1,\alpha_\NH,\alpha_\ext-\eta}\oplus\eps^{\alpha_\ext-\eta}\C\oplus\eps^{\alpha_\NH-\frac12}\C} \\
    &\qquad\qquad + \|(1-\chi_\NH)u\|_{\bar H_{\qop,\eps}^{s_0+1,\alpha_\NH,\alpha_\ext-\eta}} \\
    &\qquad\qquad + |c_1|_{\eps^{\alpha_\ext-\eta}\C}\,.
  \end{align*}
  Here we take $\chi_\NH\in\CI(\wt X)$ to be equal to $1$ near $X_\NH$ and such that $\chi_\NH u_0^\sharp=0$ and $(1-\chi_\NH)w_0^\sharp=0$ for all small $\eps$ (used below), and $\eta>0$ is such that $\gamma+\eta\in(-\frac12,\frac12)$ still. The last two lines can be absorbed into the left hand side. Indeed, the norm of the penultimate term is $\lesssim\|u\|_{\bar H_{\qop,\eps}^{s_0+1,\alpha_\NH-1,\alpha_\ext-\eta}}$ (indeed, with arbitrary $X_\NH$-decay order); and the final term is $\eps^\eta|c_1|_{\eps^{\alpha_\ext}\C}$.

  Next, using the estimate~\eqref{EqPfPNNHEst} for $P_\NH^\aug(\sigma)$, we obtain the first bound in
  \begin{align*}
    &\|(\chi_\NH u,0,c_2)\|_{\bar H_{\qop,\eps}^{s_0+1,\alpha_\NH,\alpha_\ext-\eta}\oplus\eps^{\alpha_\ext-\eta}\C\oplus\eps^{\alpha_\NH-\frac12}\C} \\
    &\qquad \lesssim
        \left\|
          \begin{pmatrix} \wh{P_\NH}(\sigma) & 0 & w_0^\flat \\ 0 & 0 & 0 \\ \la\cdot,w_0^\sharp\ra_{L^2(X_\NH)} & 0 & 0 \end{pmatrix} \begin{pmatrix} \chi_\NH u \\ c_1 \\ c_2\end{pmatrix}
        \right\|_{\bar H_{\qop,\eps}^{s_0,\alpha_\NH,\alpha_\ext-\eta}\oplus\eps^{\alpha_\ext-\eta}\C\oplus\eps^{\alpha_\NH-\frac12}\C} \\
    &\qquad \lesssim \|\wt P_\eps^\aug(\sigma)(u,c_1,c_2)\|_{\bar H_{\qop,\eps}^{s_0,\alpha_\NH,\alpha_\ext-\eta}\oplus\eps^{\alpha_\ext-\eta}\C\oplus\eps^{\alpha_\NH-\frac12}\C} \\
    &\qquad \qquad +
        \left\|
          \begin{pmatrix}
            \wh{P_\eps}(\kappa_{\rm C,\eps}\sigma)-\wh{P_\NH}(\sigma)\chi_\NH & u_\eps^\flat & 0 \\
            \la\cdot,u_0^\sharp\ra_{L^2(X_\ext)} & 0 & 0 \\
            \la(1-\chi_\NH)\cdot,w_0^\sharp\ra_{L^2(X_\NH)} & 0 & 0
          \end{pmatrix}
          \begin{pmatrix} u \\ c_1 \\ c_2 \end{pmatrix}
        \right\|_{\bar H_{\qop,\eps}^{s_0,\alpha_\NH,\alpha_\ext-\eta}\oplus\eps^{\alpha_\ext-\eta}\C\oplus\eps^{\alpha_\NH-\frac12}\C}\,.
  \end{align*}
  We claim that the second term on the right can be absorbed. Indeed, the $(3,1)$ component of the matrix on the right vanishes for small $\eps$. The norm of the output of the $(1,1)$ component is bounded by $\|u\|_{\bar H_{\qop,\eps}^{s_0+2,\alpha_\NH-1,\alpha_\ext-\eta}}$ (cf.\ \eqref{EqPfPY3}). To bound the $(1,2)$ component, we use
  \[
    \|c_1 u_\eps^\flat\|_{\bar H_{\qop,\eps}^{s_0+1,\alpha_\NH,\alpha_\ext-\eta}} = \eps^\eta\eps^{-\alpha_\ext}|c_1| \|u_\eps^\flat\|_{\bar H_{\qop,\eps}^{s_0+1,\gamma+\eta,0}} \lesssim \eps^\eta|c_1|_{\eps^{\alpha_\ext}\C}\,.
  \]
  For the $(2,1)$ component, finally, we use that $\supp u_0^\sharp\cap X_\NH=\emptyset$ to bound 
  \[
    \eps^{-\alpha_\ext+\eta}|\la u,u_0^\sharp\ra_{L^2(X_\ext)}| \lesssim \eps^{\eta/2}\|u\|_{\bar H_{\qop,\eps}^{s_0,\alpha_\NH-1,\alpha_\ext-\eta/2}}.
  \]
  This completes the proof of~\eqref{EqPf0GrY}.
\end{proof}

For $|\sigma-\sigma_0|<2 r_0$ and $\eps\in(0,\eps_1]$, we now write
\[
  \wt P_\eps^\aug(\sigma)^{-1} = \begin{pmatrix} A_{1 1,\eps}(\sigma) & A_{1 2,\eps}(\sigma) & B_{1,\eps}(\sigma) \\ A_{2 1,\eps}(\sigma) & A_{2 2,\eps}(\sigma) & B_{2,\eps}(\sigma) \\ C_{1,\eps}(\sigma) & C_{2,\eps}(\sigma) & D_\eps(\sigma) \end{pmatrix}.
\]
The analogue of Lemma~\ref{LemmaPfPNCont} holds also in the present setting:

\begin{lemma}[Continuity of $D_\eps(\sigma)$]
\label{LemmaPf0Cont}
  $D_\eps(\sigma)$ converges uniformly to $D(\sigma)$ (see~\eqref{EqPf0NH}) in the disk $\{|\sigma-\sigma_0|\leq r_0\}$.
\end{lemma}
\begin{proof}
  We only need to prove pointwise convergence for a fixed value of $\sigma$ with $|\sigma-\sigma_0|<2 r_0$. Let thus
  \[
    (u_\eps,c_{1,\eps},c_{2,\eps}) := \wt P_\eps^\aug(\sigma)^{-1}(0,0,1).
  \]
  (Thus $c_{2,\eps}=D_\eps(\sigma)$.) Using the estimate~\eqref{EqPf0GrY} for $\alpha_\NH=\frac12$ and $\alpha_\ext\in(0,1)$, we conclude uniform (in $\eps$) bounds
  \[
    \|u_\eps\|_{\bar H_{\qop,\eps}^{s,\frac12,\alpha_\ext}(X_\eps)},\ \eps^{-\alpha_\ext}|c_{1,\eps}|,\ |c_{2,\eps}| \lesssim 1.
  \]
  Passing to a subsequence, we may assume that $\chi_\NH u_\eps$, which is uniformly bounded in $\Hbext^{s,\alpha}(X_\NH)$ where $\alpha:=\alpha_\ext-\frac12\in(-\frac12,\frac12)$, converges weakly to some $u_{\NH,0}\in\Hbext^{s,\alpha}(X_\NH)$, and that
  \begin{equation}
  \label{EqPf0Bds}
    \eps^{-\alpha_\ext}c_{1,\eps} \to c_1,\quad
    c_{2,\eps} \to c_2.
  \end{equation}

  In the equation
  \[
    0 = \wh{P_\eps}(\kappa_{\rm C,\eps}\sigma)u_\eps + c_{1,\eps}u_\eps^\flat + c_{2,\eps}w_0^\flat,
  \]
  consider now the first term; arguing as after~\eqref{EqPfPNLimComp}, it converges in distributions on $(0,\infty)_z\times\Sph^2$ to $\wh{P_\NH}(\sigma)u_{\NH,0}$. The convergence of the two remaining terms is clear, so we obtain
  \[
    \wh{P_\NH}(\sigma)u_{\NH,0} + c_2 w_0^\flat = 0.
  \]
  Moreover, $1=\la u_\eps,w_0^\sharp\ra_{L^2(X_\NH)}=\la\chi_\NH u_\eps,w_0^\sharp\ra_{L^2(X_\NH)}$ for sufficiently small $\eps>0$, and this converges to $\la u_{\NH,0},w_0^\sharp\ra_{L^2(X_\NH)}$. Altogether, we deduce
  \[
    P_\NH^\aug(\sigma) ( u_{\NH,0}, c_2 ) = ( 0, 1 ),
  \]
  and therefore $c_2=D(\sigma)$ is indeed the limit of $c_{2,\eps}=D_\eps(\sigma)$.
\end{proof}

As in~\S\ref{SssPfPN}, Rouch\'e's theorem and the Schur complement formula prove parts~\eqref{ItPf0YConv}--\eqref{ItPf0YPole} of Theorem~\ref{ThmPf0Y}. Denote the unique pole of $\wh{P_\eps}(\kappa_{\rm C,\eps}\sigma)^{-1}$ in a small disk around $\sigma_0$ by $\sigma_\eps$, so $\sigma_\eps=\sigma_0+o(1)$ as $\eps\to 0$. Analogously to Step~4 of the proof of Theorem~\ref{ThmPfPN}, the corresponding resonant state is now given by
\begin{equation}
\label{EqPf0Res}
  u_\eps^\res = u_\eps + c_{1,\eps}c_{\eps,\sigma}u_{(0)},\quad (u_\eps,c_{1,\eps},c_{2,\eps}) := \wt P_\eps^\aug(\sigma_\eps)^{-1}(0,0,1).
\end{equation}
The proof of Lemma~\ref{LemmaPf0Cont} and the compactness of the inclusion $\Hbext^{s,\alpha}(X_\NH)\hra\Hbext^{s',\alpha'}(X_\NH)$ for $s'<s$, $\alpha'<\alpha$ show that
\begin{equation}
\label{EqPf0ResConv}
  \chi_\NH u_\eps \to u_\NH^\res\ \text{in}\ \Hbext^{s,\alpha}(X_\NH)\ \forall\,s\in\R,\ \alpha<\tfrac12,\quad
  c_{2,\eps} \to c_2,
\end{equation}
where $(u_\NH^\res,c_2)=P_\NH^\aug(\sigma_0)^{-1}(0,1)$, so in particular $u_\NH^\res$ is a near-horizon resonant state associated with $\sigma_0$. However, the uniform bound $c_{1,\eps}=\cO(\eps^{\alpha_\ext})$, $\alpha_\ext\in(0,1)$, recorded in~\eqref{EqPf0Bds} is not sufficient to cancel the factor $c_{\eps,\sigma}\sim\eps^{-1}$ in the expression~\eqref{EqPf0Res} of $u_\eps^\res$. We thus need to improve~\eqref{EqPf0Bds}:

\begin{lemma}[Improved bounds]
\label{LemmaPf0Bd}
  In the notation~\eqref{EqPf0Res}, we have $|c_{1,\eps}|\lesssim\eps$.
\end{lemma}
\begin{proof}
  We first construct, by hand, an approximation to $\wt P_\eps^\aug(\sigma_\eps)^{-1}(0,0,1)$ and then use $\wt P_\eps^\aug(\sigma_\eps)^{-1}$ to solve away the remaining error. To wit, define
  \[
    (u_{\NH,\eps},c_{\NH,\eps}) := P_\NH^\aug(\sigma_\eps)^{-1}(0,1).
  \]
  Thus, $u_{\NH,\eps}\in\Hbext^{s,\alpha}(X_\NH)$ and $c_{\NH,\eps}\in\C$ are uniformly bounded; here $s\in\R$ and $\alpha<\frac12$. But since
  \[
    \wh{P_\NH}(\sigma_\eps)u_{\NH,\eps} = -c_{\NH,\eps}w_0^\flat \in \CIc(X_\NH^\circ),
  \]
  we can use a normal operator argument to conclude (using the fact that all indicial roots of $\wh{P_\NH}(\sigma_\eps)$ are $\geq 1$) that, in fact, $u_{\NH,\eps}\in\cA^1(X_\NH)$ (cf.\ the proof of Proposition~\ref{PropNHFred}), with uniform bounds.

  We now compute
  \[
    \wt P_\eps^\aug(\sigma_\eps)\begin{pmatrix}\chi_\NH u_{\NH,\eps} \\ 0 \\ c_{\NH,\eps} \end{pmatrix}
    \!=\!\begin{pmatrix}
        [\wh{P_\eps}(\kappa_{\rm C,\eps}\sigma_\eps),\chi_\NH] u_{\NH,\eps} + \chi_\NH\bigl(\wh{P_\eps}(\kappa_{\rm C,\eps}\sigma_\eps)-\wh{P_\NH}(\sigma_\eps)\bigr)u_{\NH,\eps} \\ \la\chi_\NH u_{\NH,\eps},u_0^\sharp\ra_{L^2(X_\ext)} \\ \la\chi_\NH u_{\NH,\eps},w_0^\sharp\ra_{L^2(X_\NH)}
      \end{pmatrix}
    \!=:\!\begin{pmatrix} f_\eps \\ s_{1,\eps} \\ s_{2,\eps} \end{pmatrix}.
  \]
  Choosing the cutoff $\chi_\NH$ to be supported sufficiently close to $X_\NH$, we have $s_{1,\eps}=0$ for small $\eps>0$ since $\chi_\NH u_0^\sharp=0$, and $s_{2,\eps}=1$ for small $\eps>0$ since $\chi_\NH w_0^\sharp=w_0^\sharp$. Let $\tilde\chi_\NH\in\CI(\wt X)$ be equal to $1$ near $\supp\chi_\NH$ and $0$ outside a small neighborhood thereof. The uniform bounds for $u_{\NH,\eps}\in\cA^1(X_\NH)$ imply that $\tilde\chi_\NH u_{\NH,\eps}$ is pointwise uniformly bounded by $\rho_\ext$, as are all of its q-derivatives. Since the coefficients of $\chi_\NH(\wh{P_\eps}(\kappa_{\rm C,\eps}\sigma_\eps)-\wh{P_\NH}(\sigma_\eps))$ and $[\wh{P_\eps}(\kappa_{\rm C,\eps}\sigma_\eps),\chi_\NH]$ as q-differential operators are uniformly bounded by $\rho_\NH$, we conclude that $f_\eps$ and all of its q-derivatives are pointwise bounded by $\rho_\ext\rho_\NH=\eps$; therefore,
  \[
    f_\eps \in \eps\bar H_{\qop,\eps}^{s,\alpha_\NH,0}(X_\eps) = \bar H_{\qop,\eps}^{s,\alpha_\NH+1,1}(X_\eps)
  \]
  is uniformly bounded for all $s\in\R$ and $\alpha_\NH\in(-\frac12,\frac12)$. Therefore, the second term on the right in
  \[
    \begin{pmatrix}
      u_\eps \\ c_{1,\eps} \\ c_{2,\eps}
    \end{pmatrix}
    =
    \begin{pmatrix}
      \chi_\NH u_{\NH,\eps} \\ 0 \\ c_{\NH,\eps}
    \end{pmatrix}
    -
    \wt P_\eps^\aug(\sigma_\eps)^{-1}
    \begin{pmatrix}
      f_\eps \\ 0 \\ 0
    \end{pmatrix}
  \]
  is uniformly bounded in $\bar H_{\qop,\eps}^{s,\alpha_\NH+1,1}(X_\eps)\oplus\eps\C\oplus\eps^{\alpha_\NH+\frac12}\C$; the fact that the second summand is $\eps\C$ is the crucial gain here.
\end{proof}

Combining Lemma~\ref{LemmaPf0Bd} with the formula~\eqref{EqPf0Res}, the convergence~\eqref{EqPf0ResConv}, and the uniform bounds $(1-\chi_\NH)u_\eps=\cO(\eps^{1-})$ in $\CI(X_\ext)$, we have proved part~\eqref{ItPf0YState} of Theorem~\ref{ThmPf0Y}.

%%%%%%%%%%%%%%%%%%%%%%%%%%%%%%%%%%%%%%%%%%%%%%%%%%%%%%%%%%%%%%%%%%%%%%
\bibliographystyle{alphaurl}
\newcommand{\etalchar}[1]{$^{#1}$}

%\bibliography{

\begin{thebibliography}{CCD{\etalchar{+}}18b}

\bibitem[Bes20]{BessetRNdSDecay}
Nicolas Besset.
\newblock Decay of the local energy for the charged {K}lein--{G}ordon equation
  in the exterior {D}e {S}itter--{R}eissner--{N}ordstr{\"o}m spacetime.
\newblock In {\em Annales Henri Poincar{\'e}}, volume~21, pages 2433--2484.
  Springer, 2020.

\bibitem[Bes21]{BessetRNdSScattering}
Nicolas Besset.
\newblock Scattering theory for the charged {K}lein--{G}ordon equation in the
  exterior {D}e {S}itter--{R}eissner--{N}ordstr{\"o}m spacetime.
\newblock {\em The Journal of Geometric Analysis}, 31(11):10521--10585, 2021.

\bibitem[BH08]{BonyHaefnerDecay}
Jean-Fran{\c{c}}ois Bony and Dietrich H{\"a}fner.
\newblock Decay and non-decay of the local energy for the wave equation on the
  de {S}itter--{S}chwarzschild metric.
\newblock {\em Communications in Mathematical Physics}, 282(3):697--719, 2008.
\newblock \href {https://doi.org/10.1007/s00220-008-0553-y}
  {\path{doi:10.1007/s00220-008-0553-y}}.

\bibitem[BH21]{BessetHaefnerBomb}
Nicolas Besset and Dietrich H{\"a}fner.
\newblock Existence of exponentially growing finite energy solutions for the
  charged {K}lein--{G}ordon equation on the {D}e {S}itter--{K}err--{N}ewman
  metric.
\newblock {\em Journal of Hyperbolic Differential Equations}, 18(02):293--310,
  2021.
\newblock \href {https://doi.org/10.1142/S0219891621500090}
  {\path{doi:10.1142/S0219891621500090}}.

\bibitem[CCD{\etalchar{+}}18a]{CardosoCostaDestounisHintzJansenSCC}
Vitor Cardoso, Jo\~ao~L. Costa, Kyriakos Destounis, Peter Hintz, and Aron
  Jansen.
\newblock Quasinormal modes and strong cosmic censorship.
\newblock {\em Physical Review Letters}, 120(3):031103, 2018.
\newblock URL: \url{https://doi.org/10.1103/PhysRevLett.120.031103}.

\bibitem[CCD{\etalchar{+}}18b]{CardosoCostaDestounisHintzJansenSCC2}
Vitor Cardoso, Jo\~ao~L. Costa, Kyriakos Destounis, Peter Hintz, and Aron
  Jansen.
\newblock Strong cosmic censorship in charged black-hole spacetimes: still
  subtle.
\newblock {\em Phys. Rev. D}, 98:104007, Nov 2018.
\newblock \href {https://doi.org/10.1103/PhysRevD.98.104007}
  {\path{doi:10.1103/PhysRevD.98.104007}}.

\bibitem[CM22]{CasalsMarinhoSCCRotating}
Marc Casals and C\'assio I.~S. Marinho.
\newblock {Glimpses of violation of strong cosmic censorship in rotating black
  holes}.
\newblock {\em Phys. Rev. D}, 106(4):044060, 2022.
\newblock \href {https://doi.org/10.1103/PhysRevD.106.044060}
  {\path{doi:10.1103/PhysRevD.106.044060}}.

\bibitem[CMT23]{CastroMarianiToldoNearXdS}
Alejandra Castro, Francesca Mariani, and Chiara Toldo.
\newblock {Near-extremal limits of de Sitter black holes}.
\newblock {\em JHEP}, 07:131, 2023.
\newblock \href {https://doi.org/10.1007/JHEP07(2023)131}
  {\path{doi:10.1007/JHEP07(2023)131}}.

\bibitem[Daf05]{DafermosInterior}
Mihalis Dafermos.
\newblock The interior of charged black holes and the problem of uniqueness in
  general relativity.
\newblock {\em Communications on Pure and Applied Mathematics}, 58(4):445--504,
  2005.

\bibitem[DDG24]{DaveyDiasGilSCCKNdS}
Alex Davey, Oscar J.~C. Dias, and David~Sola Gil.
\newblock {Strong Cosmic Censorship in Kerr-Newman-de Sitter}.
\newblock {\em JHEP}, 07:113, 2024.
\newblock \href {https://doi.org/10.1007/JHEP07(2024)113}
  {\path{doi:10.1007/JHEP07(2024)113}}.

\bibitem[DERS18]{DiasEperonReallSantosSCC}
Oscar J.~C. Dias, Felicity~C. Eperon, Harvey~S. Reall, and Jorge~E. Santos.
\newblock {Strong cosmic censorship in de Sitter space}.
\newblock {\em Phys. Rev. D}, 97(10):104060, 2018.
\newblock \href {https://doi.org/10.1103/PhysRevD.97.104060}
  {\path{doi:10.1103/PhysRevD.97.104060}}.

\bibitem[DR09]{DafermosRodnianskiRedShift}
Mihalis Dafermos and Igor Rodnianski.
\newblock The red-shift effect and radiation decay on black hole spacetimes.
\newblock {\em Communications on Pure and Applied Mathematics}, 62(7):859--919,
  2009.

\bibitem[DRS18]{DiasReallSantosSCCrough}
Oscar J.~C. Dias, Harvey~S. Reall, and Jorge~E. Santos.
\newblock Strong cosmic censorship: taking the rough with the smooth.
\newblock {\em Journal of High Energy Physics}, 2018(10):1, 2018.

\bibitem[DRS19]{DiasReallSantosSCCChargeddSBH}
Oscar J.~C. Dias, Harvey~S. Reall, and Jorge~E. Santos.
\newblock {Strong cosmic censorship for charged de Sitter black holes with a
  charged scalar field}.
\newblock {\em Class. Quant. Grav.}, 36(4):045005, 2019.
\newblock \href {https://doi.org/10.1088/1361-6382/aafcf2}
  {\path{doi:10.1088/1361-6382/aafcf2}}.

\bibitem[DSR18]{DafermosShlapentokhRothmanSCC}
Mihalis Dafermos and Yakov Shlapentokh-Rothman.
\newblock Rough initial data and the strength of the blue-shift instability on
  cosmological black holes with ${\Lambda}>0$.
\newblock {\em Classical and Quantum Gravity}, 35(19):195010, 2018.

\bibitem[Dya11a]{DyatlovQNMExtended}
Semyon Dyatlov.
\newblock Exponential energy decay for {K}err--de {S}itter black holes beyond
  event horizons.
\newblock {\em Mathematical Research Letters}, 18(5):1023--1035, 2011.

\bibitem[Dya11b]{DyatlovQNM}
Semyon Dyatlov.
\newblock Quasi-normal modes and exponential energy decay for the {K}err--de
  {S}itter black hole.
\newblock {\em Comm. Math. Phys.}, 306(1):119--163, 2011.

\bibitem[Dya12]{DyatlovAsymptoticDistribution}
Semyon Dyatlov.
\newblock Asymptotic distribution of quasi-normal modes for {K}err--de {S}itter
  black holes.
\newblock {\em Annales Henri Poincar{\'e}}, 13(5):1101--1166, 2012.
\newblock \href {https://doi.org/10.1007/s00023-012-0159-y}
  {\path{doi:10.1007/s00023-012-0159-y}}.

\bibitem[FW24]{FicekWarnickXRNAdSQNM}
Filip Ficek and Claude Warnick.
\newblock {Quasinormal modes of
  Reissner\textendash{}Nordstr\"om\textendash{}AdS: the approach to
  extremality}.
\newblock {\em Class. Quant. Grav.}, 41(8):085011, 2024.
\newblock \href {https://doi.org/10.1088/1361-6382/ad35a0}
  {\path{doi:10.1088/1361-6382/ad35a0}}.

\bibitem[GGH17]{GeorgescuGerardHafnerComplete}
Vladimir Georgescu, Christian G{\'e}rard, and Dietrich H{\"a}fner.
\newblock Asymptotic completeness for superradiant {K}lein--{G}ordon equations
  and applications to the {D}e {S}itter--{K}err metric.
\newblock {\em Journal of the European Mathematical Society}, 19(8):2371--2444,
  2017.

\bibitem[GW21]{GajicWarnickXRNQNM}
Dejan Gajic and Claude Warnick.
\newblock Quasinormal modes in extremal reissner--nordstr{\"o}m spacetimes.
\newblock {\em Communications in Mathematical Physics}, 385(3):1395--1498,
  2021.

\bibitem[GW24]{GajicWarnickKerrQNM}
Dejan Gajic and Claude~M. Warnick.
\newblock Quasinormal modes on {K}err spacetimes.
\newblock {\em Preprint, arXiv:2407.04098}, 2024.

\bibitem[GZ21]{GalkowskiZworskiHypo}
Jeffrey Galkowski and Maciej Zworski.
\newblock Analytic hypoellipticity of {K}eldysh operators.
\newblock {\em Proceedings of the London Mathematical Society},
  123(5):498--516, 2021.
\newblock \href {https://doi.org/https://doi.org/10.1112/plms.12405}
  {\path{doi:https://doi.org/10.1112/plms.12405}}.

\bibitem[Hin]{HintzKdSMS}
Peter Hintz.
\newblock Mode stability and shallow quasinormal modes of {K}err--de {S}itter
  black holes away from extremality.
\newblock {\em Preprint, arXiv:2112.14431. Accepted for publication in \emph{J.
  Eur. Math. Soc.}}

\bibitem[Hin16]{HintzQuasilinearDS}
Peter Hintz.
\newblock Global analysis of quasilinear wave equations on asymptotically de
  {S}itter spaces.
\newblock {\em Annales de l'Institut Fourier}, 66(4):1285--1408, 2016.
\newblock \href {https://doi.org/10.5802/aif.3039}
  {\path{doi:10.5802/aif.3039}}.

\bibitem[Hin18]{HintzKNdSStability}
Peter Hintz.
\newblock {N}on-linear {S}tability of the {K}err--{N}ewman--de {S}itter
  {F}amily of {C}harged {B}lack {H}oles.
\newblock {\em Annals of PDE}, 4(1):11, Apr 2018.
\newblock \href {https://doi.org/10.1007/s40818-018-0047-y}
  {\path{doi:10.1007/s40818-018-0047-y}}.

\bibitem[Hin24a]{HintzGlueLocIII}
Peter Hintz.
\newblock Gluing small black holes along timelike geodesics {III}: construction
  of true solutions and extreme mass ratio mergers.
\newblock {\em Preprint, arXiv:2408.06715}, 2024.

\bibitem[Hin24b]{HintzConicWave}
Peter Hintz.
\newblock Local theory of wave equations with timelike curves of conic
  singularities.
\newblock {\em Preprint, arXiv:2405.10669}, 2024.

\bibitem[Hin25]{HintzMicro}
Peter Hintz.
\newblock Microlocal analysis, 2025.
\newblock URL: \url{https://people.math.ethz.ch/~hintzp/notes/micro.pdf}.

\bibitem[HK24]{HintzKleinQuantumSCC}
Peter Hintz and Christiane Klein.
\newblock Universality of the quantum energy flux at the inner horizon of
  asymptotically de {S}itter black holes.
\newblock {\em Classical and Quantum Gravity}, 41:075006, 2024.

\bibitem[Hod08]{HodQNMNearXKN}
Shahar Hod.
\newblock {Quasinormal resonances of near-extremal Kerr-Newman black holes}.
\newblock {\em Phys. Lett. B}, 666:483--485, 2008.
\newblock \href {https://doi.org/10.1016/j.physletb.2008.08.002}
  {\path{doi:10.1016/j.physletb.2008.08.002}}.

\bibitem[Hod11]{HodQNMNearXKerrKG}
Shahar Hod.
\newblock {Quasinormal resonances of a massive scalar field in a near-extremal
  Kerr black hole spacetime}.
\newblock {\em Phys. Rev. D}, 84:044046, 2011.
\newblock \href {https://doi.org/10.1103/PhysRevD.84.044046}
  {\path{doi:10.1103/PhysRevD.84.044046}}.

\bibitem[Hod12]{HodQNMChargedRN}
Shahar Hod.
\newblock {Quasinormal resonances of a charged scalar field in a charged
  Reissner-Nordstroem black-hole spacetime: A WKB analysis}.
\newblock {\em Phys. Lett. B}, 710:349--351, 2012.
\newblock \href {https://doi.org/10.1016/j.physletb.2012.03.010}
  {\path{doi:10.1016/j.physletb.2012.03.010}}.

\bibitem[Hod17]{HodNearXRNQNM}
Shahar Hod.
\newblock Quasi-bound state resonances of charged massive scalar fields in the
  near-extremal reissner--nordstr{\"o}m black-hole spacetime.
\newblock {\em The European Physical Journal C}, 77(5):351, May 2017.
\newblock \href {https://doi.org/10.1140/epjc/s10052-017-4920-8}
  {\path{doi:10.1140/epjc/s10052-017-4920-8}}.

\bibitem[Hol12]{HolzegelAdS}
Gustav Holzegel.
\newblock Well-posedness for the massive wave equation on asymptotically
  anti-de {S}itter spacetimes.
\newblock {\em Journal of Hyperbolic Differential Equations}, 9(02):239--261,
  2012.

\bibitem[H{\"o}r07]{HormanderAnalysisPDE3}
Lars H{\"o}rmander.
\newblock {\em The analysis of linear partial differential operators. {III}}.
\newblock Classics in Mathematics. Springer, Berlin, 2007.

\bibitem[HV15]{HintzVasySemilinear}
Peter Hintz and Andr{\'a}s Vasy.
\newblock Semilinear wave equations on asymptotically de {S}itter, {K}err--de
  {S}itter and {M}inkowski spacetimes.
\newblock {\em Anal. PDE}, 8(8):1807--1890, 2015.
\newblock \href {https://doi.org/10.2140/apde.2015.8.1807}
  {\path{doi:10.2140/apde.2015.8.1807}}.

\bibitem[HV16]{HintzVasyQuasilinearKdS}
Peter Hintz and Andr{\'a}s Vasy.
\newblock {G}lobal {A}nalysis of {Q}uasilinear {W}ave {E}quations on
  {A}symptotically {K}err--de {S}itter {S}paces.
\newblock {\em International Mathematics Research Notices},
  2016(17):5355--5426, 2016.
\newblock \href {https://doi.org/http://dx.doi.org/10.1093/imrn/rnv311}
  {\path{doi:http://dx.doi.org/10.1093/imrn/rnv311}}.

\bibitem[HV17]{HintzVasyCauchyHorizon}
Peter Hintz and Andr{\'a}s Vasy.
\newblock {A}nalysis of linear waves near the {C}auchy horizon of cosmological
  black holes.
\newblock {\em Journal of Mathematical Physics}, 58(8):081509, 2017.
\newblock \href {https://doi.org/10.1063/1.4996575}
  {\path{doi:10.1063/1.4996575}}.

\bibitem[HV18]{HintzVasyKdSStability}
Peter Hintz and Andr{\'a}s Vasy.
\newblock {T}he global non-linear stability of the {K}err--de {S}itter family
  of black holes.
\newblock {\em Acta mathematica}, 220:1--206, 2018.
\newblock \href {https://doi.org/10.4310/acta.2018.v220.n1.a1}
  {\path{doi:10.4310/acta.2018.v220.n1.a1}}.

\bibitem[HX21]{HintzXiedS}
Peter Hintz and YuQing Xie.
\newblock Quasinormal modes and dual resonant states on de {S}itter space.
\newblock {\em Phys. Rev. D}, 104:064037, Sep 2021.
\newblock URL: \url{https://link.aps.org/doi/10.1103/PhysRevD.104.064037},
  \href {https://doi.org/10.1103/PhysRevD.104.064037}
  {\path{doi:10.1103/PhysRevD.104.064037}}.

\bibitem[HX22]{HintzXieSdS}
Peter Hintz and YuQing Xie.
\newblock Quasinormal modes of small {S}chwarzschild--de {S}itter black holes.
\newblock {\em Journal of Mathematical Physics}, 63(1):011509, 2022.
\newblock \href {https://doi.org/10.1063/5.0062985}
  {\path{doi:10.1063/5.0062985}}.

\bibitem[HZ24]{HitrikZworskiQNM}
Michael Hitrik and Maciej Zworski.
\newblock Overdamped qnm for schwarzschild black holes.
\newblock {\em Preprint arXiv:2406.15924}, 2024.

\bibitem[Ian17]{IantchenkoRNdSDirac}
Alexei Iantchenko.
\newblock {Quasi-normal modes for de Sitter-Reissner-Nordstr\"om Black Holes}.
\newblock {\em Math. Res. Lett.}, 24:83--117, 2017.
\newblock \href {https://doi.org/10.4310/MRL.2017.v24.n1.a5}
  {\path{doi:10.4310/MRL.2017.v24.n1.a5}}.

\bibitem[Ian18]{IantchenkoKNdSDirac}
Alexei Iantchenko.
\newblock {Quasi-normal modes for Dirac fields in the Kerr-Newman-de Sitter
  black holes.}
\newblock {\em Anal. Appl. , Singap.}, 16(4):449--524, 2018.
\newblock \href {https://doi.org/10.1142/S0219530518500057}
  {\path{doi:10.1142/S0219530518500057}}.

\bibitem[Joy22]{JoykuttyXZeroDamped}
Jason Joykutty.
\newblock Existence of zero-damped quasinormal frequencies for nearly extremal
  black holes.
\newblock {\em Annales Henri Poincar{\'e}}, 23(12):4343--4390, Dec 2022.
\newblock \href {https://doi.org/10.1007/s00023-022-01202-z}
  {\path{doi:10.1007/s00023-022-01202-z}}.

\bibitem[Joy23]{JoykuttyPhD}
Jason Joykutty.
\newblock {\em {Q}uasinormal {M}odes of {N}early {E}xtremal {B}lack {H}oles}.
\newblock PhD thesis, University of Cambridge, 2023.
\newblock URL: \url{https://www.repository.cam.ac.uk/handle/1810/365433}, \href
  {https://doi.org/10.17863/CAM.106739} {\path{doi:10.17863/CAM.106739}}.

\bibitem[KMP13]{KimMyungParkNearXRNQNM}
Yong-Wan Kim, Yun~Soo Myung, and Young-Jai Park.
\newblock Quasinormal modes and hidden conformal symmetry in the
  reissner--nordstr{\"o}m black hole.
\newblock {\em The European Physical Journal C}, 73(5):2440, May 2013.
\newblock \href {https://doi.org/10.1140/epjc/s10052-013-2440-8}
  {\path{doi:10.1140/epjc/s10052-013-2440-8}}.

\bibitem[Mel96]{MelroseDiffOnMwc}
Richard~B. Melrose.
\newblock Differential analysis on manifolds with corners.
\newblock {\em Book, in preparation, available online}, 1996.
\newblock URL: \url{https://math.mit.edu/~rbm/daomwcf.ps}.

\bibitem[MTW{\etalchar{+}}18]{MoTianWangZhangZhongSCC}
Yuyu Mo, Yu~Tian, Bin Wang, Hongbao Zhang, and Zhen Zhong.
\newblock {Strong cosmic censorship for the massless charged scalar field in
  the Reissner-Nordstrom\textendash{}de Sitter spacetime}.
\newblock {\em Phys. Rev. D}, 98(12):124025, 2018.
\newblock \href {https://doi.org/10.1103/PhysRevD.98.124025}
  {\path{doi:10.1103/PhysRevD.98.124025}}.

\bibitem[PV21a]{PetersenVasyAnalytic}
Oliver~Lindblad Petersen and Andr{\'a}s Vasy.
\newblock {A}nalyticity of quasinormal modes in the {K}err and {K}err-de
  {S}itter spacetimes.
\newblock {\em Preprint, arXiv:2104.04500}, 2021.
\newblock URL: \url{https://arxiv.org/abs/2104.04500}.

\bibitem[PV21b]{PetersenVasySubextremal}
Oliver~Lindblad Petersen and Andr{\'a}s Vasy.
\newblock Wave equations in the {K}err--de {S}itter spacetime: the full
  subextremal range.
\newblock {\em Preprint, arXiv:2112.0135}, 2021.
\newblock URL: \url{https://arxiv.org/abs/2112.0135}.

\bibitem[Rom92]{RomansEinsteinMaxwellCold}
Larry~James Romans.
\newblock {Supersymmetric, cold and lukewarm black holes in cosmological
  Einstein-Maxwell theory}.
\newblock {\em Nucl. Phys. B}, 383:395--415, 1992.
\newblock \href {https://doi.org/10.1016/0550-3213(92)90684-4}
  {\path{doi:10.1016/0550-3213(92)90684-4}}.

\bibitem[SBZ97]{SaBarretoZworskiResonances}
Ant{\^o}nio S{\'a}~Barreto and Maciej Zworski.
\newblock Distribution of resonances for spherical black holes.
\newblock {\em Mathematical Research Letters}, 4:103--122, 1997.
\newblock URL: \url{https://dx.doi.org/10.4310/MRL.1997.v4.n1.a10}.

\bibitem[SR14]{ShlapentokhRothmanBlackHoleBombs}
Yakov Shlapentokh-Rothman.
\newblock {E}xponentially growing finite energy solutions for the
  {K}lein--{G}ordon equation on sub-extremal {K}err spacetimes.
\newblock {\em Communications in Mathematical Physics}, 329(3):859--891, 2014.

\bibitem[Stu24]{StuckerKerrQNM}
Thomas Stucker.
\newblock Quasinormal modes for the {K}err black hole.
\newblock {\em Preprint, arXiv:2407.04612}, 2024.

\bibitem[Vas13]{VasyMicroKerrdS}
Andr{\'a}s Vasy.
\newblock Microlocal analysis of asymptotically hyperbolic and {K}err--de
  {S}itter spaces (with an appendix by {S}emyon {D}yatlov).
\newblock {\em Invent. Math.}, 194(2):381--513, 2013.
\newblock \href {https://doi.org/10.1007/s00222-012-0446-8}
  {\path{doi:10.1007/s00222-012-0446-8}}.

\bibitem[YZZ{\etalchar{+}}13a]{YangZhangZimmermanNicholsBertiChenQNMNearXKerr}
Huan Yang, Fan Zhang, Aaron Zimmerman, David~A. Nichols, Emanuele Berti, and
  Yanbei Chen.
\newblock {Branching of quasinormal modes for nearly extremal Kerr black
  holes}.
\newblock {\em Phys. Rev. D}, 87(4):041502, 2013.
\newblock \href {https://doi.org/10.1103/PhysRevD.87.041502}
  {\path{doi:10.1103/PhysRevD.87.041502}}.

\bibitem[YZZ{\etalchar{+}}13b]{YangZimmermanZenginogluZhangBertiChenQNMNearXKerr}
Huan Yang, Aaron Zimmerman, An\i{}l Zengino\u{g}lu, Fan Zhang, Emanuele Berti,
  and Yanbei Chen.
\newblock {Quasinormal modes of nearly extremal Kerr spacetimes: spectrum
  bifurcation and power-law ringdown}.
\newblock {\em Phys. Rev. D}, 88(4):044047, 2013.
\newblock \href {https://doi.org/10.1103/PhysRevD.88.044047}
  {\path{doi:10.1103/PhysRevD.88.044047}}.

\bibitem[ZM16]{ZimmermanMarkZeroDampedQNM}
Aaron Zimmerman and Zachary Mark.
\newblock Damped and zero-damped quasinormal modes of charged, nearly extremal
  black holes.
\newblock {\em Phys. Rev. D}, 93:044033, Feb 2016.
\newblock \href {https://doi.org/10.1103/PhysRevD.93.044033}
  {\path{doi:10.1103/PhysRevD.93.044033}}.

\bibitem[Zwo16]{ZworskiRevisitVasy}
Maciej Zworski.
\newblock Resonances for asymptotically hyperbolic manifolds: {V}asy's method
  revisited.
\newblock {\em J. Spectr. Theory}, 2016(6):1087--1114, 2016.

\end{thebibliography}
%/home/peter/Peter/research/bib/math,
%/home/peter/Peter/research/bib/mathcheck,
%/home/peter/Peter/research/bib/phys
%}

\end{document}